\documentclass[12pt,english]{article}
\usepackage{times}
\usepackage[T1]{fontenc}
\usepackage{geometry}
\usepackage{array}
\usepackage{rotating}
\usepackage{graphicx}
\usepackage{setspace}
\usepackage{amssymb}

\makeatletter

\providecommand{\tabularnewline}{\\}

\usepackage{bm}

\usepackage{babel}
\makeatother
\begin{document}

\section*{\noindent Building a Smart \emph{EM} Environment - \emph{AI}-Enhanced
Aperiodic Micro-Scale Design of Passive \emph{EM} Skins}

\noindent \vfill

\noindent G. Oliveri,$^{(1)}$ \emph{Senior Member, IEEE}, F. Zardi,$^{(1)}$
P. Rocca,$^{(1)(2)}$ \emph{Senior Member, IEEE}, M. Salucci,$^{(1)}$
\emph{Member, IEEE}, and A. Massa,$^{(1)(3)(4)}$ \emph{Fellow, IEEE}

\noindent \vfill

\noindent {\footnotesize ~}{\footnotesize \par}

\noindent {\footnotesize $^{(1)}$} \emph{\footnotesize CNIT - \char`\"{}University
of Trento\char`\"{} ELEDIA Research Unit }{\footnotesize \par}

\noindent {\footnotesize Via Sommarive 9, 38123 Trento - Italy}{\footnotesize \par}

\noindent \textit{\emph{\footnotesize E-mail:}} {\footnotesize \{}\emph{\footnotesize giacomo.oliveri,
francesco.zardi, paolo.rocca, marco.salucci, andrea.massa}{\footnotesize \}@}\emph{\footnotesize unitn.it}{\footnotesize \par}

\noindent {\footnotesize Website:} \emph{\footnotesize www.eledia.org/eledia-unitn}{\footnotesize \par}

\noindent {\footnotesize ~}{\footnotesize \par}

\noindent {\footnotesize $^{(2)}$} \emph{\footnotesize ELEDIA Research
Center} {\footnotesize (}\emph{\footnotesize ELEDIA}{\footnotesize @}\emph{\footnotesize XIDIAN}
{\footnotesize - Xidian University)}{\footnotesize \par}

\noindent {\footnotesize P.O. Box 191, No.2 South Tabai Road, 710071
Xi'an, Shaanxi Province - China }{\footnotesize \par}

\noindent {\footnotesize E-mail:} \emph{\footnotesize paolo.rocca@xidian.edu.cn}{\footnotesize \par}

\noindent {\footnotesize Website:} \emph{\footnotesize www.eledia.org/eledia-xidian}{\footnotesize \par}

\noindent {\footnotesize ~}{\footnotesize \par}

\noindent {\footnotesize $^{(3)}$} \emph{\footnotesize ELEDIA Research
Center} {\footnotesize (}\emph{\footnotesize ELEDIA}{\footnotesize @}\emph{\footnotesize UESTC}
{\footnotesize - UESTC)}{\footnotesize \par}

\noindent {\footnotesize School of Electronic Engineering, Chengdu
611731 - China}{\footnotesize \par}

\noindent \textit{\emph{\footnotesize E-mail:}} \emph{\footnotesize andrea.massa@uestc.edu.cn}{\footnotesize \par}

\noindent {\footnotesize Website:} \emph{\footnotesize www.eledia.org/eledia}{\footnotesize -}\emph{\footnotesize uestc}{\footnotesize \par}

\noindent {\footnotesize ~}{\footnotesize \par}

\noindent {\footnotesize $^{(4)}$} \emph{\footnotesize ELEDIA Research
Center} {\footnotesize (}\emph{\footnotesize ELEDIA@TSINGHUA} {\footnotesize -
Tsinghua University)}{\footnotesize \par}

\noindent {\footnotesize 30 Shuangqing Rd, 100084 Haidian, Beijing
- China}{\footnotesize \par}

\noindent {\footnotesize E-mail:} \emph{\footnotesize andrea.massa@tsinghua.edu.cn}{\footnotesize \par}

\noindent {\footnotesize Website:} \emph{\footnotesize www.eledia.org/eledia-tsinghua}{\footnotesize \par}

\noindent \vfill

\noindent \emph{This work has been submitted to the IEEE for possible
publication. Copyright may be transferred without notice, after which
this version may no longer be accessible.}

\noindent \vfill

\newpage
\section*{Building a Smart \emph{EM} Environment - \emph{AI}-Enhanced Aperiodic
Micro-Scale Design of Passive \emph{EM} Skins}

~

~

~

\begin{flushleft}G. Oliveri, F. Zardi, P. Rocca, M. Salucci, and A.
Massa\end{flushleft}

\vfill

\begin{abstract}
\noindent An innovative process for the design of \emph{static passive
smart skins} (\emph{SPSS}s) is proposed to take into account, within
the synthesis, the electromagnetic (\emph{EM}) interactions due to
their finite (macro-level) size and aperiodic (micro-scale) layouts.
Such an approach leverages on the combination of an inverse source
(\emph{IS}) formulation, to define the \emph{SPSS} surface currents,
and of an instance of the \emph{System-by-Design} paradigm, to synthesize
the unit cell (\emph{UC}) descriptors suitable for supporting these
currents. As for this latter step, an enhanced Artificial Intelligence
(\emph{IA})-based digital twin (\emph{DT}) is built to efficiently
and reliably predict the relationships among the \emph{UC}s and the
non-uniform coupling effects arising when the \emph{UC}s are irregularly
assembled to build the corresponding \emph{SPSS}. Towards this end
and unlike state-of-the-art approaches, an aperiodic finite small-scale
model of the \emph{SPSS} is derived to generate the training database
for the \emph{DT} implementation. A set of representative numerical
experiments, dealing with different radiation objectives and smart
skin apertures, is reported to assess the reliability of the conceived
design process and to illustrate the radiation features of the resulting
layouts, validated with accurate full-wave simulations, as well.

\vfill
\end{abstract}
\noindent \textbf{Key words}: Smart Skins; \emph{EM} Holography; Next-Generation
Communications; Iterative Projection Method; System-by-Design; Metasurfaces;
Metamaterials.

\newpage
\section{Introduction and Rationale\label{sec:Introduction}}

\noindent The \emph{Smart Electromagnetic Environment} (\emph{SEE})
vision is at the core of a revolutionary approach currently emerging
in the design and the implementation of next generation wireless cellular
systems \cite{Basar 2019}-\cite{Liaskos 2018}. The academic and
industrial interest in such a transformative paradigm is motivated
by the unprecedented \emph{SEE} potentialities in terms of blind spot
mitigation, coverage improvement, and electromagnetic (\emph{EM})
propagation control enabled by the opportunistic exploitation of the
wireless power already available within the propagation scenario \cite{Di Renzo 2019}\cite{Massa 2021}\cite{Huang 2019}.
Moreover, a minimization of the power consumption of the wireless
infrastructure and a reduction of the overall \emph{EM} pollution
are also assured thanks to the improved wireless efficiency without
the need of installing new base stations \cite{Basar 2019}-\cite{Liaskos 2018}.

\noindent Within this framework, \emph{static passive smart skins}
(\emph{SPSS}s) have been conceived to guarantee the minimum impact
in terms of fabrication, installation, maintenance, costs, and power
consumption \cite{Massa 2021}\cite{Oliveri 2021c}. A \emph{SPSS}
is an artificial passive structure that {}``manipulates'' (i.e.,
reflection/focusing) the incident power radiated by one or more base
stations towards a desired coverage region \cite{Massa 2021}\cite{Oliveri 2021c}.
Towards this end, \emph{SPSS} implementations usually leverage on
the capabilities of passive modulated thin metasurfaces to control
the \emph{EM} reflection, thanks to their micro-scale physical structure
\cite{Oliveri 2021c}\cite{Yang 2019}, rather than using active elements
such as diodes, varactors, phase shifters, or amplifiers \cite{Yang 2019}\cite{Oliveri 2015b}.
Therefore, there are not running costs after the installation and
no electrical nor processing power is required to operate, thus making
\emph{SPSS}s particularly attractive for an inexpensive and fast deployment
in large-scale scenarios \cite{Oliveri 2021c}\@. Indeed, despite
the technological simplicity and the relatively low cost implementation,
\emph{SPSS}s can yield an excellent \emph{EM} propagation control
as already evidenced in several and different {}``\emph{Surface EM}''
applications including lenses, beam splitters, wave polarizers, and
reflect/transmit arrays \cite{Yang 2019}. 

\noindent However, the design of practically feasible \emph{SPSS}s
in \emph{SEE} must fulfil very strict and contrasting requirements
often not occurring in standard \emph{Surface EM} problems \cite{Oliveri 2021c}.
On the one hand, the use of multi-layer/complex shapes or expensive
materials in the unit cells (\emph{UC}s) of the \emph{SPSS} is prevented
for costs and weight reasons. On the other hand, there is the need
of high-performance in terms of both pattern shaping and independence
on the polarization of the incident field. To address such challenges,
a customization of the \emph{System-by-Design} (\emph{SbD}) paradigm
\cite{Massa 2021b}\cite{Oliveri 2015}-\cite{Massa 2014} has been
recently proposed in \cite{Oliveri 2021c}. By formulating the complex
multi-scale \emph{EM} design problem at hand according to the \emph{Generalized
Sheet Transition Condition} (\emph{GSTC}) technique \cite{Yang 2019}\cite{Ricoy 1990}-\cite{Achouri 2015},
the approach in \cite{Oliveri 2021c} combines a phase-only inverse
source (\emph{IS}) method, to find the reference surface currents
induced on the structure with the \emph{Iterative Projection Technique}
(\emph{IPT}) \cite{Rocca 2009b}, together with an integration between
a global optimization algorithm and a \emph{UC} \emph{Digital Twin}
(\emph{UC-DT}) to determine the descriptors of the \emph{SPSS} that
supports the \emph{IPT}-synthesized currents \cite{Oliveri 2021c}.
In order to significantly speed-up the generation of the training
set for learning the \emph{UC-DT}, the \emph{UC} response is modeled
by assuming local periodicity conditions \cite{Oliveri 2021c} as
usually done in the literature on surface electromagnetics \cite{Yang 2019}\cite{Oliveri 2015}\cite{Oliveri 2020}\cite{Oliveri 2017}-\cite{Salucci 2018c}.
Unfortunately, the aperiodicity and the edge effects of finite layouts
introduce a degree of approximation which is very difficult to compensate
in the final full-scale \emph{SPSS}. Such an issue is particularly
challenging when dealing with electrically large apertures since the
optimization of the finite layout becomes computationally unmanageable.
This paper then proposes an innovative approach to face with these
challenges by extending and generalizing the method conceived in \cite{Oliveri 2021c}.
More in detail, the \emph{SPSS} synthesis is still split into two-steps,
but unlike \cite{Oliveri 2021c}, the \emph{UC-DT} of the second step
(i.e., the definition of the micro-scale descriptors of the \emph{UC}
so that the arising modulated metasurface supports the reference currents
synthesized at the first \emph{IS} step) avoids periodic assumptions
and it predicts the behavior of each \emph{UC} by taking into account
a finite set of neighbouring cells as well as different locations
of the same \emph{UC} within the so-called \emph{small-scale model}.
The local susceptibility tensor of this latter is then computed with
a full-wave commercial solver \cite{HFSS 2021} and afterwards exploited
in an Artificial Intelligence (\emph{AI})-driven learning process
to build the {}``Local \emph{UC-DT}''\emph{.}

\noindent To the best of the authors' knowledge, the main innovative
contributions of this work include (\emph{i}) the introduction of
a \emph{Local} \emph{UC-DT} for the reliable design of effective large-scale
\emph{SPSS}s by taking into account the local aperiodicity and the
edge effects of the finite structure, (\emph{ii}) an efficient integration
of such a \emph{DT} within the \emph{SbD}-based \emph{SPSS} synthesis
process, and (\emph{iii}) the full-wave modeling/simulation of the
arising \emph{SPSS} layouts to carefully assess their performance
in realistic operative scenarios.

\noindent The outline of the paper is as follows. The design problem
at hand is formulated in Sect. \ref{sec:Problem-Formulation}, while
the proposed synthesis process is detailed in Sect. \ref{sec:Method}.
A selected set of numerical examples is reported and the synthesis
outcomes are assessed with full-wave simulations in realistic operative
conditions (Sect. \ref{sec:Numerical-Analysis-and}). Finally, come
conclusions follow (Sect. \ref{sec:Conclusions-and-Remarks}).

\section{\noindent \emph{SPSS} Problem Formulation \label{sec:Problem-Formulation} }

\noindent Let us consider the \emph{SEE} scenario in Fig. 1 where
a \emph{SPSS} of $P\times Q$ \emph{UC}s, arranged on the $xy$-plane
according to a regular lattice with spacings $\Delta x$ and $\Delta y$,
is illuminated by a plane wave of \emph{TE/TM} complex components
$E_{TE}^{inc}$/$E_{TM}^{inc}$ that impinges from the angular direction
$\left(\theta^{inc},\,\varphi^{inc}\right)$. The design of the \emph{SPSS}
can be stated as follows

\begin{quotation}
\noindent \textbf{\emph{SPSS}} \textbf{}\textbf{\emph{Design Problem}}
- Given $\left(\theta^{inc},\varphi^{inc}\right)$, $E_{TE}^{inc}$,
and $E_{TM}^{inc}$, find $\mathbf{G}$ such that $\mathcal{O}\left[\mathbf{F}\left(\mathbf{r};\mathbf{G}\right)\right]$
is minimized
\end{quotation}
\noindent where $\mathbf{G}$ is the vector of the descriptors of
the \emph{SPSS} (i.e., the set of $P\times Q$ \emph{UC}s), ($\mathbf{G}\triangleq\left\{ \mathbf{g}_{pq};\, p=1,...,P,\, q=1,...,Q\right\} $),
whose $D$-sized ($p$, $q$)-th ($p=1,...,P$; $q=1,...,Q$) entry,
$\mathbf{g}_{pq}\triangleq\left\{ g_{pq}^{\left(d\right)};\, d=1,...,D\right\} $,
lists the geometrical/physical micro-scale parameters of the corresponding
\emph{UC}. Moreover, $\mathbf{F}\left(\mathbf{r};\mathbf{G}\right)$
is the electric field reflected by the \emph{SPSS} in a far-field
point with \emph{local} coordinates $\mathbf{r}=\left(x,y,z\right)$,
while $\mathcal{O}\left[\mathbf{F}\left(\mathbf{r};\mathbf{G}\right)\right]$
is the implicit form of the macro-scale radiation objectives set on
the far-field pattern, $\mathbf{F}\left(\mathbf{r};\mathbf{G}\right)$,
and whose explicit expression is defined case-by-case according to
the specific applicative context.

\noindent Regardless of $\mathcal{O}\left[\mathbf{F}\left(\mathbf{r}\right)\right]$,
the solution of the \emph{SPSS Design Problem} requires a reliable
approach to compute $\mathbf{F}\left(\mathbf{r};\mathbf{G}\right)$
starting from $\left(\theta^{inc},\varphi^{inc}\right)$, $E_{TE}^{inc}$,
$E_{TM}^{inc}$ , and $\mathbf{G}$. In principle, such a task may
be accomplished by modeling the entire \emph{SPSS} layout and computing
the solution with a full-wave numerical solver. This strategy is practically
prevented because of the unfeasible computational costs related to
the need of performing an expensive computation of $\mathbf{F}\left(\mathbf{r};\mathbf{G}\right)$
for each guess configuration of $\mathbf{G}$. On the other hand,
it is worth pointing out that a careful modelling of the \emph{SPSS}
allows the designer to exploit the intrinsic non-uniqueness of the
underlying \emph{IS} problem (i.e., the deduction of the \emph{SPSS}
surface currents from the radiated far-field pattern) that gives a
greater flexibility in the synthesis of wave manipulation devices
\cite{Salucci 2018}.

\noindent By taking into account these considerations, a different
approach is adopted by exploiting the concept of equivalent surface
currents withing the \emph{GSTC} framework \cite{Oliveri 2021c}\cite{Yang 2019}.
More in detail, the far field pattern reflected by the \emph{SPSS}
is defined as $\mathbf{F}\left(\mathbf{r};\mathbf{G}\right)\triangleq\mathcal{L}\left[\mathbf{J}^{tot}\left(\mathbf{r};\mathbf{G}\right)\right]$,
$\mathcal{L}\left[\cdot\right]$ being the far-field Green's operator,
and it computed as \cite{Oliveri 2021c}\cite{Osipov 2017}\begin{equation}
\mathbf{F}\left(\mathbf{r};\mathbf{G}\right)=\frac{jk_{0}}{4\pi}\frac{\exp\left(-jk_{0}\left|\mathbf{r}\right|\right)}{\left|\mathbf{r}\right|}\int_{\Xi}\left\{ \mathbf{J}^{tot}\left(\widetilde{\mathbf{r}};\mathbf{G}\right)\exp\left(jk_{0}\widehat{\mathbf{r}}\cdot\widetilde{\mathbf{r}}\right)\right\} \mathrm{d}\widetilde{\mathbf{r}}\label{eq:far field}\end{equation}
where $\widehat{\mathbf{r}}\triangleq\frac{\mathbf{r}}{\left|\mathbf{r}\right|}$,
$k_{0}$ is the free-space wavenumber and $\Xi$ is the \emph{SPSS}
surface aperture, while $\mathbf{J}^{tot}$ is the surface equivalent
source induced on $\Xi$ () and computed as the superposition of the
electric, $\mathbf{J}^{e}$, and the magnetic, $\mathbf{J}^{m}$,
effective current components\begin{equation}
\mathbf{J}^{tot}\left(\mathbf{r};\mathbf{G}\right)=\widehat{\mathbf{r}}\times\left[\eta_{0}\widehat{\mathbf{r}}\times\mathbf{J}^{e}\left(\mathbf{r};\mathbf{G}\right)+\mathbf{J}^{m}\left(\mathbf{r};\mathbf{G}\right)\right],\label{eq:surface currents}\end{equation}
$\eta_{0}$ being the free-space impedance ($\eta_{0}\triangleq\sqrt{\frac{\mu_{0}}{\varepsilon_{0}}}$,
$\varepsilon_{0}$ and $\mu_{0}$ being the free-space permittivity
and permeability, respectively).

\noindent According to the \emph{GSTC} technique \cite{Yang 2019}\cite{Achouri 2015},
the current components $\mathbf{J}^{e}$ and $\mathbf{J}^{m}$ are
functions of the electric, $\mathbf{S}^{e}\left(\mathbf{r};\mathbf{G}\right)$,
and the magnetic, $\mathbf{S}^{m}\left(\mathbf{r};\mathbf{G}\right)$,
polarization surface densities so that $\mathbf{J}^{e}\left(\mathbf{r};\mathbf{G}\right)=j\omega\mathbf{S}_{t}^{e}\left(\mathbf{r};\mathbf{G}\right)-\widehat{\mathbf{n}}\times\nabla_{t}S_{n}^{m}\left(\mathbf{r};\mathbf{G}\right)$
and $\mathbf{J}^{m}\left(\mathbf{r};\mathbf{G}\right)=j\omega\mu_{0}\mathbf{S}_{t}^{m}\left(\mathbf{r};\mathbf{G}\right)+\frac{1}{\varepsilon_{0}}\widehat{\mathbf{n}}\times\nabla_{t}S_{n}^{e}\left(\mathbf{r};\mathbf{G}\right)$,
$\widehat{\mathbf{n}}$ being the normal to the smart skin surface
$\Xi$, while $\mathbf{S}_{t}^{e/m}\left(\mathbf{r};\mathbf{G}\right)\triangleq\mathbf{S}^{e/m}\left(\mathbf{r};\mathbf{G}\right)\times\widehat{\mathbf{n}}$
and $S_{n}^{e/m}\left(\mathbf{r};\mathbf{G}\right)\triangleq\mathbf{S}^{e/m}\left(\mathbf{r};\mathbf{G}\right)\cdot\widehat{\mathbf{n}}$.
Moreover, for (sufficiently) symmetric \emph{UC}s, $\mathbf{S}^{e}$
and $\mathbf{S}^{m}$ are given by\begin{equation}
\mathbf{S}^{e}\left(\mathbf{r};\mathbf{G}\right)=\sum_{p=1}^{P}\sum_{q=1}^{Q}\left[\varepsilon_{0}\overline{\overline{\psi}}_{pq}^{e}\left(\mathbf{G}\right)\cdot\mathbf{E}_{pq}^{ave}\left(\mathbf{G}\right)\right]\Psi^{pq}\left(\mathbf{r}\right)\label{eq:expansion surface density E}\end{equation}
and\begin{equation}
\mathbf{S}^{m}\left(\mathbf{r};\mathbf{G}\right)=\sum_{p=1}^{P}\sum_{q=1}^{Q}\left[\overline{\overline{\psi}}_{pq}^{m}\left(\mathbf{G}\right)\cdot\mathbf{H}_{pq}^{ave}\left(\mathbf{G}\right)\right]\Psi^{pq}\left(\mathbf{r}\right)\label{eq:expansion surface density M}\end{equation}
where $\Psi^{pq}\left(\mathbf{r}\right)\triangleq\left\{ 1\, if\,\mathbf{r}\in\Xi_{pq},\,0\, if\,\mathbf{r}\notin\Xi_{pq}\right\} $,
$\Xi_{pq}$ being the support of the ($p$, $q$)-th ($p=1,...,P$;
$q=1,...,Q$) \emph{UC} so that $\Xi=\sum_{p=1}^{P}\sum_{q=1}^{Q}\Xi_{pq}$.

\noindent The computation of (\ref{eq:expansion surface density E})
and (\ref{eq:expansion surface density M}) only requires to determine
the electric/magnetic local surface susceptibility diagonal tensors
$\overline{\overline{\psi}}_{pq}^{e/m}\left(\mathbf{G}\right)$ of
the ($p$, $q$)-th ($p=1,...,P$; $q=1,...,Q$) \emph{UC} since the
surface averaged fields $\bm{\Phi}_{pq}^{ave}\left(\mathbf{G}\right)$
($\bm{\Phi}=\left\{ \mathbf{E},\mathbf{H}\right\} $) are given by
\cite{Oliveri 2021c}\cite{Yang 2019}

\noindent \begin{equation}
\bm{\Phi}_{pq}^{ave}\left(\mathbf{G}\right)\triangleq\frac{1}{2\times\Delta x\times\Delta y}\int_{\Xi_{pq}}\left\{ \mathbf{1}+\overline{\overline{R}}\left[\overline{\overline{\psi}}_{pq}^{e/m}\left(\mathbf{G}\right)\right]\right\} \cdot\bm{\Phi}^{inc}\left(\mathbf{r}\right)d\mathbf{r}\quad\bm{\Phi}=\left\{ \mathbf{E},\mathbf{H}\right\} \label{eq:field average}\end{equation}
where the incident electric,\begin{equation}
\mathbf{E}^{inc}\left(\mathbf{r}\right)\triangleq\left(E_{TE}^{inc}\widehat{\mathbf{e}}_{TE}+E_{TM}^{inc}\widehat{\mathbf{e}}_{TM}\right)\exp\left(-j\mathbf{k}^{inc}\cdot\mathbf{r}\right),\label{eq:incident wave}\end{equation}
and magnetic,\begin{equation}
\mathbf{H}^{inc}\left(\mathbf{r}\right)\triangleq\frac{1}{\eta_{0}k_{0}}\mathbf{k}^{inc}\times\mathbf{E}^{inc}\left(\mathbf{r}\right),\label{eq:}\end{equation}
fields are known quantities, $\mathbf{k}^{inc}$ and $\widehat{\mathbf{e}}_{TE/TM}$
being the incident wave vector ($\mathbf{k}^{inc}$ $\triangleq$
$-$ $k_{0}$ {[}$\sin\left(\theta^{inc}\right)\cos\left(\varphi^{inc}\right)\widehat{\mathbf{x}}$
$+$ $\sin\left(\theta^{inc}\right)\sin\left(\varphi^{inc}\right)\widehat{\mathbf{y}}$
$+$ $\cos\left(\theta^{inc}\right)\widehat{\mathbf{z}}${]}) and
the \emph{TE}/\emph{TM} mode unit vectors ($\widehat{\mathbf{e}}_{TE}=\frac{\mathbf{k}^{inc}\times\widehat{\mathbf{n}}}{\left|\mathbf{k}^{inc}\times\widehat{\mathbf{n}}\right|}$;
$\widehat{\mathbf{e}}_{TM}=\frac{\widehat{\mathbf{e}}_{TE}\times\mathbf{k}^{inc}}{\left|\widehat{\mathbf{e}}_{TE}\times\mathbf{k}^{inc}\right|}$),
respectively, while $\mathbf{1}$ is a diagonal unitary tensor, and
$\overline{\overline{R}}\left[\overline{\overline{\psi}}_{pq}^{e/m}\left(\mathbf{G}\right)\right]$
$\triangleq$ $R^{TE,TE}\left[\overline{\overline{\psi}}_{pq}^{e/m}\left(\mathbf{G}\right)\right]$
$\widehat{\mathbf{e}}_{TE}\widehat{\mathbf{e}}_{TE}$ $+$ $R^{TE,TM}\left[\overline{\overline{\psi}}_{pq}^{e/m}\left(\mathbf{G}\right)\right]$
$\widehat{\mathbf{e}}_{TE}\widehat{\mathbf{e}}_{TM}$ $+$ $R^{TM,TE}\left[\overline{\overline{\psi}}_{pq}^{e/m}\left(\mathbf{G}\right)\right]$
$\widehat{\mathbf{e}}_{TM}\widehat{\mathbf{e}}_{TE}$ $+$ $R^{TM,TM}\left[\overline{\overline{\psi}}_{pq}^{e/m}\left(\mathbf{G}\right)\right]$
$\widehat{\mathbf{e}}_{TM}\widehat{\mathbf{e}}_{TM}$ is the local
reflection tensor, which can be derived from $\overline{\overline{\psi}}_{pq}^{e/m}\left(\mathbf{G}\right)$
\cite{Yang 2019}.

\noindent It is worthwhile to point out that here, unlike \cite{Oliveri 2021c},
the electric/magnetic local surface susceptibility diagonal tensors
$\overline{\overline{\psi}}_{pq}^{e/m}$ of the ($p$, $q$)-th ($p=1,...,P$;
$q=1,...,Q$) \emph{UC} depends on the whole finite structure of the
\emph{SPSS} {[}i.e., $\overline{\overline{\psi}}_{pq}^{e/m}=\overline{\overline{\psi}}_{pq}^{e/m}\left(\mathbf{G}\right)${]}
instead of on the $D$ descriptors of the same ($p$, $q$)-th \emph{UC}
{[}i.e., $\overline{\overline{\psi}}_{pq}^{e/m}=\overline{\overline{\psi}}_{pq}^{e/m}\left(\mathbf{g}_{pq}\right)${]}.
The interested readers should also notice that such an \emph{IS}-based
formulation of the \emph{SPSS} problem {[}i.e., equation (\ref{eq:far field})
through (\ref{eq:expansion surface density E}) and (\ref{eq:expansion surface density M})
with the intermediate step (\ref{eq:surface currents}) for the computation
of the induced surface current, $\mathbf{J}^{tot}\left(\mathbf{r};\mathbf{G}\right)${]}
allows one to exploit the multiplicity of the induced currents \cite{Salucci 2018}
to identify the most proper (i.e., physically-admissible and easy-to-build)
\emph{SPSS} layout for the scenario/objectives at hand. On the other
hand, it is evident the multi-scale nature of the \emph{SPSS} synthesis
problem since the fulfilment of the macro-scale objectives, $\mathcal{O}\left[\mathbf{F}\left(\mathbf{r};\mathbf{G}\right)\right]$,
is obtained by optimizing the micro-scale descriptors of the \emph{UC}s,
$\mathbf{G}$. Last but not least, the design of a \emph{SPSS} intrinsically
features complexity and high-dimensionality since the number of degrees-of-freedom
(\emph{DoF}s), $\mathcal{N}\triangleq P\times Q\times D$, rapidly
increases with the \emph{SPSS} aperture, $\Xi$, and the number of
descriptors of the \emph{UC}, $D$.

\section{\noindent \emph{SPSS} Synthesis Procedure\label{sec:Method}}

\noindent According to the mathematical formulation in Sect. \ref{sec:Problem-Formulation}
and referring to (\ref{eq:far field}), the \emph{SPSS} design is
split into two steps \cite{Oliveri 2021c}, as usually done also in
reflectarray engineering \cite{Oliveri 2020}\cite{Huang 2008}, concerned
with a macro-scale constrained \emph{IS} problem followed by a local
current matching step. More specifically, the \emph{SPSS} layout is
synthesized by first computing the optimal reference current, $\mathbf{J}_{opt}^{tot}$,
that radiates a far field pattern, $\mathbf{F}_{opt}\left(\mathbf{r}\right)\triangleq\mathcal{L}\left\{ \mathbf{J}_{opt}^{tot}\left(\mathbf{r}\right)\right\} $,
minimizing $\mathcal{O}\left[\mathbf{F}_{opt}\left(\mathbf{r}\right)\right]$,
then finding the optimal setup of the \emph{SPSS} descriptors, $\mathbf{G}_{opt}$,
such that\begin{equation}
\mathbf{G}_{opt}=\arg\left[\min_{\mathbf{G}}\left\{ \Delta\left(\mathbf{G}\right)\right\} \right],\label{eq:cost function G}\end{equation}
where $\Delta\left(\mathbf{G}\right)$ is the surface current mismatch
($\Delta\left(\mathbf{G}\right)\triangleq\left\Vert \mathbf{J}^{tot}\left(\mathbf{r};\mathbf{G}\right)-\mathbf{J}_{opt}^{tot}\left(\mathbf{r}\right)\right\Vert _{\Xi}^{2}$,
$\left\Vert \cdot\right\Vert _{\Xi}$ being the $\ell_{2}$-norm over
$\Xi$ given by $\left\Vert .\right\Vert _{\Xi}\triangleq\sqrt{\int_{\Xi}\left|.\right|^{2}\mathrm{d}\mathbf{r}}$).

\noindent The first step is carried out as in \cite{Oliveri 2021c}
by choosing an \emph{IPT}-based approach \cite{Rocca 2009b} to find
the optimal \emph{SPSS} currents. Towards this end, the {}``pattern''
feasibility space $\mathcal{W}\left\{ \mathbf{F}\left(\mathbf{r}\right)\right\} $
and the {}``current'' feasibility space $\mathcal{W}\left\{ \mathbf{J}^{tot}\left(\mathbf{r}\right)\right\} $
as well as the mutual projection operators are defined. The former,
$\mathcal{W}\left\{ \mathbf{F}\left(\mathbf{r}\right)\right\} $,
is cast into the following mask-matching form\begin{equation}
\mathcal{W}\left\{ \mathbf{F}\left(\mathbf{r}\right)\right\} \triangleq\left\{ \mathbf{F}\left(\mathbf{r}\right):\,\left|\mathbf{F}\left(\mathbf{r}\right)\right|^{2}\geq\mathcal{M}\left(\mathbf{r}\right);\,\mathbf{r}\in\Theta\right\} \label{eq:feasibility E}\end{equation}
where $\mathcal{M}\left(\mathbf{r}\right)$ is the user-defined lower
footprint power mask in the coverage region $\Theta$. Moreover, let
us describe the feasibility space of the {}``current'' as $\mathcal{W}\left\{ \mathbf{J}^{tot}\left(\mathbf{r}\right)\right\} $
$\triangleq$ \{$\mathbf{J}^{tot}\left(\mathbf{r}\right)$ : $\mathbf{J}^{tot}\left(\mathbf{r}\right)$
$=$ $C$ $\exp\left[j\chi\left(\mathbf{r}\right)\right]$; $\mathbf{r}\in\Xi$\},
$C$ and $\chi\left(\mathbf{r}\right)$ being the magnitude constant
and the locally-controlled phase distribution of the surface current,
respectively.

\noindent The \emph{IPT} synthesis process iteratively updates the
$i$-th ($i=1,...,I$) guess current, $\mathbf{J}_{i}^{tot}$ ($i$
being the iteration index), starting from a random distribution, $\mathbf{J}_{0}^{tot}$,
and alternatively computing the $i$-th ($i=1,...,I$) {}``projected
pattern'', $\widetilde{\mathbf{F}}_{i}$ ($\mathbf{F}_{i}\left(\mathbf{r}\right)\triangleq\mathcal{L}\left[\mathbf{J}_{i}^{tot}\left(\mathbf{r}\right)\right]$),\begin{equation}
\widetilde{\mathbf{F}}_{i}\left(\mathbf{r}\right)=\left\{ \begin{array}{ll}
\sqrt{\mathcal{M}\left(\mathbf{r}\right)} & \textnormal{if}\,\,\,\left|\mathbf{F}_{i}\left(\mathbf{r}\right)\right|^{2}\leq\mathcal{M}\left(\mathbf{r}\right)\\
\mathbf{F}_{i}\left(\mathbf{r}\right) & \mathrm{otherwise}\end{array}\right.\label{eq:}\end{equation}
and the {}``projected current'' at the next ($i+1$)-th iteration\begin{equation}
\mathbf{J}_{i+1}^{tot}\left(\mathbf{r}\right)=\frac{\mathbf{J}_{i}^{MN}\left(\mathbf{r}\right)}{\left\Vert \mathbf{J}_{i}^{MN}\left(\mathbf{r}\right)\right\Vert }\label{eq:}\end{equation}
where $\mathbf{J}_{i}^{MN}$ is the minimum norm solution of the integral
equation (\ref{eq:far field}) (i.e., $\mathbf{J}_{i}^{MN}\left(\mathbf{r}\right)$
$\triangleq$ $\mathcal{L}_{MN}^{-1}$ $\left\{ \widetilde{\mathbf{F}}_{i}\left(\mathbf{r}\right)\right\} $),
which is determined with the truncated singular value decomposition
\cite{Salucci 2018}.

\noindent The process is stopped when either the maximum number of
\emph{IPT} iterations has been reached (i.e., $i=I$) or the value
of the \emph{pattern matching index}, $\Gamma_{i}$ ($i<I$), given
by\begin{equation}
\Gamma_{i}=\frac{\left\Vert \widetilde{\mathbf{F}}_{i}\left(\mathbf{r}\right)-\mathbf{F}_{i}\left(\mathbf{r}\right)\right\Vert _{\Theta}}{\left\Vert \mathbf{F}_{i}\left(\mathbf{r}\right)\right\Vert _{\Theta}},\label{eq:pattern convergence}\end{equation}
is smaller than a user-defined convergence threshold, $\gamma$, and
the estimate of the surface current distribution is outputted by setting
$\mathbf{J}_{opt}^{tot}\left(\mathbf{r}\right)=\mathbf{J}_{i}^{tot}\left(\mathbf{r}\right)$,
$\mathbf{r}\in\Xi$.

\noindent As for the second step towards the \emph{SPSS} synthesis,
solving (\ref{eq:cost function G}) to find $\mathbf{G}_{opt}$ is
a computationally challenging optimization task owing to the number
and the heterogeneity of the \emph{DoF}s and the cost function at
hand, $\Delta\left(\mathbf{G}\right)$. According to the guidelines
in the global optimization literature (see for instance \cite{Rocca 2009}\cite{Massa 2021b}
and the reference therein), an iterative \emph{SbD}-based strategy
is applied by generating a succession of $N$ trial sets, \{$\mathcal{P}^{\left(n\right)}$;
$n=1,...,N$\} ($n$ being the iteration index during the optimization),
with the \emph{Particle Swarm} mechanism \cite{Rocca 2009} and computing
the current mismatch, $\Delta\left(\left.\mathbf{G}^{a}\right|^{\left(n\right)}\right)$,
for each $a$-th ($a=1,...,A$) guess \emph{SPSS} layout of each $n$-th
($n=1,...,N$) iteration/swarm being $\mathcal{P}^{\left(n\right)}$
$=$ \{$\left.\mathbf{G}^{a}\right|^{\left(n\right)}$; $a=1,...,A$\},
while $A$ is the swarm size. The calculation of $\Delta\left(\left.\mathbf{G}^{a}\right|^{\left(n\right)}\right)$
requires the knowledge of the surface current $\mathbf{J}^{tot}$
in correspondence with the $a$-th ($a=1,...,A$) layout of the $n$-th
($n=1,...,N$) swarm, $\left.\mathbf{G}^{a}\right|^{\left(n\right)}$
{[}i.e., $\mathbf{J}^{tot}\left(\left.\mathbf{G}^{a}\right|^{\left(n\right)}\right)${]},
which is accomplished once the electric and the magnetic polarization
surface densities, $\mathbf{S}^{e/m}\left(\mathbf{r};\left.\mathbf{G}^{a}\right|^{\left(n\right)}\right)$,
are determined through (\ref{eq:expansion surface density E}) and
(\ref{eq:expansion surface density M}). Towards this end, the key
task is to deduce the local surface susceptibility tensors of the
($p$, $q$)-th ($p=1,...,P$; $q=1,...,Q$) \emph{UC}, $\overline{\overline{\psi}}_{pq}^{e/m}$,
but depending on the whole \emph{SPSS} arrangement {[}i.e., $\overline{\overline{\psi}}_{pq}^{e/m}=\overline{\overline{\psi}}_{pq}^{e/m}\left(\left.\mathbf{G}^{a}\right|^{\left(n\right)}\right)${]}.
Because of the computational costs, the use of a full-wave simulator
is impossible since it would imply the expensive \emph{EM} modelling
of the behavior of $P\times Q\times A\times N$ full-size \emph{SPSS}s.
To reduce the computational burden, the \emph{EM} response of the
\emph{SPSS} has been emulated with a \emph{UC-DT}, which has been
trained by exploiting a single-cell full-wave model under the assumption
of periodic boundary conditions \cite{Oliveri 2021c}. However, neglecting
the aperiodicity and the edge effects of the actual finite-size \emph{SPSS}
introduces a non-negligible degree of approximation. In order to avoid
such a drawback and unlike \cite{Oliveri 2021c}, a \emph{Local UC-DT}
is defined to substitute the actual $\overline{\overline{\psi}}_{pq}^{e/m}\left(\mathbf{G}\right)$
with its surrogate $\overline{\overline{\zeta}}_{pq}^{e/m}\left(\mathbf{G}\right)$
learned \emph{offline} from a \emph{small-scale model} of the \emph{SPSS}.
More specifically, such a \emph{Local UC-DT} is implemented according
to the following procedure:

\begin{itemize}
\item \noindent \textbf{Small-Scale} \textbf{\emph{SPSS}} \textbf{Modeling}
- An aperiodic $P'\times Q'$ (with $P'\ll P$, $Q'\ll Q$) \emph{small-scale}
\emph{SPSS} layout, which is described by the reduced vector of descriptors
$\mathbf{G}^{'}$ ($\mathbf{G}^{'}\triangleq\left\{ \mathbf{g}_{p'q'};\, p'=1,...,P';\, q'=1,...,Q'\right\} $)
whose ($p'$, $q'$)-th entry is still of size $D$ (i.e., $\mathbf{g}_{p'q'}\triangleq\left\{ g_{p'q'}^{\left(d\right)};\, d=1,...,D\right\} $,
is built;
\item \textbf{Small-Scale} \textbf{\emph{SPSS}} \textbf{Training Set Definition}
- A representative set of $B$ variations of the $P'\times Q'$ $D$-size
small-scale descriptors is considered to derive $B$ small-scale \emph{SPSS}
layouts, \{$\mathbf{G}_{b}^{'}$; $b=1,..,B$\}, then the behavior
of each $b$-th ($b=1,...,B$) layout is full-wave simulated to predict
its \emph{EM} behavior by extracting the corresponding susceptibility
tensors, \{$\overline{\overline{\psi}}_{p'q'}^{e/m}\left(\mathbf{G}_{b}^{'}\right)$;
$p'=1,...,P'$; $q'=1,...,Q'$\}, from the local scattering parameters
\cite{Yang 2019};
\item \textbf{Small-Scale} \textbf{\emph{SPSS}} \textbf{}\textbf{\emph{AI}}\textbf{-based
Surrogate Model Creation} - The \emph{AI}-based \emph{Local UC-DT}
surrogate model $\overline{\overline{\zeta}}_{p'q'}^{e/m}\left(\mathbf{G}^{'}\right)$
($p'=1,...,P'$; $q'=1,...,Q'$) is created starting from the small-scale
$B$-size \emph{SPSS} training set, \{{[}$\mathbf{G}_{b}^{'}$, $\overline{\overline{\psi}}_{p'q'}^{e/m}\left(\mathbf{\mathbf{G}_{b}^{'}}\right)${]};
$b=1,..,B$\}, by means of a statistical learning approach based on
the \emph{Ordinary Kriging} (\emph{OK}) method \cite{Oliveri 2020}\cite{Salucci 2018c}.
More in detail, the value of $\overline{\overline{\zeta}}_{p'q'}^{e/m}\left(\mathbf{G}^{'}\right)$
($p'=1,...,P'$; $q'=1,...,Q'$) is predicted as follows\begin{equation}
\begin{array}{l}
\overline{\overline{\zeta}}_{p'q'}^{e/m}\left(\mathbf{G}^{'}\right)=\wp_{3}^{reg}\left\{ \psi_{p'q'}^{e/m}\left(\mathbf{G}_{b}^{'}\right);\,\mathbf{c}\right\} \Im_{3\times3}^{unit}+\left[\wp_{B}^{corr}\left\{ \mathbf{\mathbf{G}}^{'};\,\mathbf{c}\right\} \right]^{\dagger}\left[\Im_{B\times B}^{corr}\left\{ \mathbf{G}_{b}^{'};\,\mathbf{c}\right\} \right]^{-1}\\
\times\left(\Im_{B\times3}^{train}\left\{ \psi_{p'q'}^{e/m}\left(\mathbf{G}_{b}^{'}\right)\right\} -\wp_{B}^{unit}\wp_{3}^{\left(4\right)}\left\{ \psi_{p'q'}^{e/m}\left(\mathbf{G}_{b}^{'}\right);\,\mathbf{c}\right\} \left\{ \psi_{p'q'}^{e/m}\left(\mathbf{G}_{b}^{'}\right);\,\mathbf{c}\right\} \right)\Im_{3\times3}^{unit}\end{array}\label{eq:}\end{equation}
where\begin{equation}
\begin{array}{c}
\wp_{3}^{reg}\left\{ \psi_{p'q'}^{e/m}\left(\mathbf{G}_{b}^{'}\right);\,\mathbf{c}\right\} \triangleq\left(\left[\wp_{B}^{unit}\right]^{\dagger}\left[\Im_{B\times B}^{corr}\left\{ \mathbf{G}_{b}^{'};\,\mathbf{c}\right\} \right]^{-1}\wp_{B}^{unit}\right)^{-1}\\
\left[\wp_{B}^{unit}\right]^{\dagger}\left[\Im_{B\times B}^{corr}\left\{ \mathbf{G}_{b}^{'};\,\mathbf{c}\right\} \right]^{-1}\Im_{B\times3}^{train}\left\{ \psi_{p'q'}^{e/m}\left(\mathbf{G}_{b}^{'}\right)\right\} \end{array}\label{eq:parameter vector}\end{equation}
is the \emph{}regression parameter matrix,%
\footnote{\noindent For notation simplicity, the symbol $\Im_{u\times v}$ is
used to indicate a matrix of $u$ rows and $v$ columns, while $\wp_{u}$
identifies a vector of $u$ rows.%
} $\wp_{B}^{corr}\left\{ \mathbf{\mathbf{G}}^{'};\,\mathbf{c}\right\} $
is the $B$-size exponential correlation vector ($\wp_{B}^{corr}\left\{ \mathbf{\mathbf{G}}^{'};\,\mathbf{c}\right\} $
$\triangleq$ \{$Z_{b}\left(\mathbf{G}^{'};\mathbf{c}\right)$ $=$
$\exp$ ($-$ $\sum_{p'=1}^{P'}$ $\sum_{q'=1}^{Q'}$ $\sum_{d=1}^{D}$
$c_{n}$ $\left|g_{p'q'}^{\left(d\right)}-\left.g_{p'q'}^{\left(d\right)}\right\rfloor _{b}\right|$);
$b=1,...,B$\} being $n=p'+P'\times\left(q'-1\right)+P'\times Q'\times\left(d-1\right)$),
$\Im_{B\times B}^{corr}\left\{ \mathbf{G}_{b}^{'};\,\mathbf{c}\right\} $
is a $B\times B$ matrix whose $b$-th ($b=1,...,B$) column is equal
to $\wp_{B}^{corr}\left\{ \mathbf{\mathbf{G}}^{'};\,\mathbf{c}\right\} $,
while $\mathbf{c}\triangleq\left\{ c_{n};\, n=1,...,\mathcal{N}\right\} $
is the set of $\mathcal{N}$ control parameters, which are automatically
optimized during the \emph{OK} self-tuning \cite{Salucci 2018c}.
Moreover, $\Im_{B\times3}^{train}\left\{ \psi_{p'q'}^{e/m}\left(\mathbf{G}_{b}^{'}\right)\right\} $
is the training-set matrix whose ($w$, $b$)-th ($w=x,y,z$; $b=1,...,B$)
entry is equal to the ($w$$w$)-th component of the local surface
susceptibility diagonal tensor of the ($p'$, $q'$)-th \emph{UC}
of the $b$-th small-scale \emph{SPSS} layout, $\mathbf{G}_{b}^{'}$,
(i.e., $\Im_{B\times3}^{train}\left\{ \psi_{p'q'}^{e/m}\left(\mathbf{G}_{b}^{'}\right)\right\} $
$\triangleq$ \{$\left.\psi_{p'q'}^{e/m}\left(\mathbf{G}_{b}^{'}\right)\right|_{ww}$;
$w=x,y,z$; $b=1,...,B$\}) being \cite{Oliveri 2021c}\cite{Yang 2019}\begin{equation}
\overline{\overline{\psi}}_{p'q'}^{e/m}\left(\mathbf{G}_{b}^{'}\right)\triangleq\sum_{w=x,y,z}\left.\psi_{p'q'}^{e/m}\left(\mathbf{G}_{b}^{'}\right)\right|_{ww}\widehat{w}\widehat{w},\label{eq:suscept electr}\end{equation}
while $\wp_{B}^{unit}$ is a unitary column vector of length $B$
and $\cdot^{\dagger}$ stands for the transpose operator;
\item \textbf{Small-Scale to Full-Scale} \textbf{\emph{SPSS}} \textbf{Prediction
Mapping} - The actual local surface susceptibility tensors of the
full-size \emph{SPSS}, \{$\overline{\overline{\psi}}_{pq}^{e/m}\left(\mathbf{G}\right)$;
$p=1,...,P$; $q=1,...,Q$\}, are then estimated by deriving \{$\overline{\overline{\zeta}}_{pq}^{e/m}\left(\mathbf{G}\right)$;
$p=1,...,P$; $q=1,...,Q$\} from the \emph{OK} small-scale \emph{SPSS}
predictions, \{$\overline{\overline{\zeta}}_{p'q'}^{e/m}\left(\mathbf{G}^{'}\right)$;
$p'=1,...,P'$; $q'=1,...,Q'$\} according to the mapping scheme pictorially
illustrated in Fig. 2 and performed according to the rules in \emph{Appendix}.
\end{itemize}
\noindent Once the \emph{Local UC-DT} has been implemented \emph{offline}
and it is available, the \emph{SbD}-based \emph{SPSS} synthesis process
can be efficiently executed, even multiple, times without recurring
to additional full-wave simulations of either a part or the whole
of the \emph{UC}s arrangement. Moreover, no further training is required
even if the synthesis objectives $\mathcal{O}\left[\mathbf{F}\left(\mathbf{r};\mathbf{G}\right)\right]$
are changed after the \emph{Local UC-DT} was created. Those features
assure to the arising design method high-scalability and re-usability
properties.

\section{\noindent Numerical Results\label{sec:Numerical-Analysis-and}}

\noindent In this section, a selected set of examples from a wide
numerical analysis is reported to illustrate the proposed \emph{AI}-enhanced
aperiodic micro-scale design procedure for the synthesis of \emph{SPSS}s
working in real scenarios as well as to assess the reliability of
the synthesized layouts in operative conditions through full-wave
simulations. More specifically, the numerical validation has been
carried out by considering various supports of the \emph{SPSS} and
different radiation objectives, $\mathcal{O}\left[\mathbf{F}\left(\mathbf{r};\mathbf{G}\right)\right]$.
As for this latter, the \emph{Pencil Coverage} and the \emph{Shaped
Coverage} cases have been chosen since as largely representative of
several real applications%
\footnote{\noindent It is worth remarking that the solution process is independent
on $\mathcal{O}\left[\mathbf{F}\left(\mathbf{r};\mathbf{G}\right)\right]$.%
}. In the \emph{Pencil Coverage} case, the power reflected by the \emph{SPSS}
is focused towards a desired (anomalous non-Snell) target direction,
$\left(\theta^{T},\varphi^{T}\right)$, which means setting $\mathcal{O}\left[\mathbf{F}\left(\mathbf{r};\mathbf{G}\right)\right]=\mathcal{O}^{pen}\left[\mathbf{F}\left(\mathbf{r};\mathbf{G}\right)\right]$
{[}$\mathcal{O}^{pen}\left[\mathbf{F}\left(\mathbf{r};\mathbf{G}\right)\right]\triangleq\frac{1}{\left|\mathbf{F}\left(r^{T},\theta^{T},\varphi^{T};\mathbf{G}\right)\right|^{2}}$
where $r^{T}$ is an arbitrary (far-field) distance, while $\left|\cdot\right|$
stands for the vector magnitude) and $\Theta=\left(r^{T},\theta^{T},\varphi^{T}\right)$
in (\ref{eq:feasibility E}). Otherwise, the \emph{SPSS} of the \emph{Shaped
Coverage} case maximizes the reflected power in the user-defined footprint/coverage
area $\Theta$ (Fig. 1), thus the synthesis objective is set to $\mathcal{O}\left[\mathbf{F}\left(\mathbf{r};\mathbf{G}\right)\right]=\mathcal{O}^{sha}\left[\mathbf{F}\left(\mathbf{r};\mathbf{G}\right)\right]$
($\mathcal{O}^{sha}\left[\mathbf{F}\left(\mathbf{r};\mathbf{G}\right)\right]\triangleq\frac{1}{\int_{\Omega}\left|\mathbf{F}\left(\mathbf{r}\right);\mathbf{G}\right|^{2}d\mathbf{r}}$).
Finally, $\mathcal{M}\left(\mathbf{r}\right)=0$ {[}dB{]} has been
set in both cases.

\noindent Concerning the benchmark \emph{SEE} scenario, a base station
has been assumed to illuminate the \emph{SPSS} at $f=3.5$ {[}GHz{]}
(i.e., sub-6GHz $n78$ band \cite{Ciydem 2020}) with a plane wave
featuring circular polarization (i.e., $E_{TE}^{inc}=1$ and $E_{TM}^{inc}=j$)
and impinging from broadside (i.e., $\left(\theta^{inc},\varphi^{inc}\right)=\left(0,0\right)$
{[}deg{]} $\to$ $\widehat{\mathbf{e}}_{TE}=\widehat{\mathbf{y}}$
and $\widehat{\mathbf{e}}_{TM}=\widehat{\mathbf{x}}$). Moreover,
the calibration parameters of the synthesis procedure have been set
following the guidelines in \cite{Oliveri 2021c}\cite{Oliveri 2020}:
$B=2\times10^{4}$, $\gamma=10^{-4}$, $I=10^{3}$, $A=10$, and $N=10^{4}$.
Finally, the full-wave simulations have been carried out with \emph{Ansys
HFSS} \cite{HFSS 2021}.

\noindent In order to illustrate the proposed \emph{SPSS} design method,
let start with the offline process of building the \emph{Local UC-DT},
which does not depend on the synthesis objective (i.e., $\mathcal{O}\left[\mathbf{F}\left(\mathbf{r};\mathbf{G}\right)\right]$),
to be defined for the (second) step of the \emph{SbD}-based \emph{SPSS}
layout synthesis and kept in all numerical tests. Accordingly, a \emph{UC}
consisting of a metallic square patch of side $\ell$ ($D=1$) printed
on a \emph{Rogers RT/duroid 5870} laminate with thickness $\tau=3.175\times10^{-3}$
{[}m{]} {[}Fig. 3(\emph{a}){]} has been selected as the benchmark
\emph{SPSS} element. It can be noticed {[}Figs. 3(\emph{c})-3(\emph{d}){]}
that the phase coverage of such a \emph{UC}, which presents a structural
resonance centered around $\ell=2.5\times10^{-2}$ {[}m{]}, is not
complete {[}i.e., $\approx330$ {[}deg{]} at $f=3.5$ GHz - Fig. 3(\emph{c}){]}
as a consequence of the simple shape and the single-layer design.
The choice of a basic/elementary \emph{UC} has been made to assess
the effectiveness of the proposed approach without the bias of a high-performance
\emph{UC}. Then, a \emph{small-scale} \emph{SPSS} layout with $P'\times Q'=5\times5$
\emph{UC}s {[}$\left.g_{p'q'}^{\left(d\right)}\right\rfloor _{d=1}=\ell_{p'q'}$,
Fig. 3(\emph{b}){]}, arranged on a uniform lattice with periodicity
$\Delta x=\Delta y=4.28\times10^{-2}$ {[}m{]}, has been defined ({}``\emph{Small-Scale}
SPSS \emph{Modeling}'') and simulated in a full-wave fashion taking
into account the copper thickness ($\nu=35\times10^{-6}$ {[}m{]}),
as well. The $B$-sized \emph{OK} training dataset has been built
by varying the $P'\times Q'$ patch sides {[}i.e., $\ell_{p'q'}$
($p'=1,...,P'$; $q'=1,...,Q'$){]} of the small-scale \emph{SPSS}
layout ({}``\emph{Small-Scale} SPSS \emph{Training Set Definition}'').
Starting from this training set, an \emph{OK}-based \textbf{}surrogate
of the Small-Scale \emph{SPSS} model has been built by deriving $\overline{\overline{\zeta}}_{p'q'}^{e/m}\left(\mathbf{G}^{'}\right)$
($p'=1,...,P'$; $q'=1,...,Q'$). Some insights on the accuracy of
the arising predictor can be inferred by the plots of the phase {[}Fig.
3(\emph{c}){]} and of the magnitude {[}Fig. 3(\emph{d}){]} of the
average {}``\emph{TE-TE}'' reflection coefficient of a representative
element {[}i.e., the central one located at ($p'$, $q'$) $=$ ($\frac{P'+1}{2}$,
$\frac{Q'+1}{2}$){]} of the small-scale \emph{}SPSS \emph{}layout\begin{equation}
\widetilde{R}_{p'q'}^{TE,TE}=\frac{1}{B}\sum_{b=1}^{B}R^{TE,TE}\left[\overline{\overline{\psi}}_{p'q'}^{e/m}\left(\mathbf{G}_{b}^{'}\right)\right]_{p'=\frac{P'+1}{2}}^{q'=\frac{Q'+1}{2}}.\label{eq:average R}\end{equation}
As it can be observed, the predicted curve (i.e., $\widetilde{R}_{p'q'}^{TE,TE}$
$=$ $\frac{1}{B}\sum_{b=1}^{B}R^{TE,TE}\left[\overline{\overline{\zeta}}_{p'q'}^{e/m}\left(\mathbf{G}_{b}^{'}\right)\right]_{p'=\frac{P'+1}{2}}^{q'=\frac{Q'+1}{2}}$)
faithfully matches the actual one passing throughout the {}``training''
samples as expected from the \emph{OK} theory.

\noindent Once the {}``Small-Scale \emph{SPSS} \emph{AI}-based Surrogate
Model'' has been created, it is then possible to deal with the \emph{SEE}
problem at hand, which is user-defined by setting the macro-scale
radiation objectives (i.e., $\mathcal{O}\left[\mathbf{F}\left(\mathbf{r};\mathbf{G}\right)\right]$)
as well as the \emph{SPSS} support and position in the global reference
system of coordinates ($x_{glob}$, $y_{glob}$, $z_{glob}$) (Fig.
1).

\noindent The first test case of the numerical assessment is concerned
with the synthesis of a \textbf{}\emph{SPSS} installed at $H=5$ {[}m{]}
over the ground (Fig. 1) with $P\times Q=30\times30$ \emph{UC}s (i.e.,
$\Xi\approx1.28\times1.28$ {[}$\mathrm{m}^{2}${]}) to yield a \emph{Pencil
Coverage} with focusing direction $\left(\theta^{T},\varphi^{T}\right)=\left(50,-8\right)$
{[}deg{]}, which corresponds to a footprint spot at $x_{glob}=-35.7$
{[}m{]}, $y_{glob}=30.14$ {[}m{]}. By following the procedure described
in Sect. \ref{sec:Method}, the first step has been carried out with
the \emph{IPT} {[}Fig. 4(\emph{a}){]} to compute the optimal reference
current, $\mathbf{J}_{opt}^{tot}$ {[}Fig. 4(\emph{b})%
\footnote{\noindent For symmetry reasons, the two components of the surface
current are identical. Thus, only the $x$-component, $J_{x}^{tot}\left(\mathbf{r}\right)$,
will be shown.%
}{]}, that radiates the far field pattern $\mathbf{F}_{opt}\left(\mathbf{r}\right)\triangleq\mathcal{L}\left\{ \mathbf{J}_{opt}^{tot}\left(\mathbf{r}\right)\right\} $
{[}Fig. 4(\emph{c}){]} by minimizing $\mathcal{O}^{pen}\left[\mathbf{F}_{opt}\left(\mathbf{r}\right)\right]$.
Figure 4(\emph{a}) shows that there is a quick reduction of the value
of the pattern matching index $\Gamma_{i}$ (\ref{eq:pattern convergence})
($i=1,...,I$), which confirms the effectiveness of the \emph{IPT}
when applied to fulfil macro-scale objectives (i.e., reflection performance)
by optimizing micro-scale \emph{DoF}s (i.e., the surface currents).
It is also worth pointing out that the synthesized current distribution
only depends on the project targets and not on the \emph{UC} geometry,
hence the same surface current can be kept also varying the \emph{UC}
materials, shapes, layers, etc .... In the second step of the \emph{SPSS}
synthesis, the layout in Fig. 5(\emph{a}) has been derived from the
\emph{SbD}-driven optimization of the \emph{UC} descriptors to match
the reference current profile in Fig. 4(\emph{b}). As a matter of
fact, the arising surface current, $\mathbf{J}_{N}^{tot}$ {[}Fig.
5(\emph{b}){]}, turns out to be very close to the reference one, the
same similarity being present between the radiated far-field patterns
in the ($u$, $v$)-domain ($u\triangleq\sin\theta\cos\varphi$ and
$v\triangleq\sin\theta\sin\varphi$) {[}Fig. 5(\emph{c}) vs. Fig.
4(\emph{c}){]}. Both outcomes confirm the effectiveness of the optimization
process that allows the designer to analytically synthesize a large
\emph{SPSS} (i.e., $30\times30$ \emph{UCs}) with a careful control
of the resulting reflection/focusing properties. As for this latter
item, it turns out that there is the need of a non-uniform arrangement
of the \emph{UC}s {[}Fig. 5(\emph{a}){]} to afford a {}``double anomalous''
(i.e., $\theta^{T}\neq\theta^{Snell}$ and $\varphi^{T}\neq\varphi^{Snell}$
being $\theta^{Snell}=\varphi^{Snell}=0$ {[}deg{]}) reflection of
the incident beam. 

\noindent In order to assess the reliability of the synthesis process
as well as the effectiveness of the arising design, the \emph{SPSS}
layout has been also \emph{HFSS}-modeled and the results have been
compared in terms of surface currents (i.e., the target of the second
step) {[}Fig. 6(\emph{a}) vs. Fig. 5(\emph{b}){]} and far-field reflected
power patterns {[}Fig. 6(\emph{b}) vs. Fig. 5(\emph{c}){]} as well
as footprint patterns (i.e., the objective of the whole synthesis
process) {[}Fig. 7(\emph{b}) vs. Fig. 7(\emph{a}){]}. As expected,
the plot of the full-wave simulated pattern {[}Fig. 6(\emph{b}){]}
confirms that the synthesized \emph{SPSS} is able to focus the reflected
beam along the desired anomalous direction {[}i.e., ($u^{T}$, $v^{T}$)
$=$ ($7.58\times10^{-1}$, $-1.06\times10^{-1}$) being $u^{T}=\sin\theta^{T}\cos\varphi^{T}$
and $v^{T}=\sin\theta^{T}\sin\varphi^{T}${]}. Moreover, the \emph{HFSS}
plot indirectly proves the accuracy of the \emph{SPSS} surrogate since
the far-field distributions are very similar {[}Fig. 6(\emph{b}) vs.
Fig. 5(\emph{c}){]} within the whole visible domain except for a minor
secondary lobe located at $u\in\left[-0.5,-0.4\right]$, $v\in\left[-0.3,-0.2\right]$
{[}Fig. 6(\emph{b}){]} possibly caused by the higher-order modes arising
in the \emph{SPSS} \emph{UC}s patches and not taken into account in
Sect. \ref{sec:Problem-Formulation}. The reliability of the designed
\emph{SPSS} is further confirmed by the analytically-computed {[}Fig.
7(\emph{a}){]} and \emph{HFSS}-simulated {[}Fig. 7(\emph{b}){]} footprint
power densities analyzed in an area of extension $120\times60$ {[}$m^{2}]$
in front of the smart skin. Both indicate that the peak of the reflected
power is maximum within the coverage spot centered at $x_{glob}=-35.7$
{[}m{]} and $y_{glob}=30.14$ {[}m{]} with close values of the directivity
index (i.e., $\xi^{pen}=32.52$ {[}dB{]} {[}Fig. 7(\emph{a}){]} vs.
$\xi_{HFSS}^{pen}=31.91$ {[}dB{]} {[}Fig. 7(\emph{b}){]} being $\xi^{pen}\triangleq\frac{4\pi r^{2}\left|\mathbf{F}\left(\mathbf{r};\mathbf{G}\right)\right|^{2}}{\int_{0}^{2\pi}\int_{0}^{\pi}\left|\mathbf{F}\left(\mathbf{r};\mathbf{G}\right)\right|^{2}r^{2}\sin\left(\theta\right)\mathrm{d}\theta\mathrm{d}\varphi}$
the \emph{average directivity in the footprint region} for the \emph{SEE}
pencil coverage). Moreover, despite the approximations of the analytic
approach (Sect. \ref{sec:Problem-Formulation}) in modeling the surface
currents on $\Xi$, there is a good agreement between the current
distributions {[}Fig. 6(\emph{a}) vs. Fig. 5(\emph{b}){]} and the
surface averaged fields (\ref{eq:field average}), \{$\mathbf{E}_{pq}^{ave}\left(\mathbf{G}\right)$;
$p=1,...,P$; $q=1,...,Q$\}, {[}Fig. 7(\emph{c}) vs. Fig. 7(\emph{d}){]}.
This further assesses the effectiveness of the {}``\emph{Small-Scale
to Full-Scale} \emph{SPSS} \emph{Prediction Mapping}{}`` step in
predicting the actual local surface susceptibility tensors of the
full-size \emph{SPSS}, \{$\overline{\overline{\psi}}_{pq}^{e/m}\left(\mathbf{G}\right)$;
$p=1,...,P$; $q=1,...,Q$\}, both $\mathbf{J}^{tot}\left(\mathbf{r};\mathbf{G}\right)$
and $\mathbf{E}_{pq}^{ave}\left(\mathbf{G}\right)$ being related
to $\overline{\overline{\psi}}_{pq}^{e/m}\left(\mathbf{G}\right)$
through (\ref{eq:surface currents})-(\ref{eq:expansion surface density M})
and (\ref{eq:field average}), respectively.

\noindent The synthesis of a larger $P\times Q=50\times50$ layout
(\emph{i.e.}, $\Xi\approx2.14\times2.14$ {[}$\mathrm{m}^{2}${]})
has been performed next to analyse the method robustness when dealing
with higher dimensionalities of the optimization problem, $\mathcal{N}$
being proportional to $P\times Q$, as well as the dependence of the
focalization features on the \emph{}size of the \emph{SPSS} support.
Let us take a look to the \emph{SbD}-synthesized layout in Fig. 8(\emph{a}).
Analogously to the $P=Q=30$ case {[}Fig. 5(\emph{a}){]}, once again
the \emph{UC}s are non-uniform and the patterned surface appears (similarly)
irregular as actually expected since the same anomalous coherent reflection
of the previous test case is required-and-obtained here {[}Fig. 8(\emph{b}){]}
even though by exploiting a wider aperture. The larger \emph{SPSS}
support implies that the majority of the reflected power along the
same desired direction ($u^{T}=7.58\times10^{-1}$ and $v=-1.06\times10^{-1}$)
is focused in a narrower beam {[}Fig. 8(\emph{b}) vs. Fig. 5(\emph{c}){]}
and a more confined coverage footprint {[}Fig. 8(\emph{c}) vs. Fig.
7(\emph{a}){]}. Quantitatively, the increase of the \emph{SPSS} area
{[}$\frac{\Xi_{P=Q=50}}{\Xi_{P=Q=30}}\approx2.77$ $\to$ $\delta\Xi=4.42$
{[}dB{]} - Fig. 8(\emph{a}) vs. Fig. 5(\emph{a}){]} corresponds to
a proportional improvement of the focusing efficiency ($\delta\xi^{pen}\approx4.5$
{[}dB{]} ) being $\left.\xi^{pen}\right\rfloor _{P=Q=50}=36.96$ {[}dB{]}
{[}Fig. 8(\emph{c}){]} and $\left.\xi^{pen}\right\rfloor _{P=Q=30}=32.52$
{[}dB{]} {[}Fig. 7(\emph{a}){]}). Such a result ($\delta\Xi\approx\delta\xi^{pen}$)
can provide useful guidelines on how to size and design the \emph{SPSS}s
for obtaining a desired (selective/broad) coverage in a \emph{SEE}
scenario.

\noindent The same conclusions drawn in the first example on the reliability
of the synthesis results hold true also for this wider \emph{SPSS}
as confirmed by the comparisons with the \emph{HFSS} simulations,
at micro-scale, of the spatial distributions of both the surface current,
$\mathbf{J}^{tot}\left(\mathbf{r};\mathbf{G}\right)$ {[}Fig. 9(\emph{c})
vs. Fig. 8(\emph{d}){]}, and the averaged electric field, $\mathbf{E}_{pq}^{ave}\left(\mathbf{G}\right)$
{[}Fig. 9(\emph{d}) vs. Fig. 8(\emph{e}){]}, while at macro-scale,
of the radiated patterns {[}Fig. 9(\emph{a}) vs. Fig. 8(\emph{b}){]}
and footprints {[}Fig. 9(\emph{b}) vs. Fig. 8(\emph{c}){]}. For instance,
it turns out that $\xi_{HFSS}^{pen}=36.08$ {[}dB{]} $\approx$ $\xi^{pen}=36.96$
{[}dB{]} so that the rule $\delta\Xi\approx\delta\xi_{HFSS}^{pen}$
is verified also here.

\noindent But what about more complex footprints? The design of \textbf{}\emph{SPSS}s
for shaped coverages is thus discussed by considering two different
applicative scenarios, each with a different setup of the pattern-mask
region $\Theta$ in (\ref{eq:feasibility E}), while keeping the same
\emph{SPSS} aperture ($P\times Q=50\times50$). More specifically,
the former ({}``\emph{Two-Squares Footprint}'' scenario) mimics
the realistic case where the \emph{SPSS} is requested to afford two
separate and asymmetric beams that focus the power in very narrow,
but shaped, regions, which emulate two small town-squares in a urban
environment. Numerically, $\Theta$ consists of two regions of $20\times10$
{[}$\mathrm{m}^{2}${]} centered at ($x_{glob}^{(1)}$, $y_{glob}^{(1)}$)
$=$ ($-30$, $15$) {[}m{]} and ($x_{glob}^{(2)}$, $y_{glob}^{(2)}$)
$=$ ($28$, $17$) {[}m{]}, respectively. Otherwise, the coverage
region of the {}``\emph{Street-Square Footprint}'' case maps a $10\times120$
{[}$\mathrm{m}^{2}${]} street centered at ($x_{glob}^{(1)}$, $y_{glob}^{(1)}$)
$=$ ($-35$, $80$) {[}m{]} that opens on a square of size $30\times30$
{[}$\mathrm{m}^{2}${]} located at $x_{glob}^{(2)}=-y_{glob}^{(2)}=-45$
{[}m{]}. From a methodological viewpoint, such a test case is devoted
to assess the potentialities of the \emph{SPSS} in focusing the power
in a region characterized by both complex contours and very low grazing
angles, which are almost parallel to the ground surface, due to both
the street position and length ($L_{street}=140$ {[}m{]}) and the
\emph{SPSS} height above the ground ($H=5$ {[}m{]}).

\noindent The layouts of the synthesized \emph{SPSS}s {[}Fig. 10(\emph{a})
- {}``\emph{Two-Squares Footprint}''; Fig. 10(\emph{b}) - {}``\emph{Street-Square
Footprint}''{]} are still non-uniform, but less regular than in the
{}``\emph{Pencil Coverage}'' case {[}Fig. 5(\emph{a}) and Fig. 8(\emph{a}){]}
owing to the complexity of the shaped-beam. As a consequence, the
{}``periodicity'' of the spatial behavior of the surface currents
in Fig. 8(\emph{d}) is completely lost {[}Figs. 10(\emph{c})-10(\emph{d}){]}.
On the other hand, the analytical implementation based on the \emph{Local
UC-DT} (Sect. \ref{sec:Problem-Formulation}) is still very reliable,
as confirmed by the full-wave simulations, at both micro- {[}Figs.
10(\emph{c})-10(\emph{d}) vs. Figs. 10(\emph{e})-10(\emph{f}){]} and
macro- {[}Figs. 11(\emph{a})-11(\emph{b}) vs. Figs. 11(\emph{c})-11(\emph{d})
and Figs. 12(\emph{a})-12(\emph{b}) vs. Figs. 12(\emph{c})-12(\emph{d}){]}
scale. Moreover, the plots of the far-field reflection patterns in
Fig. 11 point out, on the one hand, the challenging nature of the
addressed \emph{Shaped Coverage} \textbf{}\emph{SPSS}s problems while,
on the other hand, they prove the feasibility of \emph{SPSS}s that
focus selectively in narrow angular regions. The effectiveness of
using a \emph{SPSS} in a \emph{SEE} context can be better appreciated
when analyzing the footprint power densities (Fig. 12). With reference
to the \emph{HFSS} simulations of the full-scale layouts, the reflected
power is properly directed towards the user-defined coverage regions,
regardless of their {}``dual beam'' {[}Fig. 12(\emph{a}) and Fig.
12(\emph{c}){]} or {}``low grazing'' {[}Fig. 12(\emph{b}) and Fig.
12(\emph{d}){]} nature, with limited power losses outside $\Theta$
(Fig. 11). The quantitative assessment of such a behavior is given
by the corresponding values of the efficiency index $\xi_{HFSS}^{sha}$
($\xi^{sha}\triangleq\frac{4\pi}{\Theta}\frac{\int_{\Theta}\left|\mathbf{F}\left(\mathbf{r};\mathbf{G}\right)\right|^{2}r^{2}\sin\left(\theta\right)\mathrm{d}\theta\mathrm{d}\varphi}{\int_{0}^{2\pi}\int_{0}^{\pi}\left|\mathbf{F}\left(\mathbf{r};\mathbf{G}\right)\right|^{2}r^{2}\sin\left(\theta\right)\mathrm{d}\theta\mathrm{d}\varphi}$
being the \emph{average directivity in the footprint region} for the
\emph{SEE} shaped coverage) being $\xi_{HFSS}^{sha\,(Two-Squares)}=18.81$
{[}dB{]} {[}Fig. 12(\emph{c}){]} and $\xi_{HFSS}^{sha\,(Street+Square)}=20.92$
{[}dB{]} {[}Fig. 12(\emph{d}){]}.

\noindent The next example is aimed at answering to the following
questions: {}``\emph{Widening the aperture, the patterned surface
similarity observed in the pencil coverage is still maintained when
dealing with complex shaped footprints?}\char`\"{} and {}``\emph{Does
the design rule} $\delta\Xi\approx\delta\xi$ \emph{apply also for
complex shaped coverages?}''. Towards this end, a $P\times Q=75\times75$
square lattice comprising $\approx5600$ \emph{UC}s with an extension
of $\Xi\approx3.2\times3.2$ {[}$\mathrm{m}^{2}${]} {[}Fig. 13(\emph{a}){]}
has been chosen to radiate the {}``\emph{Street-Square}'' footprint
and the performance of the arising layout have been compared with
those of the smaller (i.e., $P\times Q=50\times50$) arrangement in
Fig. 10(\emph{b}). By taking a look to the sketch of the \emph{SPSS}
in Fig. 13(\emph{a}), the answer to the first question is that, unlike
the \emph{Pencil Coverage} case, the increase of the size seems to
imply a {}``stretching'' of the \emph{UC}s distribution {[}Fig.
13(\emph{a}) vs. Fig. 10(\emph{b}){]} rather than a {}``repetition''
of the same spatial behavior {[}Fig. 8(\emph{a}) vs. Fig. 5(\emph{a}){]}
probably because of the need of affording more complex wave manipulation
phenomena to generate the complex footprint at hand. As for the second
question, let us compare the radiation behavior of the two \emph{SPSS}s
in Fig. 10(\emph{b}) and Fig. 13(\emph{a}) pictorially described by
the footprint color-maps in Fig. 12(\emph{d}) and Fig. 13(\emph{b}),
respectively. Both layouts fulfil the requirement of reflecting the
incident power towards the desired region $\Theta$. This happens
more and more as the \emph{SPSS} support enlarges according to the
rule-of-thumb $\delta\xi^{sha}\approx\delta\Xi$. Indeed, $\delta\Xi\approx3.5$
{[}dB{]} and $\delta\xi_{HFSS}^{sha}\approx3.2$ {[}dB{]} (i.e., $\left.\xi_{HFSS}^{sha}\right\rfloor _{P=Q=75}=23.94$
{[}dB{]} {[}Fig. 13(\emph{b}){]} and $\left.\xi_{HFSS}^{sha}\right\rfloor _{P=Q=50}=20.92$
{[}dB{]} {[}Fig. 12(\emph{d}){]}).

\noindent Whether the feasibility and the effectiveness of a \emph{SPSS}
deployment are key items to be addressed towards the implementation
of a \emph{SEE}, certainly the sustainable installation in a living
environment (e.g., a town-square of a city or a room in a building
floor) is a relevant issue, as well. Without pretending to give {}``the''
solution to this problem, but just for adding some insights on the
topic, some tests in the underlying numerical analysis have been devoted
to this line of reasoning. More specifically, a representative example
dealt with the possibility to realize, on a standard {}``Junior Poster
Billboard'' support {[}i.e., $\Xi\approx3.6\times1.8$ {[}$\mathrm{m}^{2}${]}
- Fig. 14(\emph{a}){]}, a \emph{SPSS} focusing the beam on the same
{}``Street-Square'' region $\Theta$ of the previous examples. Subject
to the size constraint, the \emph{SPSS} has been designed by optimizing
the descriptors of $P\times Q=84\times42$ \emph{UC}s. Figure 14 shows
the synthesized layout {[}Fig. 14(\emph{a}){]} along with the \emph{HFSS}-simulated
footprint pattern {[}Fig. 14(\emph{b}){]}. As hoped, the radiated
beam fulfils the mask requirements with a good accuracy being $\left.\xi_{HFSS}^{sha}\right\rfloor _{P\times Q=84\times42}=21.30$
{[}dB{]} {[}Fig. 14(\emph{b}){]}, even though, unlike Fig. 13(\emph{b}),
some \emph{spill-over} effects appear at low grazing angles (i.e.,
$y_{glob}>140$ {[}m{]}) {[}Fig. 14(\emph{b}){]} due to the smaller
vertical size of the \emph{SPSS} {[}Fig. 14(\emph{a}) vs. Fig. 13(\emph{a}){]}.

\noindent The last test case is not concerned with a realistic \emph{SEE}
problem, but it is more aimed at giving the flavour of what it can
be done with just a static and passive structure provided you have
at disposal a suitable tool for managing the huge computational complexity
of a high-dimension optimization problem. Accordingly, the proposed
\emph{SPSS} design has been applied to optimize a $P\times Q=75\times75$
\emph{UC}s arrangement, located $H=20$ {[}m{]} on the floor, for
beaming a {}``\emph{Olympic Flag}''-shaped region of extension $-150$
{[}m{]} $\le$ $x_{glob}$ $\le$ $70$ {[}m{]} and $50$ {[}m{]}
$\le$ $y_{glob}$ $\le$ $150$ {[}m{]} (Fig. 1). The plots of the
synthesized layout {[}Fig. 15(\emph{a}){]} and of the \emph{HFSS}-computed
distribution of the power reflected on the ground {[}Fig. 15(\emph{b}){]}
prove the feasibility of a \emph{SPSS} matching hard pattern-mask
requirements {[}$\xi_{HFSS}^{sha}\approx18.3$ {[}dB{]} - Fig. 15(\emph{b}){]}
as well as the capability of the proposed design method to efficiently
deal with large-scale optimization problems.

\section{\noindent Conclusions\label{sec:Conclusions-and-Remarks}}

\noindent The feasibility of \emph{SPSS}s with advanced wave manipulation
properties has been addressed. A two-step design process, which combines
the solution of an \emph{IS} problem to determine the surface currents
affording user-desired pattern-mask constraints together with a \emph{SbD}-based
optimization of the corresponding patterned layout, has been proposed.
Towards this end, an innovative scheme for building an \emph{AI}-based
\emph{DT} of the \emph{UC}s of the \emph{SPSS} for predicting the
\emph{EM} response of this latter has been introduced. The arising
synthesis method has been validated by considering different apertures,
radiation objectives, and various operative \emph{SEE} scenarios.
Full-wave finite element simulations \cite{HFSS 2021} of the synthesized
layouts have been performed to assess the reliability of the synthesis
results.

\noindent The outcomes from the numerical validation have demonstrated
that the proposed \emph{AI}-driven method allows the designer to reliably
synthesize large \emph{SPSS}s with excellent beam control capabilities
by efficiently solving high-dimension optimization problems. The exploitation
of a \emph{Local UC-DT} within the \emph{SbD}-based optimization of
the \emph{SPSS} layout assures a faithful prediction of the \emph{EM}
behavior of the non-uniform modulated patterned surface, while minimizing
the computational costs of the solution of the forward problem at
hand. The synthesis layouts fulfil challenging radiation objectives
despite the choice of a basic \emph{UC} with a limited phase control.

\noindent Future works, beyond the scope of the present paper, will
be aimed at exploring the potentialities of the proposed method when
using more complex \emph{UC}s (e.g., a higher number of per-element
\emph{DoF}s or multi-layer structures) and tiled architectures instead
of single-panel layouts. Moreover, the extension of the proposed method
to real-time reconfigurable smart skins is currently under development.

\section*{\noindent Acknowledgements}

\noindent This work benefited from the networking activities carried
out within the Project {}``Cloaking Metasurfaces for a New Generation
of Intelligent Antenna Systems (MANTLES)'' (Grant No. 2017BHFZKH)
funded by the Italian Ministry of Education, University, and Research
under the PRIN2017 Program (CUP: E64I19000560001). Moreover, it benefited
from the networking activities carried out within the Project {}``SPEED''
(Grant No. 61721001) funded by National Science Foundation of China
under the Chang-Jiang Visiting Professorship Program, the Project
'Inversion Design Method of Structural Factors of Conformal Load-bearing
Antenna Structure based on Desired EM Performance Interval' (Grant
no. 2017HZJXSZ) funded by the National Natural Science Foundation
of China, and the Project 'Research on Uncertainty Factors and Propagation
Mechanism of Conformal Loab-bearing Antenna Structure' (Grant No.
2021JZD-003) funded by the Department of Science and Technology of
Shaanxi Province within the Program Natural Science Basic Research
Plan in Shaanxi Province. A. Massa wishes to thank E. Vico for her
never-ending inspiration, support, guidance, and help.

\section*{\noindent Appendix}

\subsection*{\noindent Small-Scale/Full-Scale Mapping Rules}

\noindent The mapping rules between the predictions of the local surface
susceptibility tensors of the small-scale \emph{SPSS}, \{$\overline{\overline{\zeta}}_{p'q'}^{e/m}\left(\mathbf{G}^{'}\right)$;
$p'=1,...,P'$; $q'=1,...,Q'$\}, and the predictions of the local
surface susceptibility tensors of the full-scale \emph{SPSS}, \{$\overline{\overline{\zeta}}_{pq}^{e/m}\left(\mathbf{G}\right)$;
$p=1,...,P$; $q=1,...,Q$\}, which estimate the corresponding actual
values, \{$\overline{\overline{\psi}}_{pq}^{e/m}\left(\mathbf{G}\right)$;
$p=1,...,P$; $q=1,...,Q$\}, are defined as follows\begin{equation}
\overline{\overline{\zeta}}_{pq}^{e/m}\left(\mathbf{G}\right)=\left\{ \begin{array}{ll}
\left.\overline{\overline{\zeta}}_{p'q'}^{e/m}\left(\mathbf{G}^{'}\right)\right]_{p'=p}^{q'=q} & if\,\left(p,\, q\right)\in\mathcal{C}^{\left(0\right)}\\
\left.\overline{\overline{\zeta}}_{p'q'}^{e/m}\left(\mathbf{G}^{'}\right)\right]_{p'=p-P+P'}^{q'=q-Q+Q'} & if\,\left(p,\, q\right)\in\mathcal{C}^{\left(1\right)}\\
\left.\overline{\overline{\zeta}}_{p'q'}^{e/m}\left(\mathbf{G}^{'}\right)\right]_{p'=p}^{q'=q-Q+Q'} & if\,\left(p,\, q\right)\in\mathcal{C}^{\left(2\right)}\\
\left.\overline{\overline{\zeta}}_{p'q'}^{e/m}\left(\mathbf{G}^{'}\right)\right]_{p'=p-P+P'}^{q'=q} & if\,\left(p,\, q\right)\in\mathcal{C}^{\left(3\right)}\\
\left.\overline{\overline{\zeta}}_{p'q'}^{e/m}\left(\mathbf{G}^{'}\right)\right]_{p'=p}^{q'=3} & if\,\left(p,\, q\right)\in\mathcal{C}^{\left(4\right)}\\
\left.\overline{\overline{\zeta}}_{p'q'}^{e/m}\left(\mathbf{G}^{'}\right)\right]_{p'=3}^{q'=q} & if\,\left(p,\, q\right)\in\mathcal{C}^{\left(5\right)}\\
\left.\overline{\overline{\zeta}}_{p'q'}^{e/m}\left(\mathbf{G}^{'}\right)\right]_{p'=3}^{q'=3} & if\,\left(p,\, q\right)\in\mathcal{C}^{\left(6\right)}\end{array}\right.\label{eq:}\end{equation}

\noindent where $\mathcal{C}^{\left(0\right)}=\left\{ \left(p,q\right):\, p\in\left[1,2\right],q\in\left[1,2\right]\right\} $,
$\mathcal{C}^{\left(1\right)}=\left\{ \left(p,q\right):\, p\in\left[P-1,P\right],q\in\left[Q-1,Q\right]\right\} $,
$\mathcal{C}^{\left(2\right)}=\left\{ \left(p,q\right):\, p\in\left[1,2\right],q\in\left[Q-1,Q\right]\right\} $,
$\mathcal{C}^{\left(3\right)}=\left\{ \left(p,q\right):\, p\in\left[P-1,P\right],q\in\left[1,2\right]\right\} $,
$\mathcal{C}^{\left(4\right)}=\left\{ \left(p,q\right):\, p=\left\{ 1,2,P-1,P\right\} ,q\in\left[3,Q-2\right]\right\} $,
$\mathcal{C}^{\left(5\right)}=\left\{ \left(p,q\right):\, p\in\left[3,P-2\right],q=\left\{ 1,2,Q-1,Q\right\} \right\} $,
and $\mathcal{C}^{\left(6\right)}=\left\{ \left(p,q\right):\, p\in\left[3,P-2\right],q\in\left[3,Q-2\right]\right\} $
(Fig. 2).

\newpage
\section*{FIGURE CAPTIONS}

\begin{itemize}
\item \textbf{Figure 1.} \emph{Problem geometry}. Sketch of the smart \emph{EM}
environment (\emph{SEE}) scenario.
\item \textbf{Figure 2.} \emph{Illustrative Example} - Example of \emph{Local
UC-DT} mapping between the small-scale model ($P'\times Q'=5\times5$)
and the full-scale \emph{SPSS} arrangement ($P\times Q=10\times10$).
\item \textbf{Figure 3.} \emph{Illustrative Example} - Model of (\emph{a})
the unit cell (\emph{UC}) of the \emph{SPSS} and of \emph{}(\emph{b})
\emph{}the small-scale \emph{SPSS} layout. Plot of (\emph{c}) the
phase, $\angle\widetilde{R}_{TE,TE}$, and of (\emph{d}) the magnitude,
$\left|\widetilde{R}_{TE,TE}\right|$, of the average {}``\emph{TE-TE}''
reflection coefficient of the central element {[}($p'$, $q'$) $=$
($\frac{P'+1}{2}$, $\frac{Q'+1}{2}$){]} of the small-scale \emph{SPSS}
layout.
\item \textbf{Figure 4.} \emph{Illustrative Example} (\emph{Pencil Coverage},
$P=Q=30$, $\theta^{T}=50$ {[}deg{]}, $\varphi^{T}=-8$ {[}deg{]},
$H=5$ {[}m{]}) - Plot of (\emph{a}) the evolution of the pattern
matching index, $\Gamma_{i}$ ($i=1,...,I$) and of the (\emph{b})
reference surface current distribution ($x$-component), $J_{x}^{opt}\left(\mathbf{r}\right)\triangleq\mathbf{J}_{opt}^{tot}\left(\mathbf{r}\right)\cdot\widehat{\mathbf{x}}$
($\mathbf{r}\in\Xi$), with (\emph{c}) the radiated far-field pattern
in the ($u$, $v$) plane, $\left|\mathbf{F}_{opt}\left(\mathbf{r}\right)\right|$
(\ref{eq:far field}).
\item \textbf{Figure 5.} \emph{Illustrative Example} (\emph{Pencil Coverage},
$P=Q=30$, $\theta^{T}=50$ {[}deg{]}, $\varphi^{T}=-8$ {[}deg{]},
$H=5$ {[}m{]}) - Plot of (\emph{a}) the synthesized \emph{SPSS} layout
and of the corresponding (\emph{b}) surface current distribution ($x$-component),
$J_{x}^{N}\left(\mathbf{r}\right)\triangleq\mathbf{J}_{N}^{tot}\left(\mathbf{r}\right)\cdot\widehat{\mathbf{x}}$
($\mathbf{r}\in\Xi$), and (\emph{c}) radiated far-field pattern in
the ($u$, $v$) plane, $\left|\mathbf{F}_{N}\left(\mathbf{r}\right)\right|$
($\mathbf{F}_{N}\left(\mathbf{r};\mathbf{G}\right)\triangleq\mathcal{L}\left[\mathbf{J}_{N}^{tot}\left(\mathbf{r};\mathbf{G}\right)\right]$)
(\ref{eq:far field}).
\item \textbf{Figure 6.} \emph{Numerical Validation} (\emph{Pencil Coverage},
$P=Q=30$, $\theta^{T}=50$ {[}deg{]}, $\varphi^{T}=-8$ {[}deg{]},
$H=5$ {[}m{]}) - Plot of the \emph{HFSS}-simulated (\emph{a}) surface
current distribution ($x$-component), $J_{x}^{HFSS}\left(\mathbf{r}\right)\triangleq\mathbf{J}_{HFSS}^{tot}\left(\mathbf{r}\right)\cdot\widehat{\mathbf{x}}$
($\mathbf{r}\in\Xi$), and (\emph{c}) radiated far-field pattern in
the ($u$, $v$) plane, $\left|\mathbf{F}_{HFSS}\left(\mathbf{r}\right)\right|$
($\mathbf{F}_{HFSS}\left(\mathbf{r};\mathbf{G}\right)\triangleq\mathcal{L}\left[\mathbf{J}_{HFSS}^{tot}\left(\mathbf{r};\mathbf{G}\right)\right]$)
(\ref{eq:far field}).
\item \textbf{Figure 7.} \emph{Numerical Validation} (\emph{Pencil Coverage},
$P=Q=30$, $\theta^{T}=50$ {[}deg{]}, $\varphi^{T}=-8$ {[}deg{]},
$H=5$ {[}m{]}) - Plots of (\emph{a})(\emph{b}) the \emph{SPSS} footprint
pattern and of (\emph{c})(\emph{d}) a component of the surface averaged
electric field (\ref{eq:field average}), \{$\mathbf{E}_{pq}^{ave}\left(\mathbf{G}\right)$;
$p=1,...,P$; $q=1,...,Q$\}, (\emph{a})(\emph{c}) analytically-computed
or (\emph{b})(\emph{d}) \emph{HFSS}-simulated.
\item \textbf{Figure 8.} \emph{Numerical Validation} (\emph{Pencil Coverage},
$P=Q=50$, $\theta^{T}=50$ {[}deg{]}, $\varphi^{T}=-8$ {[}deg{]},
$H=5$ {[}m{]}) - Plot of (\emph{a}) the synthesized \emph{SPSS} layout
and of the corresponding (\emph{b}) far-field pattern in the ($u$,
$v$) plane, $\left|\mathbf{F}_{N}\left(\mathbf{r}\right)\right|$
($\mathbf{F}_{N}\left(\mathbf{r};\mathbf{G}\right)\triangleq\mathcal{L}\left[\mathbf{J}_{N}^{tot}\left(\mathbf{r};\mathbf{G}\right)\right]$)
(\ref{eq:far field}), (\emph{c}) \emph{SPSS} footprint pattern, (\emph{d})
surface current distribution ($x$-component), $J_{x}^{N}\left(\mathbf{r}\right)\triangleq\mathbf{J}_{N}^{tot}\left(\mathbf{r}\right)\cdot\widehat{\mathbf{x}}$
($\mathbf{r}\in\Xi$), and (\emph{e}) $y$-component of the surface
averaged electric field (\ref{eq:field average}), \{$\mathbf{E}_{pq}^{ave}\left(\mathbf{G}\right)$;
$p=1,...,P$; $q=1,...,Q$\}.
\item \textbf{Figure 9.} \emph{Numerical Validation} (\emph{Pencil Coverage},
$P=Q=50$, $\theta^{T}=50$ {[}deg{]}, $\varphi^{T}=-8$ {[}deg{]},
$H=5$ {[}m{]}) - Plot of the \emph{HFSS}-simulated (\emph{a}) far-field
pattern in the ($u$, $v$) plane, $\left|\mathbf{F}_{HFSS}\left(\mathbf{r}\right)\right|$
($\mathbf{F}_{HFSS}\left(\mathbf{r};\mathbf{G}\right)\triangleq\mathcal{L}\left[\mathbf{J}_{HFSS}^{tot}\left(\mathbf{r};\mathbf{G}\right)\right]$)
(\ref{eq:far field}), (\emph{b}) \emph{SPSS} footprint pattern, (\emph{c})
surface current distribution ($x$-component), $J_{x}^{HFSS}\left(\mathbf{r}\right)\triangleq\mathbf{J}_{HFSS}^{tot}\left(\mathbf{r}\right)\cdot\widehat{\mathbf{x}}$
($\mathbf{r}\in\Xi$), and (\emph{e}) $y$-component of the surface
averaged electric field (\ref{eq:field average}), \{$\mathbf{E}_{pq}^{ave}\left(\mathbf{G}\right)$;
$p=1,...,P$; $q=1,...,Q$\}.
\item \textbf{Figure 10.} \emph{Numerical Validation} (\emph{Shaped Coverage},
$P=Q=50$, $H=5$ {[}m{]}) - Plots of (\emph{a})(\emph{b}) the synthesized
\emph{SPSS} layouts along with the corresponding (\emph{c})(\emph{d})
analytically-computed or (\emph{e})(\emph{f}) \emph{HFSS}-simulated
surface current distributions ($x$-component), $J_{x}^{HFSS}\left(\mathbf{r}\right)\triangleq\mathbf{J}_{HFSS}^{tot}\left(\mathbf{r}\right)\cdot\widehat{\mathbf{x}}$
($\mathbf{r}\in\Xi$) for (\emph{a})(\emph{c})(\emph{e}) the {}``\emph{Two-Square}''
and (\emph{b})(\emph{d})(\emph{f}) the {}``\emph{Street-Square}''
footprints.
\item \textbf{Figure 11.} \emph{Numerical Validation} (\emph{Shaped Coverage},
$P=Q=50$, $H=5$ {[}m{]}) - Plots of (\emph{a})(\emph{b}) the analytically-computed
and (\emph{c})(\emph{d}) the \emph{HFSS}-simulated far-field patterns
in the ($u$, $v$) plane, $\left|\mathbf{F}\left(\mathbf{r}\right)\right|$
($\mathbf{F}\left(\mathbf{r};\mathbf{G}\right)\triangleq\mathcal{L}\left[\mathbf{J}^{tot}\left(\mathbf{r};\mathbf{G}\right)\right]$)
(\ref{eq:far field}) for (\emph{a})(\emph{c}) the {}``\emph{Two-Square}''
and (\emph{b})(\emph{d}) the {}``\emph{Street-Square}'' footprints.
\item \textbf{Figure 12.} \emph{Numerical Validation} (\emph{Shaped Coverage},
$P=Q=50$, $H=5$ {[}m{]}) - Plots of (\emph{a})(\emph{b}) the analytically-computed
and (\emph{c})(\emph{d}) the \emph{HFSS}-simulated \emph{SPSS} footprint
patterns in correspondence with (\emph{a})(\emph{c}) the {}``\emph{Two-Square}''
and (\emph{b})(\emph{d}) the {}``\emph{Street-Square}'' footprint
targets.
\item \textbf{Figure 13.} \emph{Numerical Validation} (\emph{Shaped Coverage,}
{}``\emph{Street-Square}'' footprint, $P=Q=75$, $H=5$ {[}m{]})
- Plot of (\emph{a}) the synthesized \emph{SPSS} layout and of (\emph{b})
the \emph{HFSS}-simulated \emph{SPSS} footprint pattern.
\item \textbf{Figure 14.} \emph{Numerical Validation} (\emph{Shaped Coverage,}
{}``\emph{Street-Square}'' footprint, $P\times Q=84\times42$, $H=5$
{[}m{]}) - Plot of (\emph{a}) the synthesized \emph{SPSS} layout and
of (\emph{b}) the \emph{HFSS}-simulated \emph{SPSS} footprint pattern.
\item \textbf{Figure 15.} \emph{Numerical Validation} (\emph{Shaped Coverage},
{}``\emph{Olympic Flag}'' footprint, $P=Q=75$, $H=20$ {[}m{]})
- Plot of (\emph{a}) the synthesized \emph{SPSS} layout and of (\emph{b})
the \emph{HFSS}-simulated \emph{SPSS} footprint pattern.
\end{itemize}
\newpage
\begin{center}~\vfill\end{center}

\begin{center}\includegraphics[%
  clip,
  width=0.95\columnwidth,
  keepaspectratio]{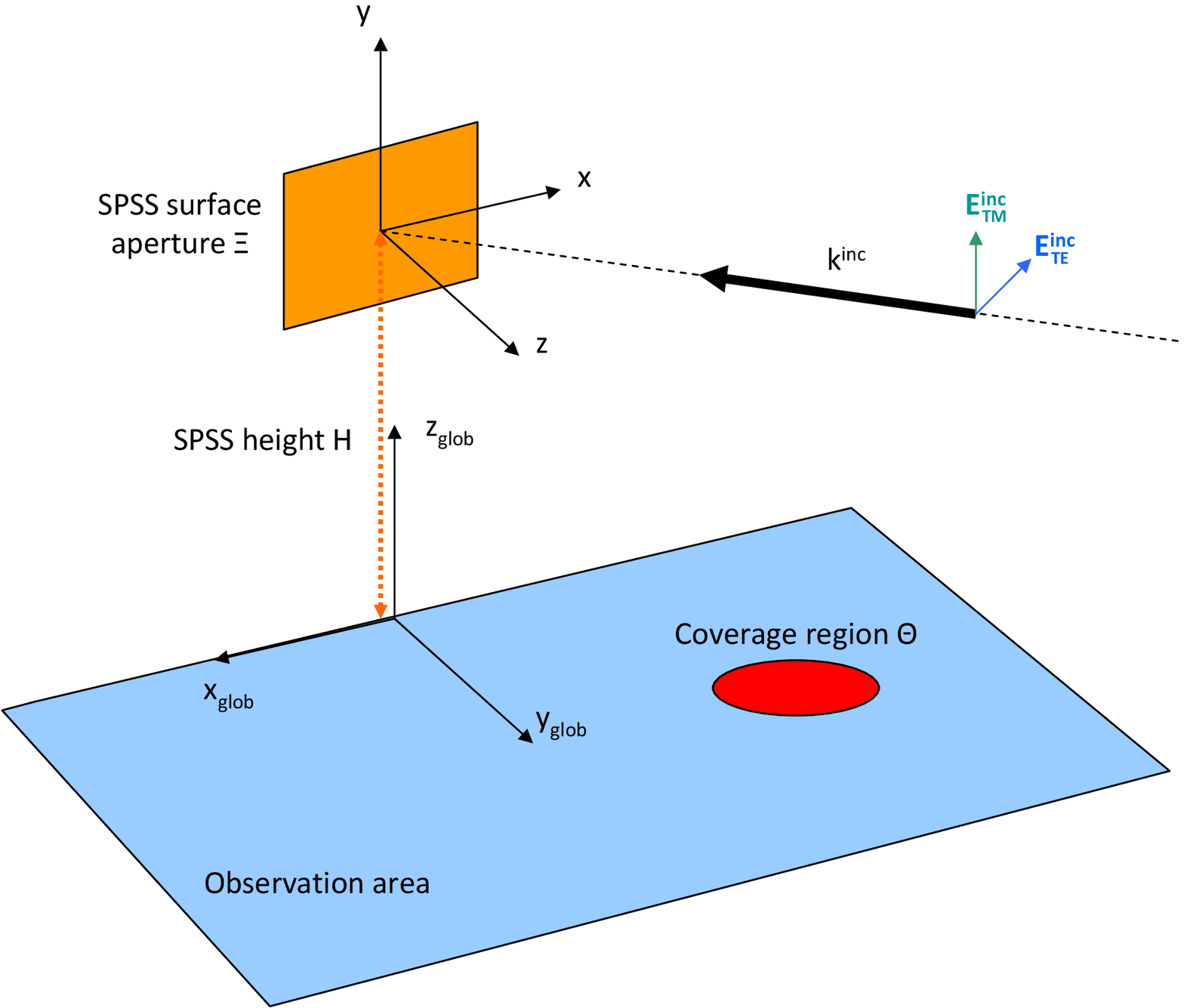}\end{center}

\begin{center}~\vfill\end{center}

\begin{center}\textbf{Fig. 1 - G. Oliveri et} \textbf{\emph{al.}}\textbf{,}
{}``Building a Smart \emph{EM} Environment - \emph{AI}-Enhanced Aperiodic
...''\end{center}
\newpage

\begin{center}~\vfill\end{center}

\begin{center}\includegraphics[%
  clip,
  width=0.95\columnwidth,
  keepaspectratio]{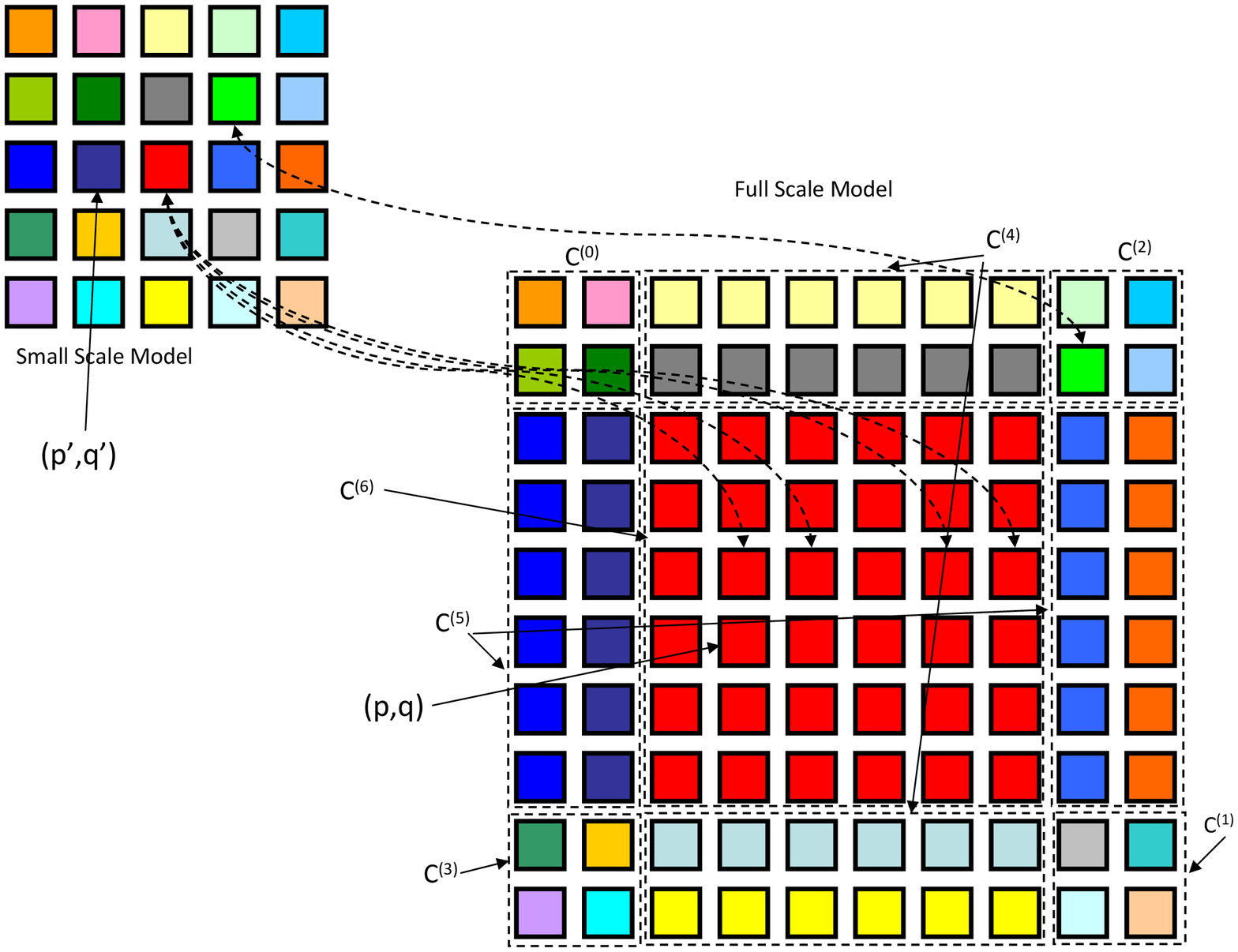}\end{center}

\begin{center}~\vfill\end{center}

\begin{center}\textbf{Fig. 2 - G. Oliveri et} \textbf{\emph{al.}}\textbf{,}
{}``Building a Smart \emph{EM} Environment - \emph{AI}-Enhanced Aperiodic
...''\end{center}
\newpage

\begin{center}\begin{tabular}{cc}
\includegraphics[%
  clip,
  width=0.48\columnwidth,
  keepaspectratio]{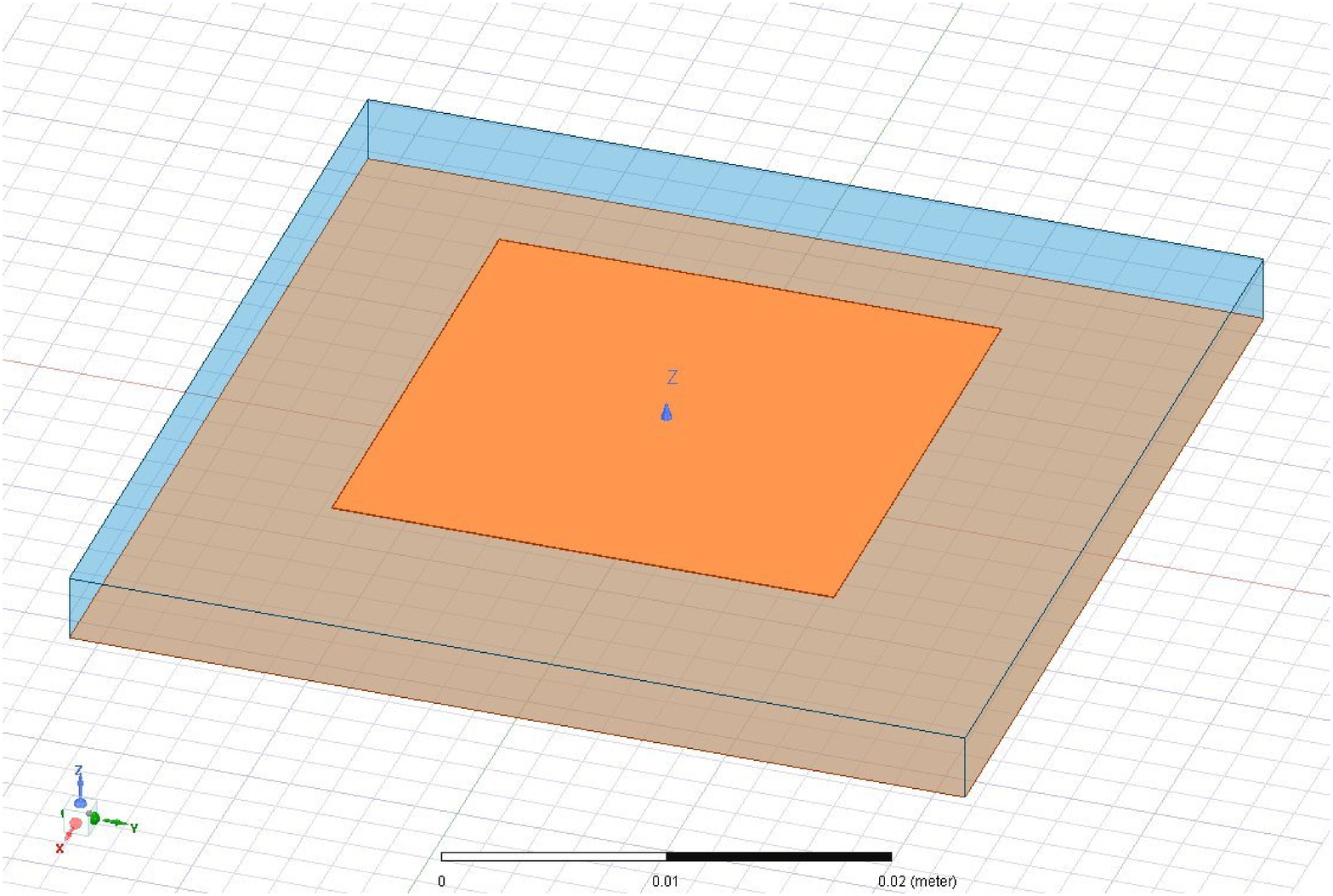}&
\includegraphics[%
  clip,
  width=0.48\columnwidth,
  keepaspectratio]{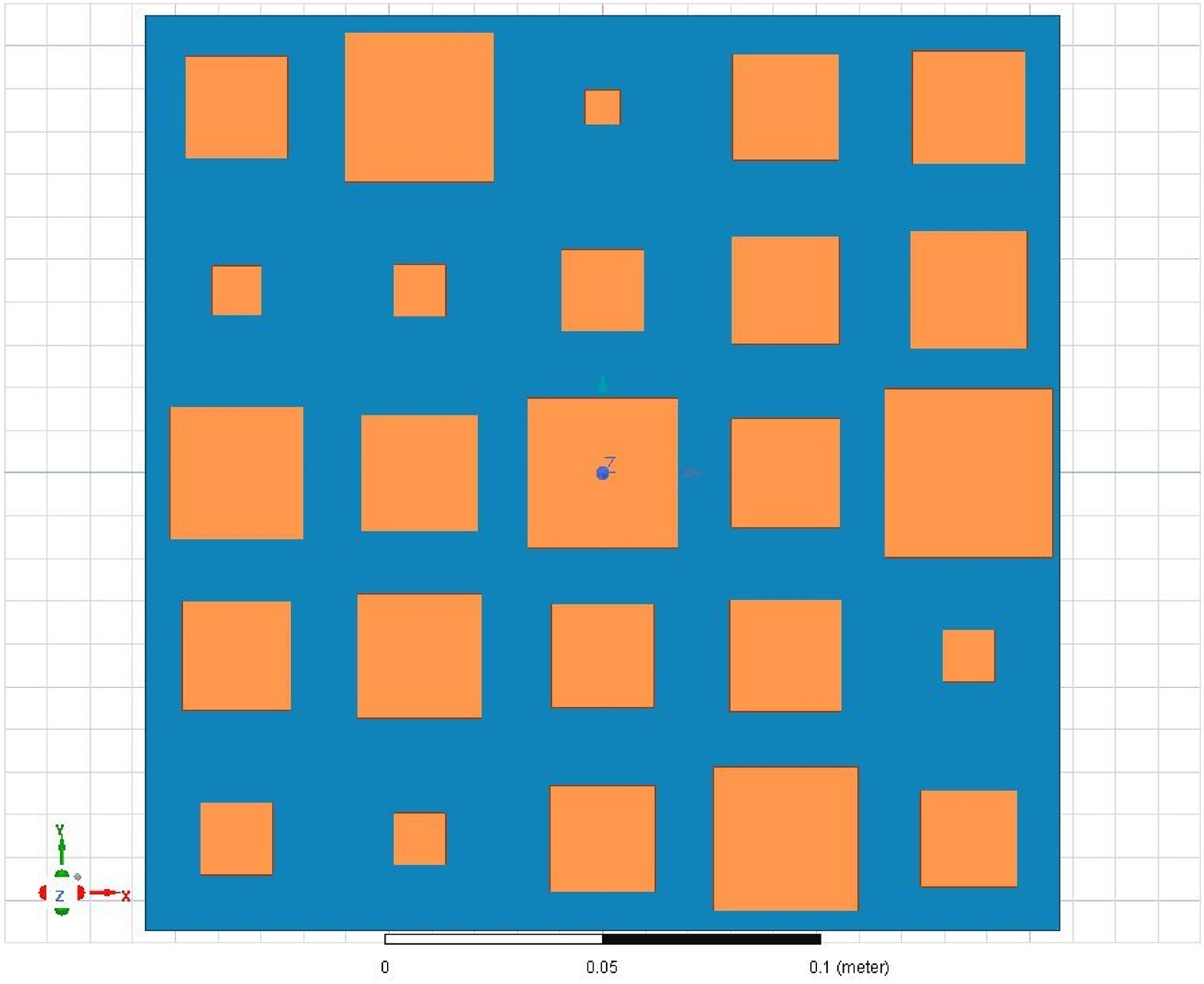}\tabularnewline
(\emph{a})&
(\emph{b})\tabularnewline
\multicolumn{2}{c}{\includegraphics[%
  clip,
  width=0.75\columnwidth,
  keepaspectratio]{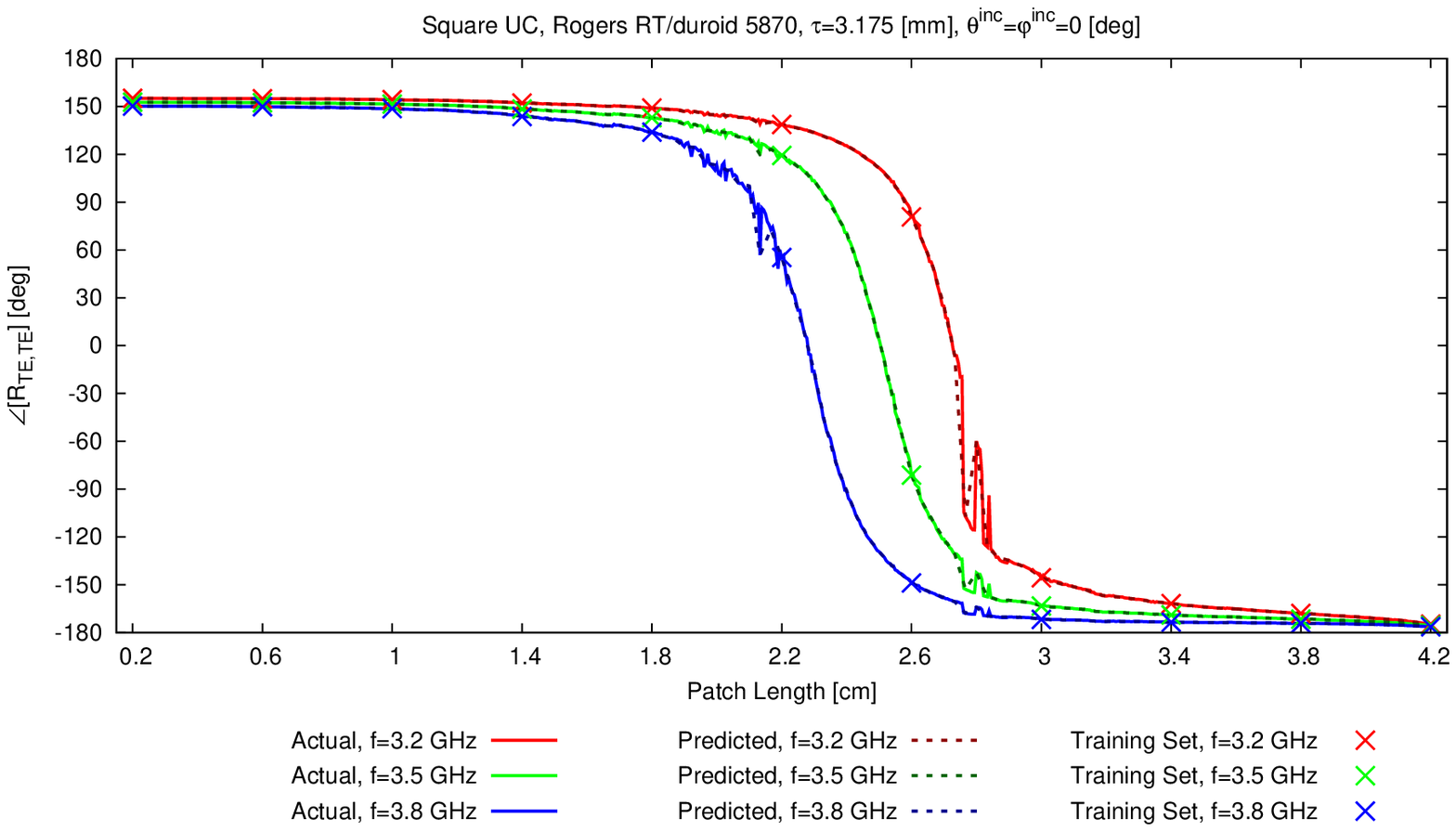}}\tabularnewline
\multicolumn{2}{c}{(\emph{c})}\tabularnewline
\multicolumn{2}{c}{\includegraphics[%
  clip,
  width=0.75\columnwidth,
  keepaspectratio]{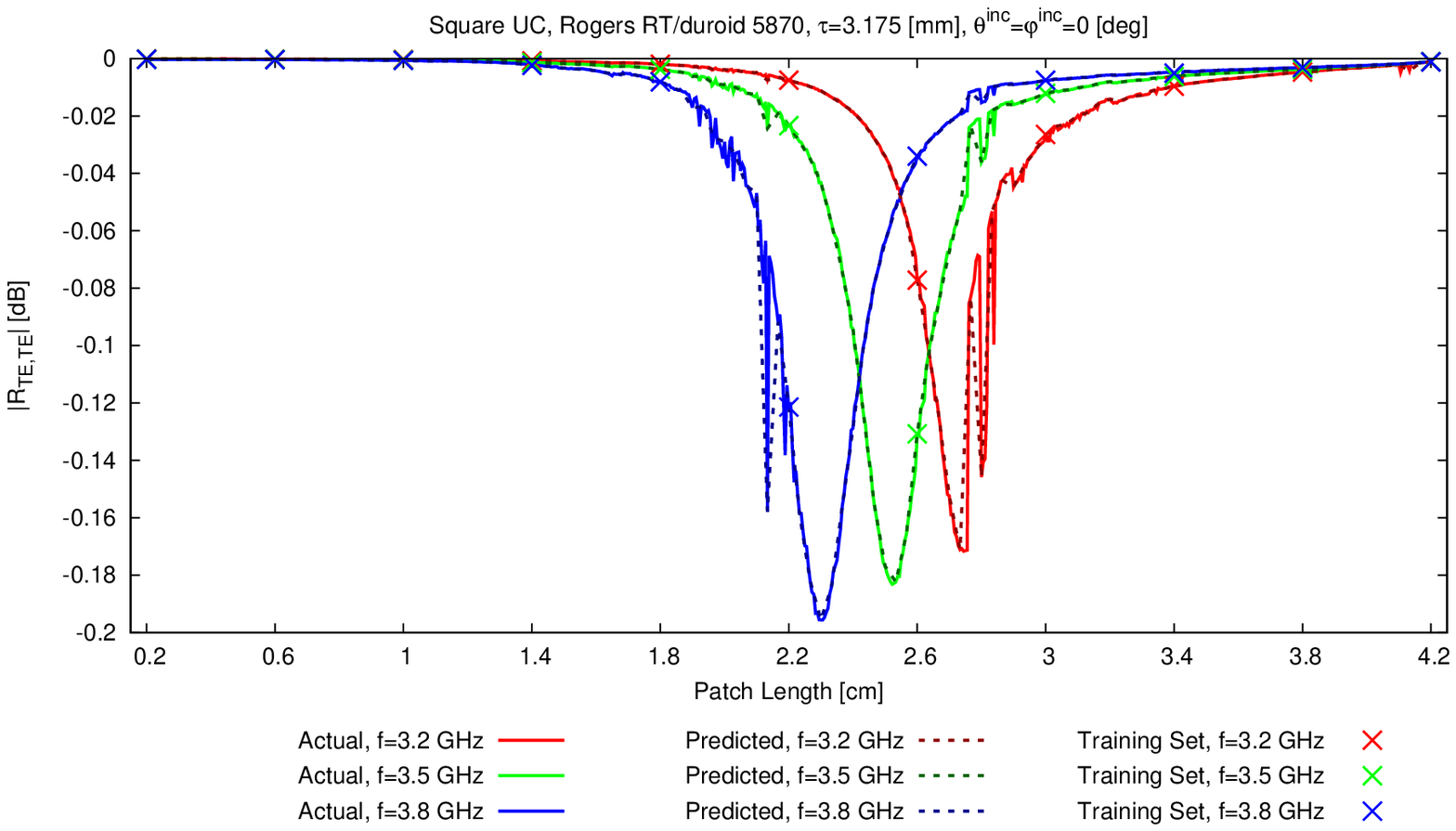}}\tabularnewline
\multicolumn{2}{c}{(\emph{d})}\tabularnewline
\end{tabular}\end{center}

\begin{center}\textbf{Fig. 3 - G. Oliveri et} \textbf{\emph{al.}}\textbf{,}
{}``Building a Smart \emph{EM} Environment - \emph{AI}-Enhanced Aperiodic
...''\end{center}
\newpage

\begin{center}\begin{tabular}{c}
\includegraphics[%
  clip,
  width=0.55\columnwidth,
  keepaspectratio]{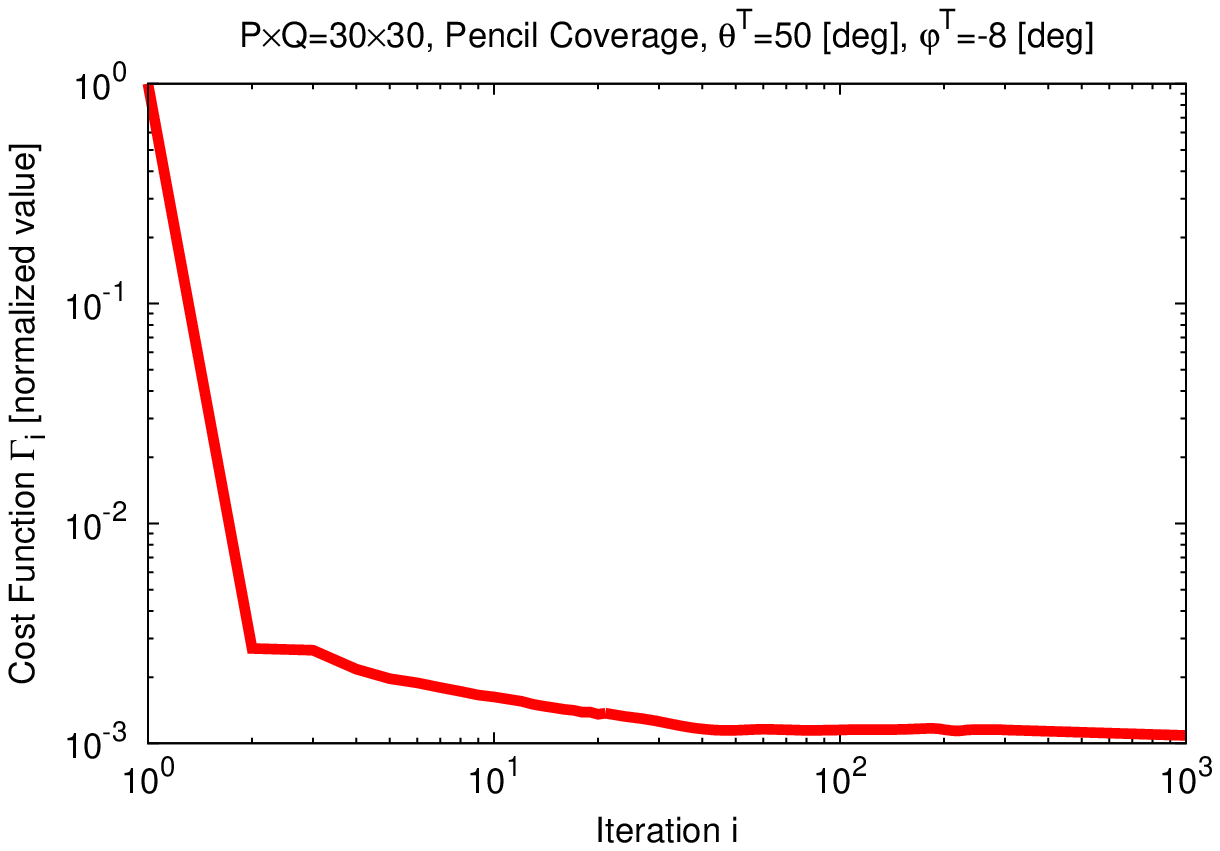}\tabularnewline
(\emph{a})\tabularnewline
\includegraphics[%
  clip,
  width=0.54\columnwidth,
  keepaspectratio]{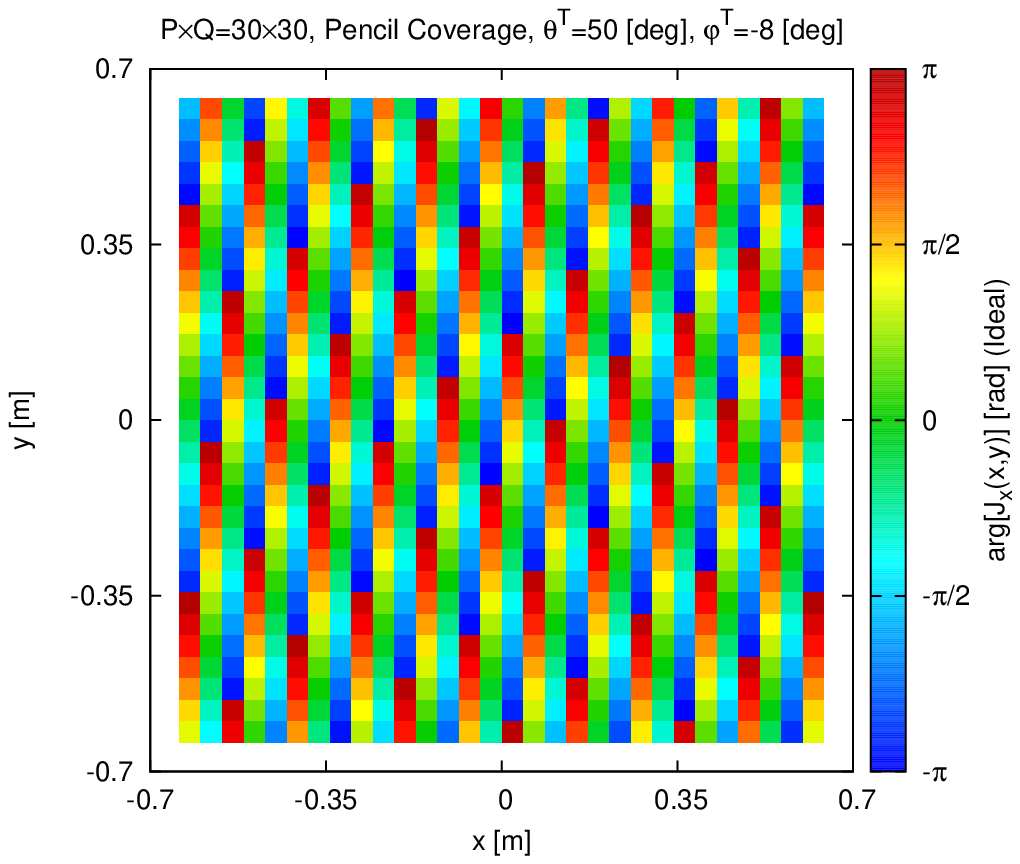}\tabularnewline
(\emph{b})\tabularnewline
\includegraphics[%
  clip,
  width=0.54\columnwidth,
  keepaspectratio]{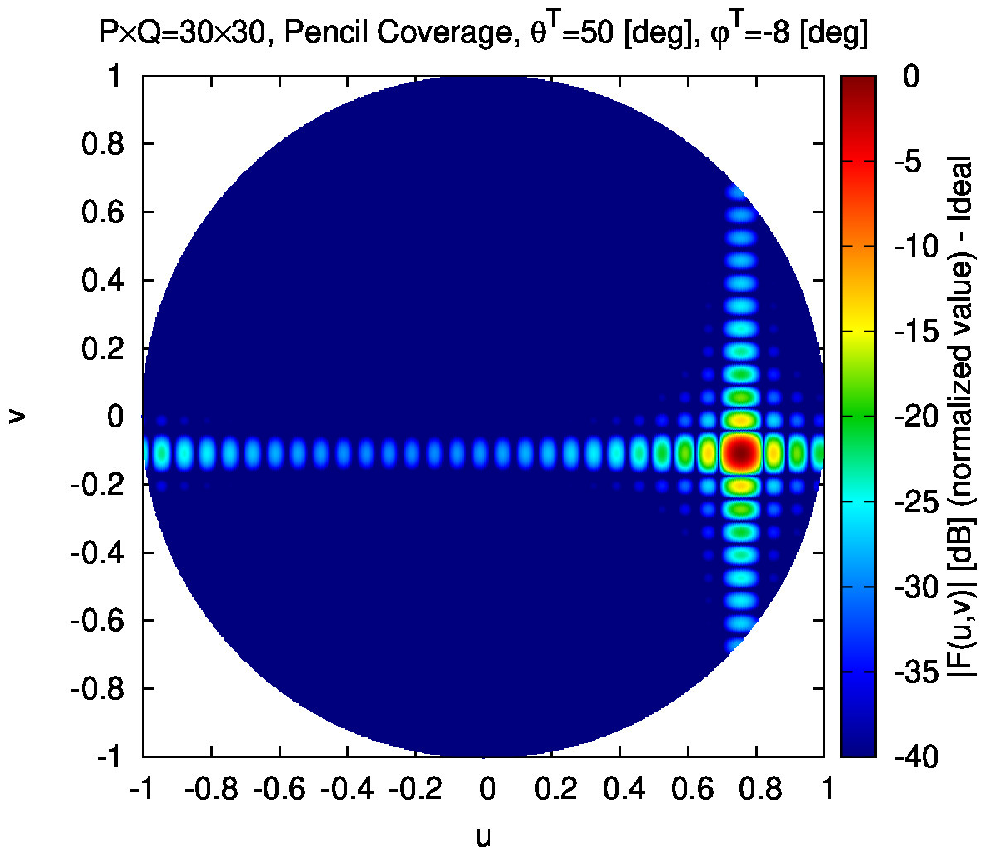}\tabularnewline
(c)\tabularnewline
\end{tabular}\end{center}

\begin{center}\textbf{Fig. 4 - G. Oliveri et} \textbf{\emph{al.}}\textbf{,}
{}``Building a Smart \emph{EM} Environment - \emph{AI}-Enhanced Aperiodic
...''\end{center}
\newpage

\begin{center}\begin{tabular}{c}
\includegraphics[%
  clip,
  width=0.55\columnwidth,
  keepaspectratio]{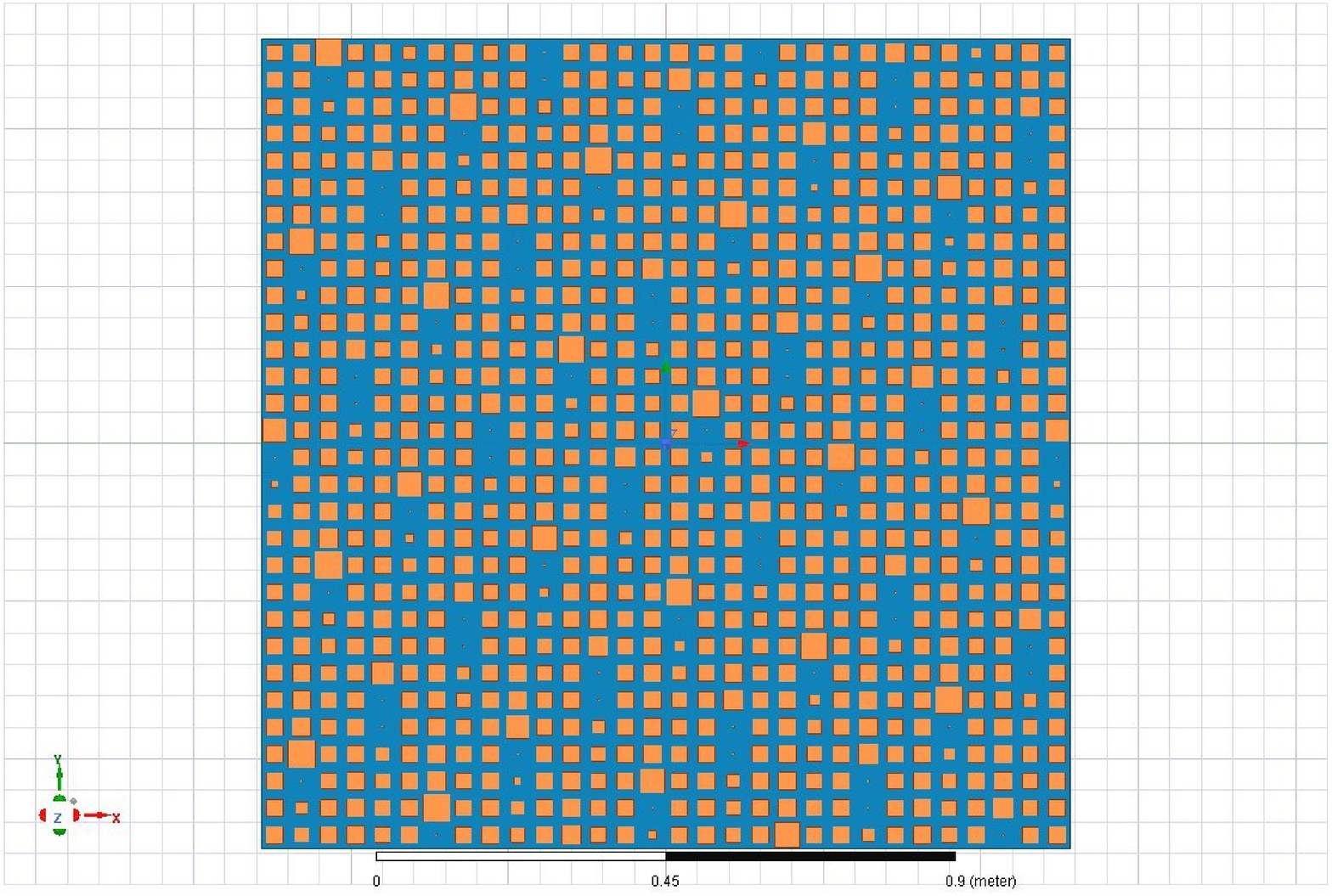}\tabularnewline
(\emph{a})\tabularnewline
\includegraphics[%
  clip,
  width=0.54\columnwidth,
  keepaspectratio]{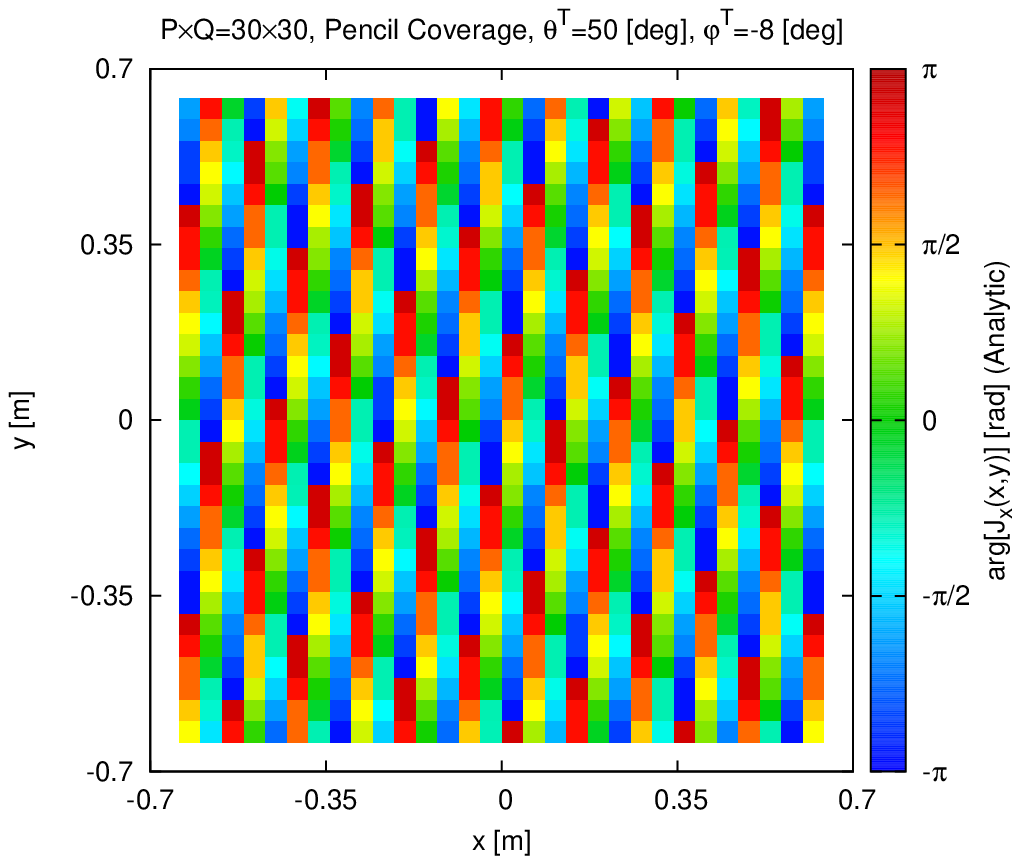}\tabularnewline
(\emph{b})\tabularnewline
\includegraphics[%
  clip,
  width=0.54\columnwidth,
  keepaspectratio]{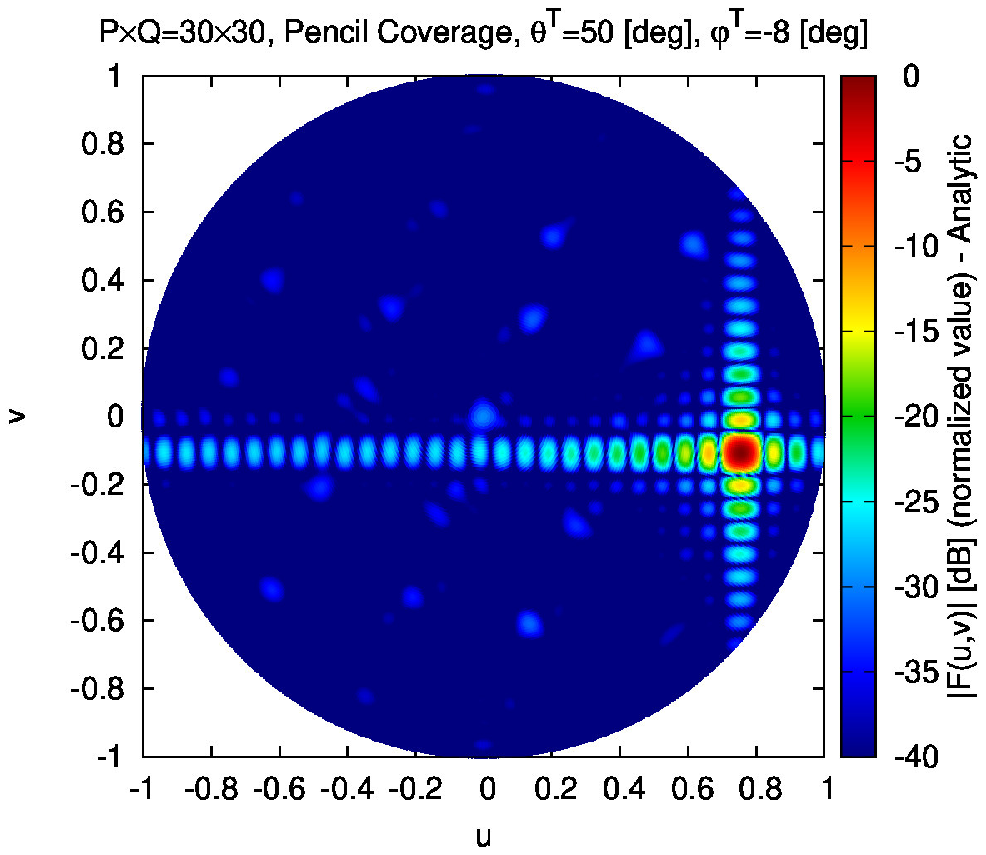}\tabularnewline
(\emph{c})\tabularnewline
\end{tabular}\end{center}

\begin{center}\textbf{Fig. 5 - G. Oliveri et} \textbf{\emph{al.}}\textbf{,}
{}``Building a Smart \emph{EM} Environment - \emph{AI}-Enhanced Aperiodic
...''\end{center}
\newpage

\begin{center}~\vfill\end{center}

\begin{center}\begin{tabular}{c}
\includegraphics[%
  clip,
  width=0.54\columnwidth,
  keepaspectratio]{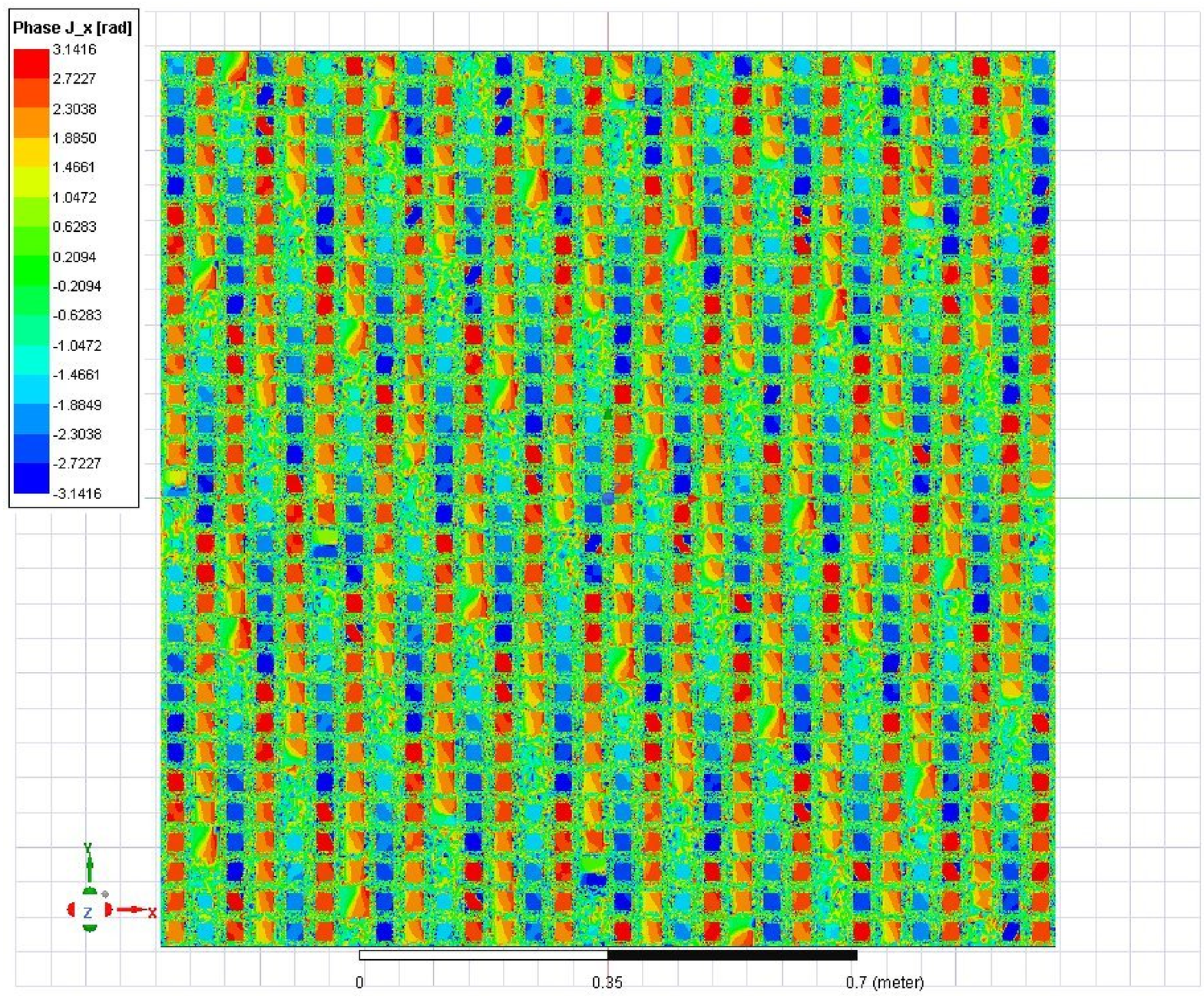}\tabularnewline
(\emph{a})\tabularnewline
\tabularnewline
\includegraphics[%
  clip,
  width=0.55\columnwidth,
  keepaspectratio]{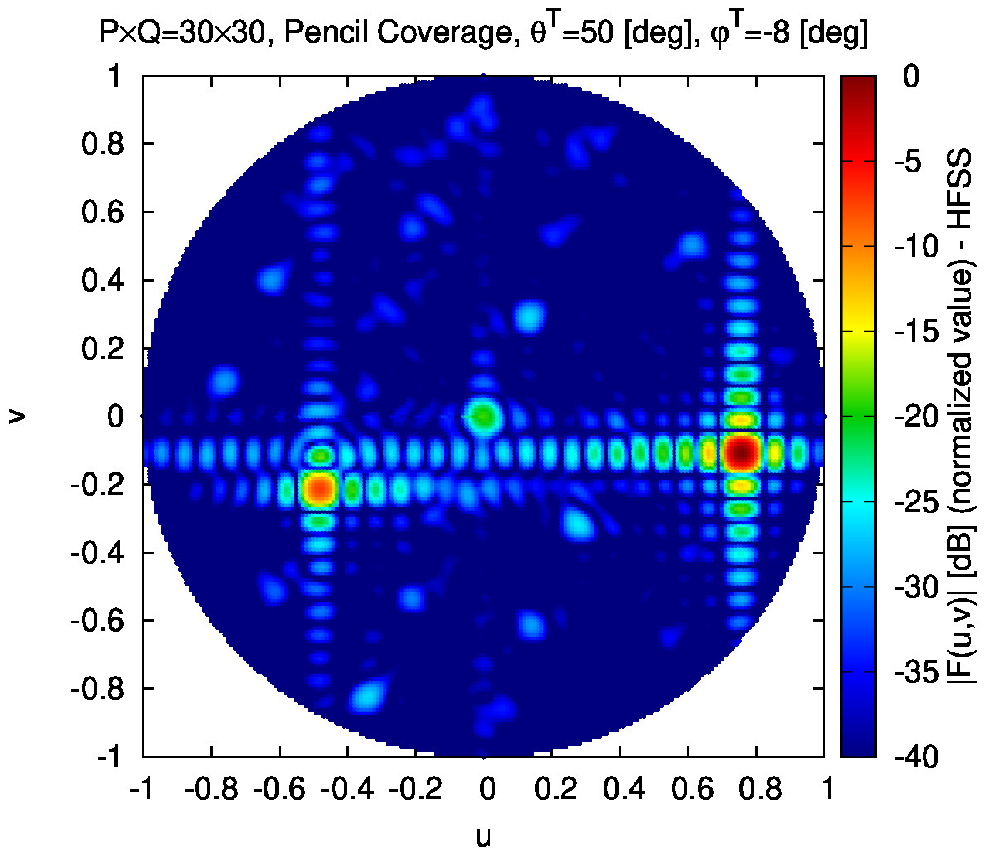}\tabularnewline
(\emph{b})\tabularnewline
\end{tabular}\end{center}

\begin{center}~\vfill\end{center}

\begin{center}\textbf{Fig. 6 - G. Oliveri et} \textbf{\emph{al.}}\textbf{,}
{}``Building a Smart \emph{EM} Environment - \emph{AI}-Enhanced Aperiodic
...''\end{center}
\newpage

\begin{center}\begin{sideways}
\begin{tabular}{cc}
\includegraphics[%
  clip,
  width=0.80\columnwidth,
  keepaspectratio]{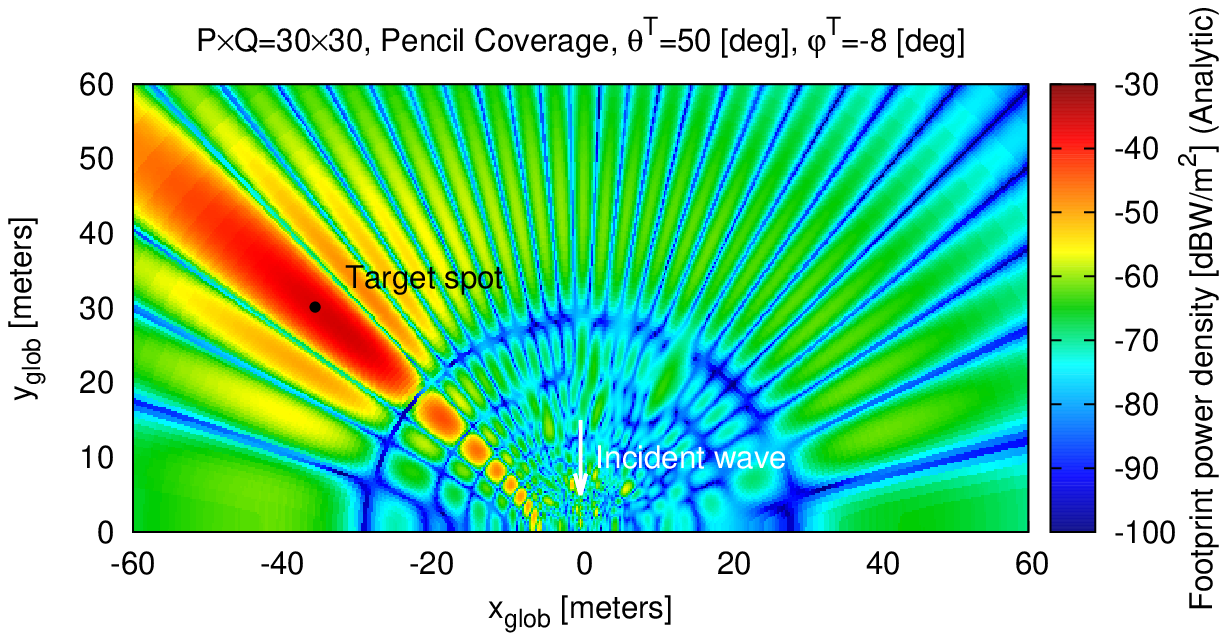}&
\includegraphics[%
  clip,
  width=0.55\columnwidth,
  keepaspectratio]{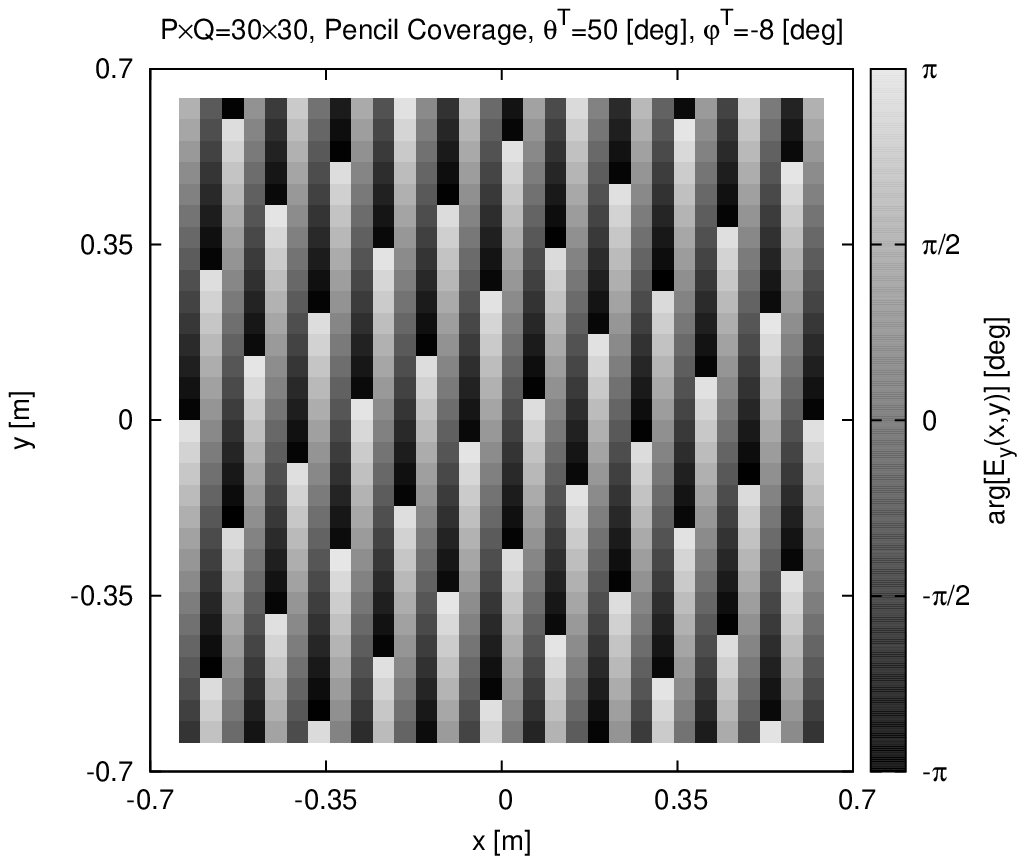}\tabularnewline
(\emph{a})&
(\emph{c})\tabularnewline
\includegraphics[%
  clip,
  width=0.80\columnwidth,
  keepaspectratio]{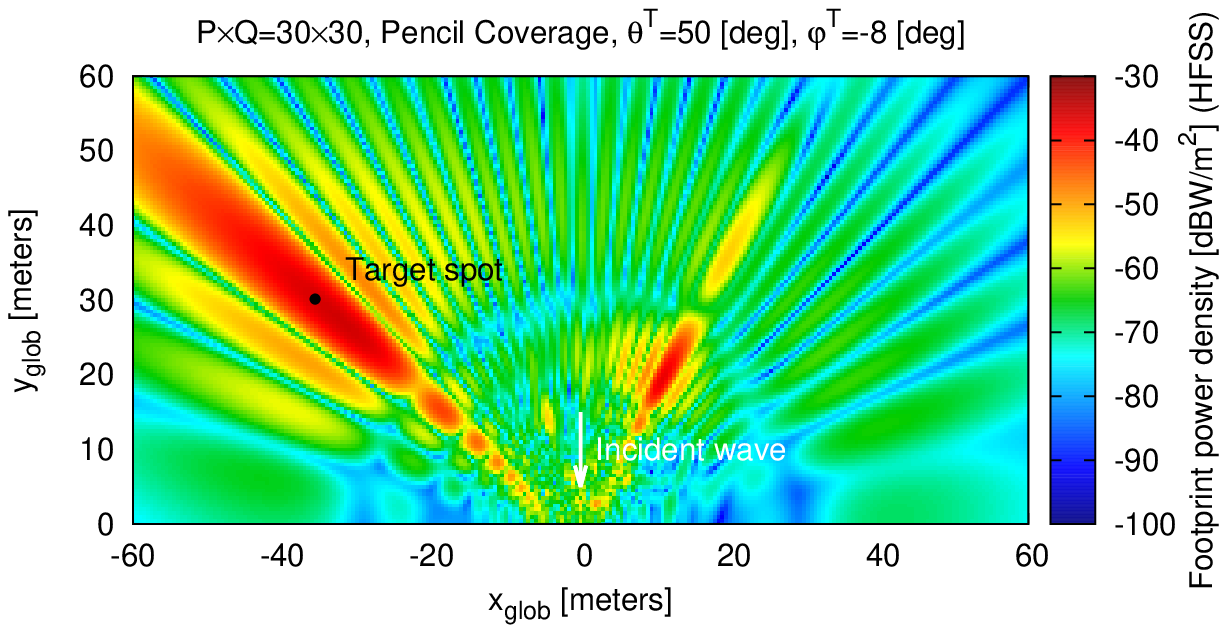}&
\includegraphics[%
  clip,
  width=0.48\columnwidth,
  keepaspectratio]{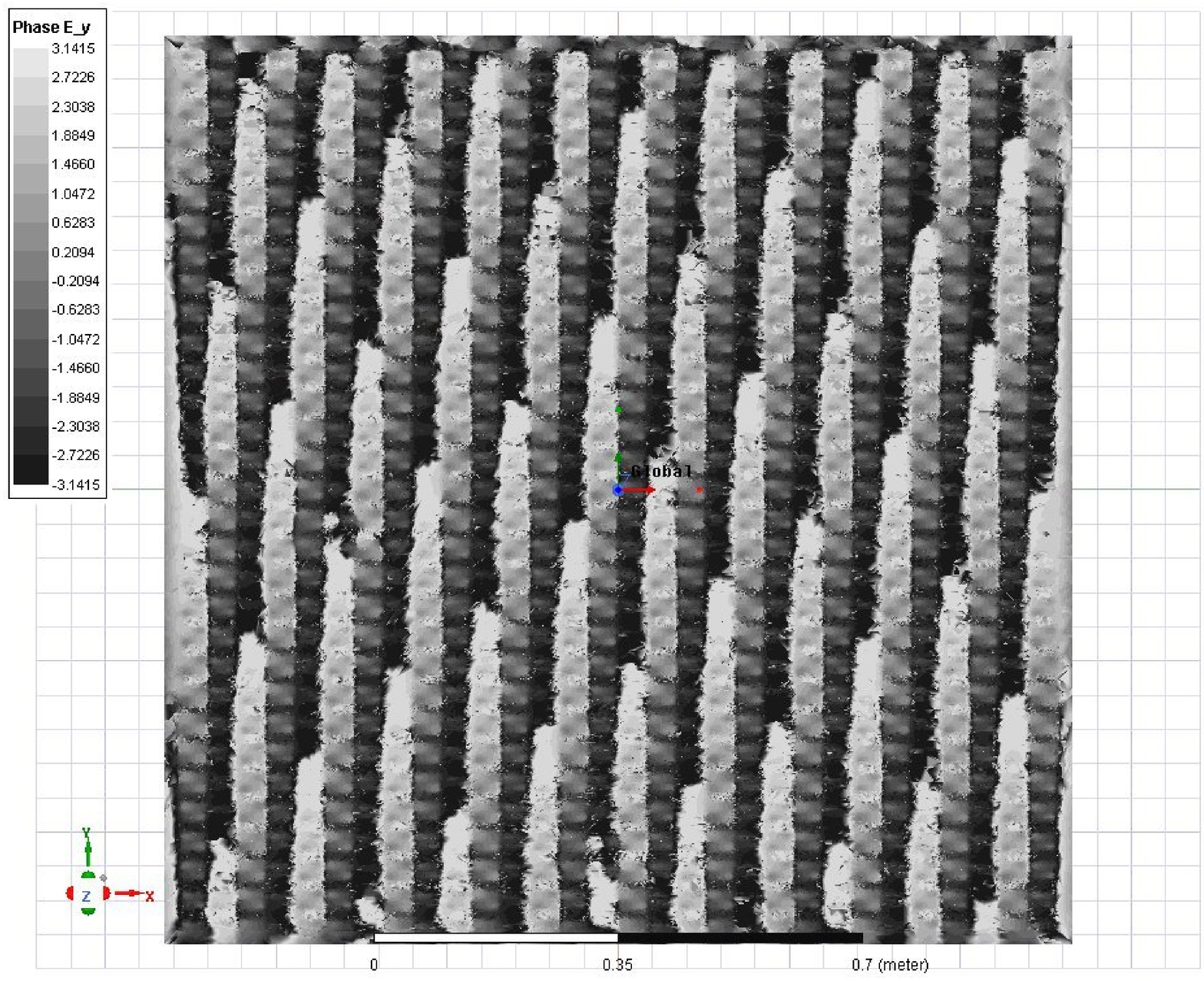}\tabularnewline
(\emph{b})&
(d)\tabularnewline
\end{tabular}
\end{sideways}\end{center}

\begin{center}\textbf{Fig. 7 - G. Oliveri et} \textbf{\emph{al.}}\textbf{,}
{}``Building a Smart \emph{EM} Environment - \emph{AI}-Enhanced Aperiodic
...''~\vfill\end{center}
\newpage

\begin{center}\begin{tabular}{cc}
\includegraphics[%
  clip,
  width=0.50\columnwidth,
  keepaspectratio]{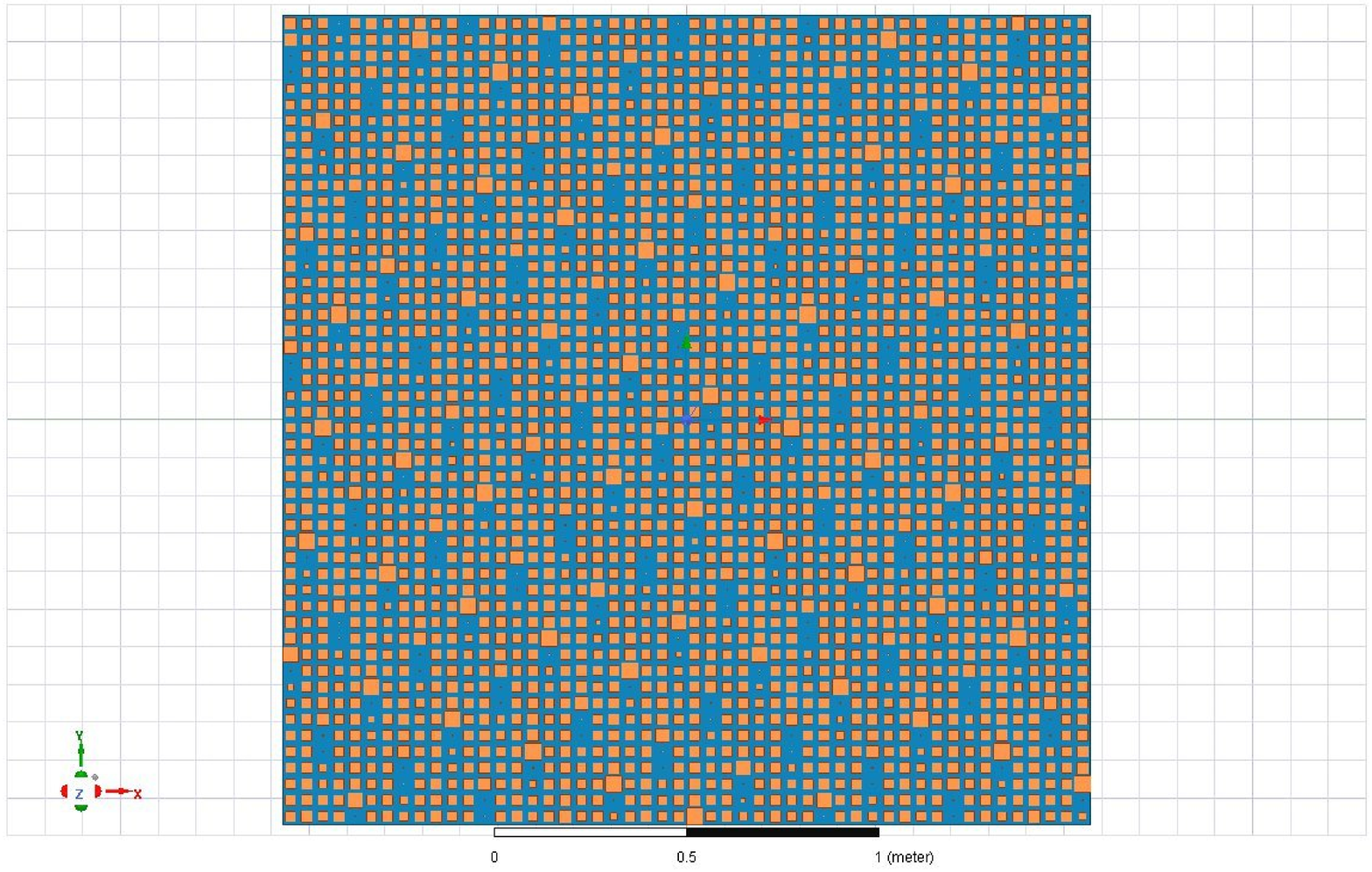}&
\includegraphics[%
  clip,
  width=0.45\columnwidth,
  keepaspectratio]{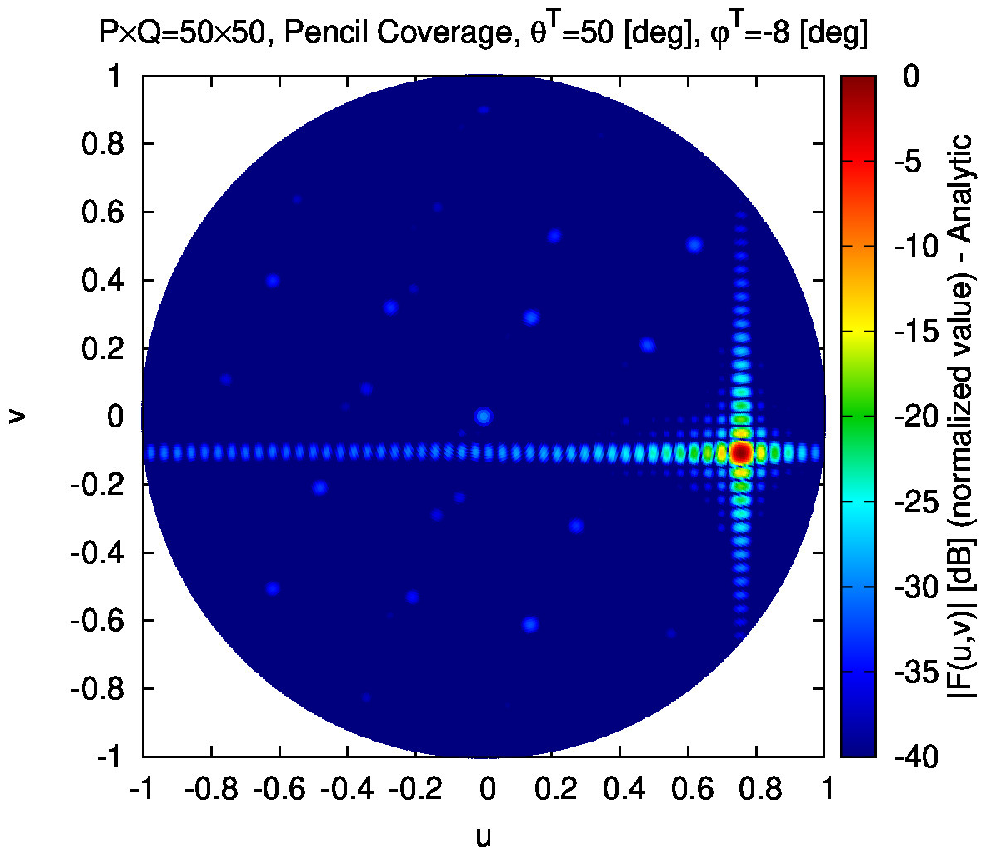}\tabularnewline
(\emph{a})&
(\emph{b})\tabularnewline
\includegraphics[%
  clip,
  width=0.50\columnwidth,
  keepaspectratio]{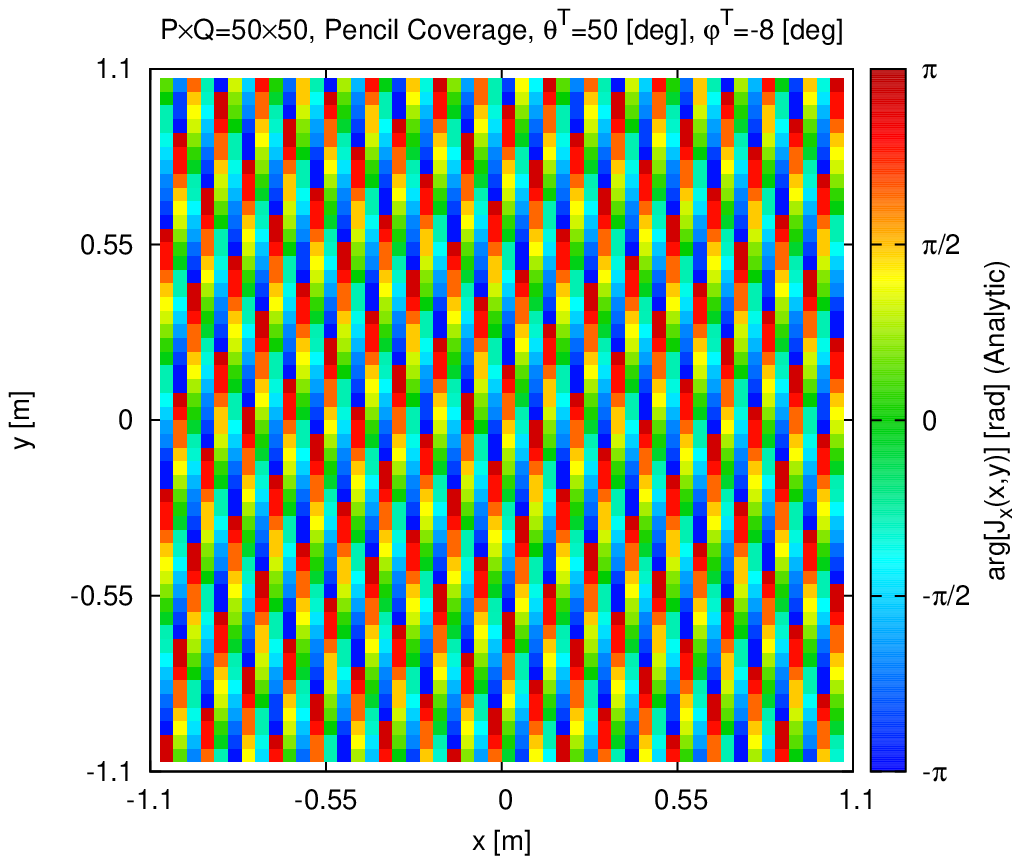}&
\includegraphics[%
  clip,
  width=0.50\columnwidth,
  keepaspectratio]{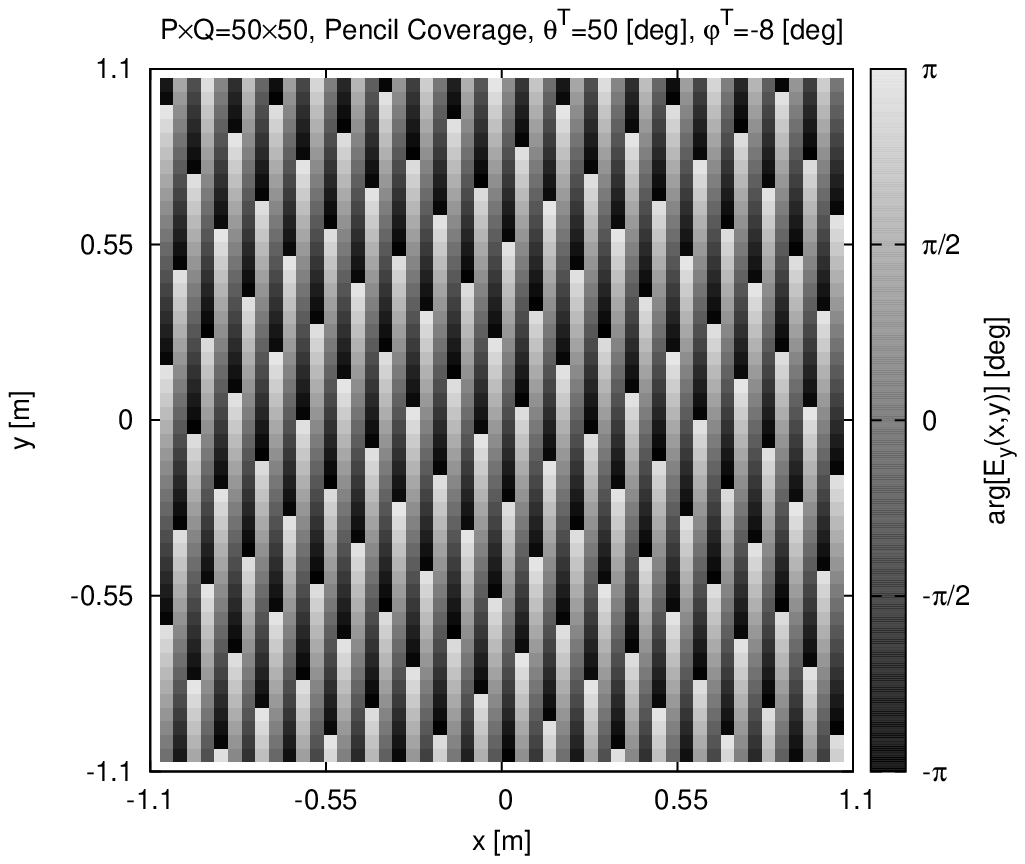}\tabularnewline
(\emph{d})&
(\emph{e})\tabularnewline
\multicolumn{2}{c}{\includegraphics[%
  width=0.80\columnwidth,
  keepaspectratio]{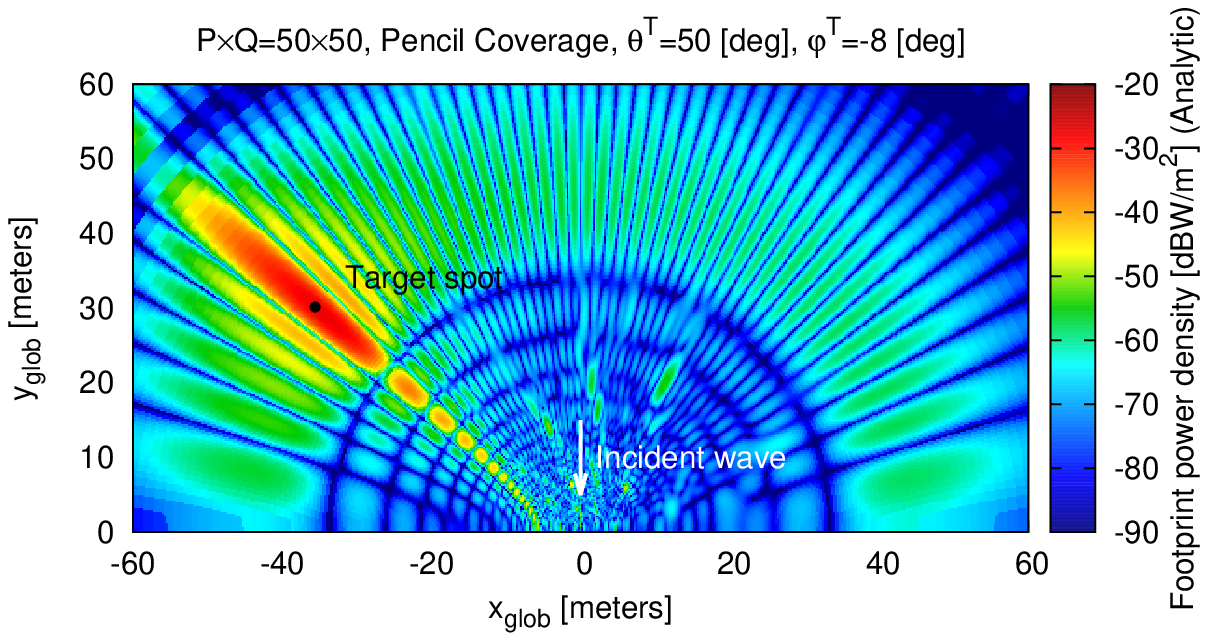}}\tabularnewline
\multicolumn{2}{m{0.50\columnwidth}}{(\emph{c})}\tabularnewline
\end{tabular}\end{center}

\begin{center}\textbf{Fig. 8 - G. Oliveri et} \textbf{\emph{al.}}\textbf{,}
{}``Building a Smart \emph{EM} Environment - \emph{AI}-Enhanced Aperiodic
...''\end{center}
\newpage

\begin{center}\begin{tabular}{cc}
\multicolumn{2}{c}{\includegraphics[%
  clip,
  width=0.50\columnwidth,
  keepaspectratio]{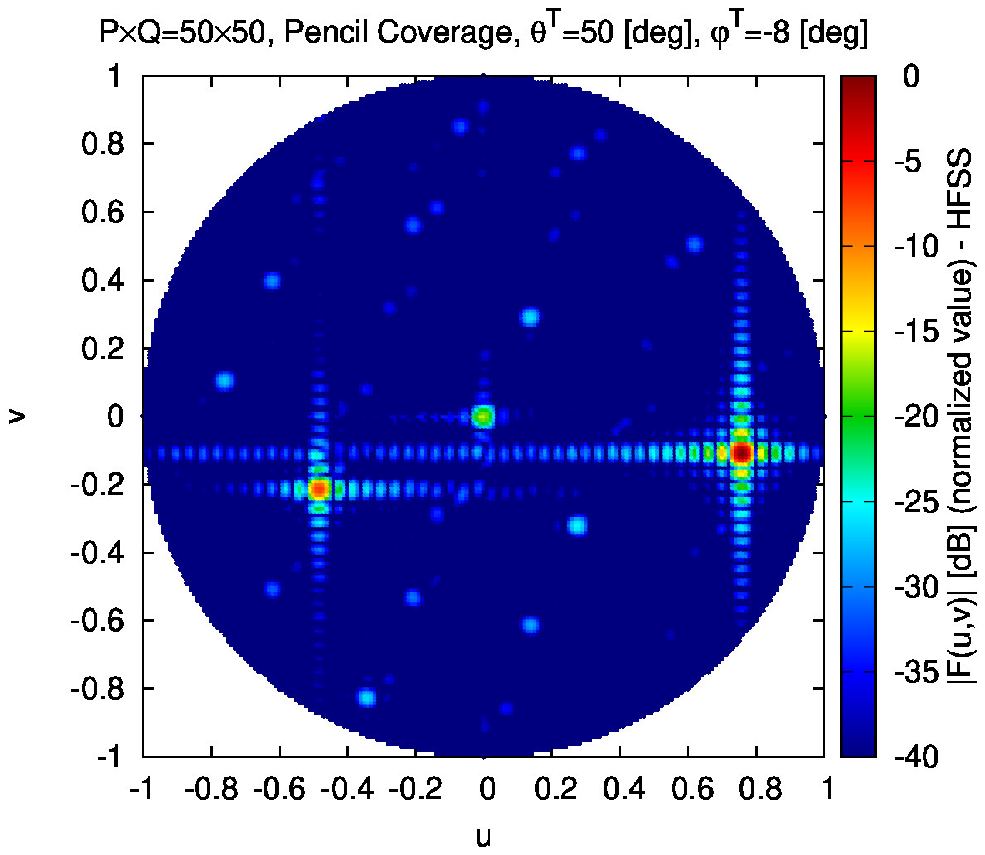}}\tabularnewline
\multicolumn{2}{c}{(\emph{a})}\tabularnewline
\includegraphics[%
  clip,
  width=0.50\columnwidth,
  keepaspectratio]{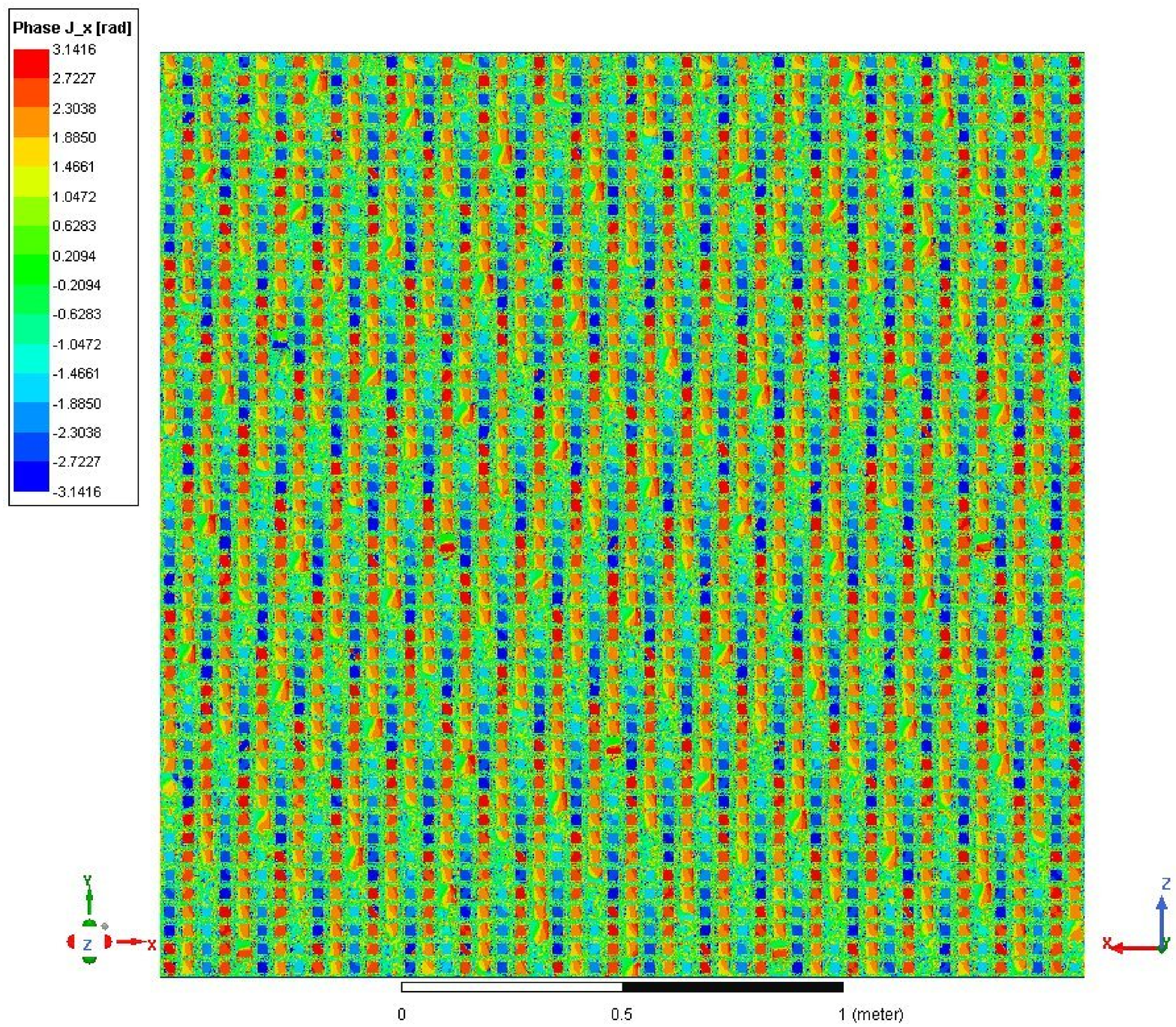}&
\includegraphics[%
  clip,
  width=0.49\columnwidth,
  keepaspectratio]{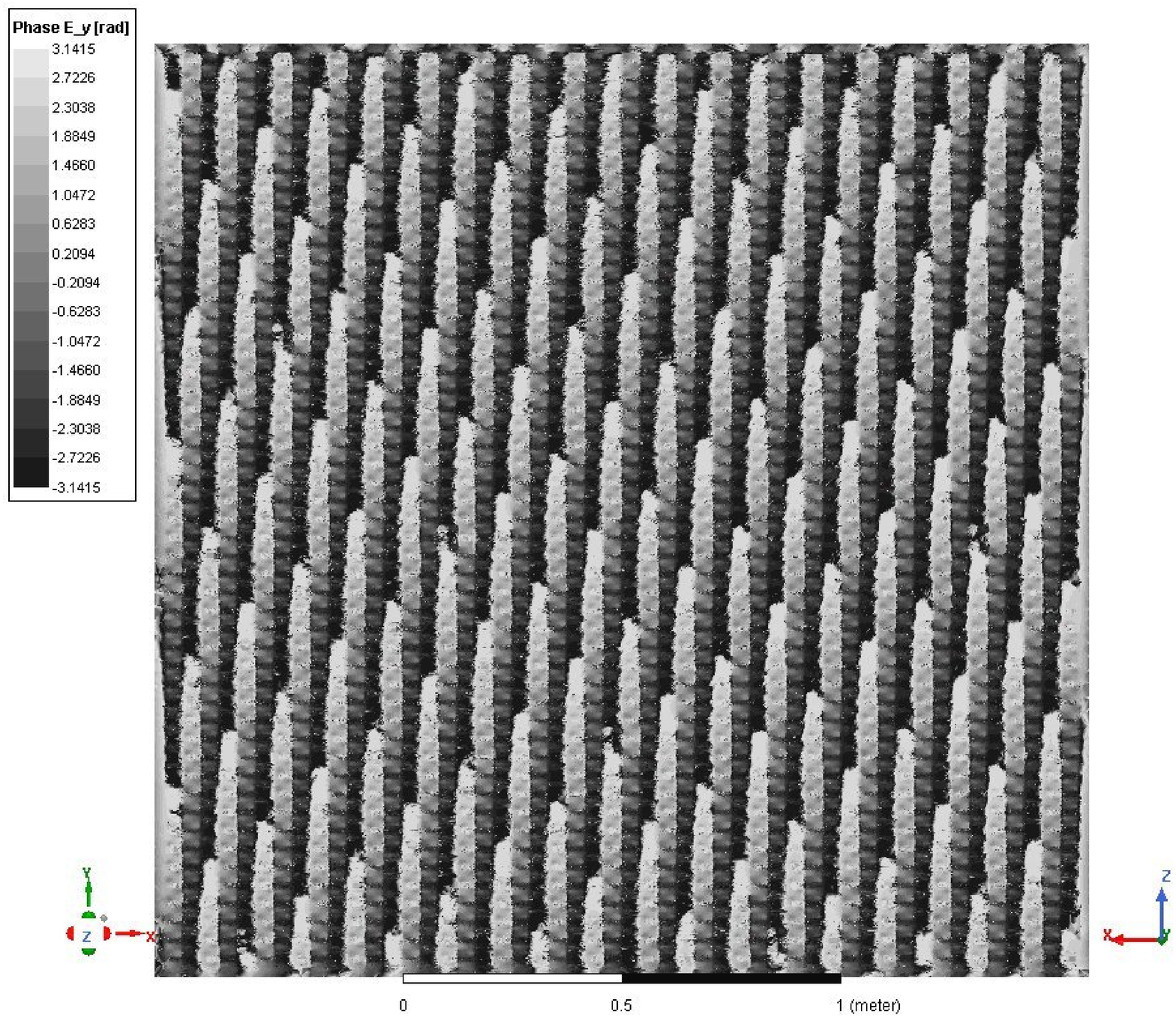}\tabularnewline
(\emph{c})&
(\emph{d})\tabularnewline
\multicolumn{2}{c}{\includegraphics[%
  clip,
  width=0.80\columnwidth,
  keepaspectratio]{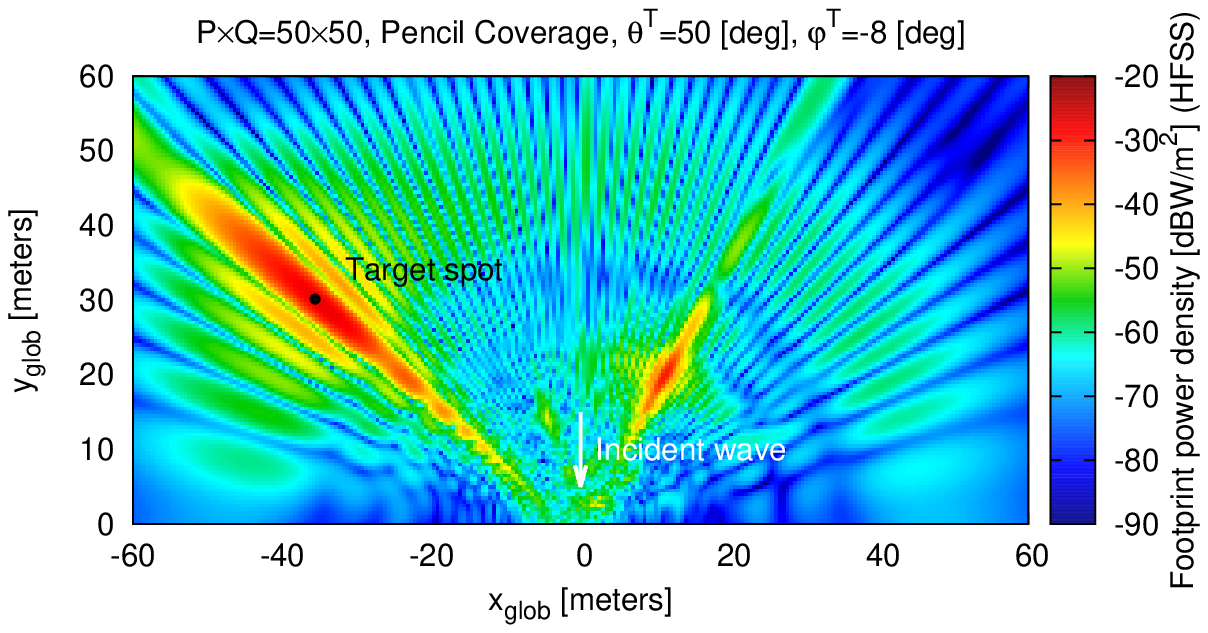}}\tabularnewline
\multicolumn{2}{c}{(\emph{b})}\tabularnewline
\end{tabular}\end{center}

\begin{center}\textbf{Fig. 9 - G. Oliveri et} \textbf{\emph{al.}}\textbf{,}
{}``Building a Smart \emph{EM} Environment - \emph{AI}-Enhanced Aperiodic
...''\end{center}
\newpage

\begin{center}\begin{tabular}{cc}
\includegraphics[%
  clip,
  width=0.48\columnwidth,
  keepaspectratio]{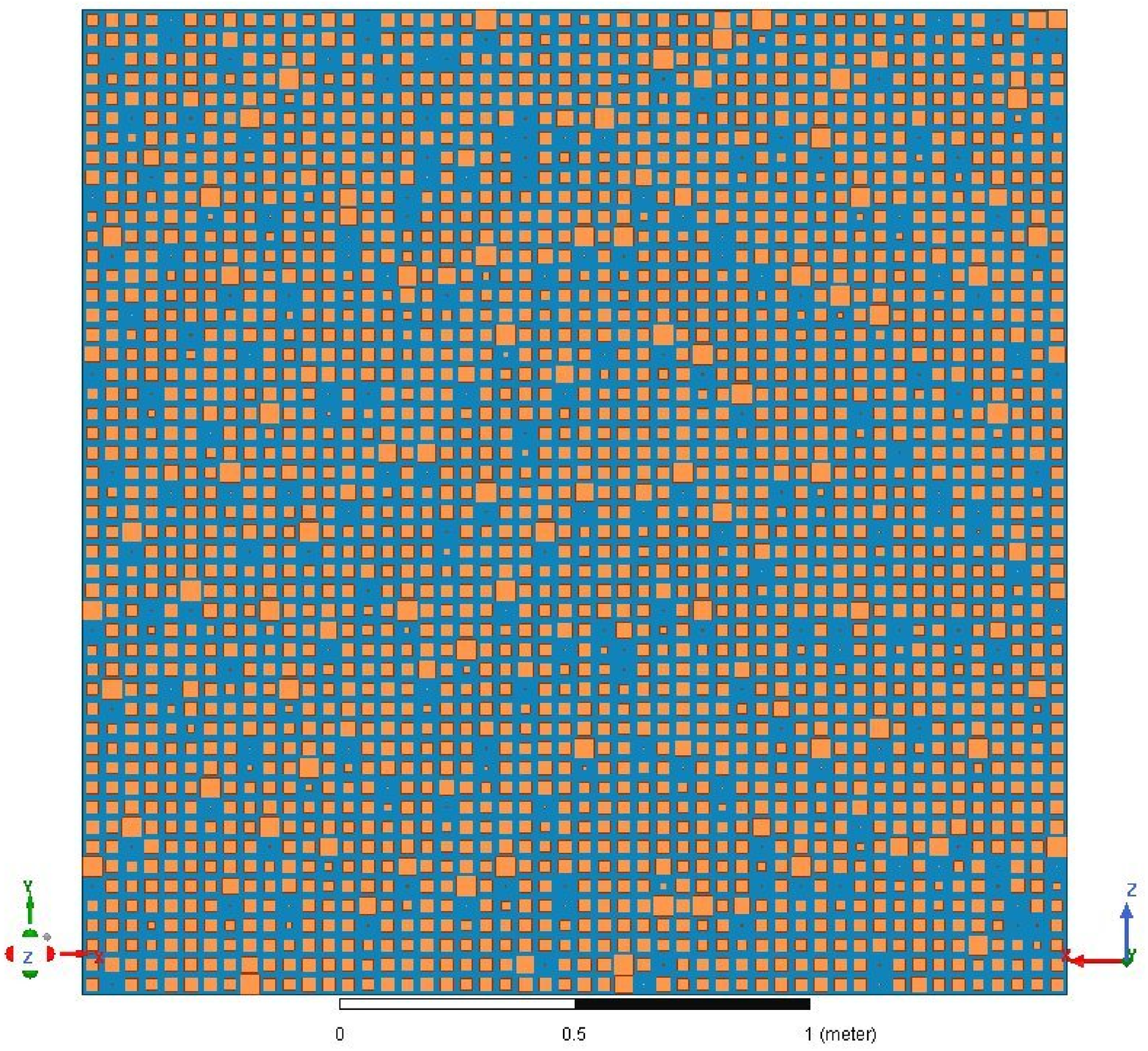}&
\includegraphics[%
  clip,
  width=0.50\columnwidth,
  keepaspectratio]{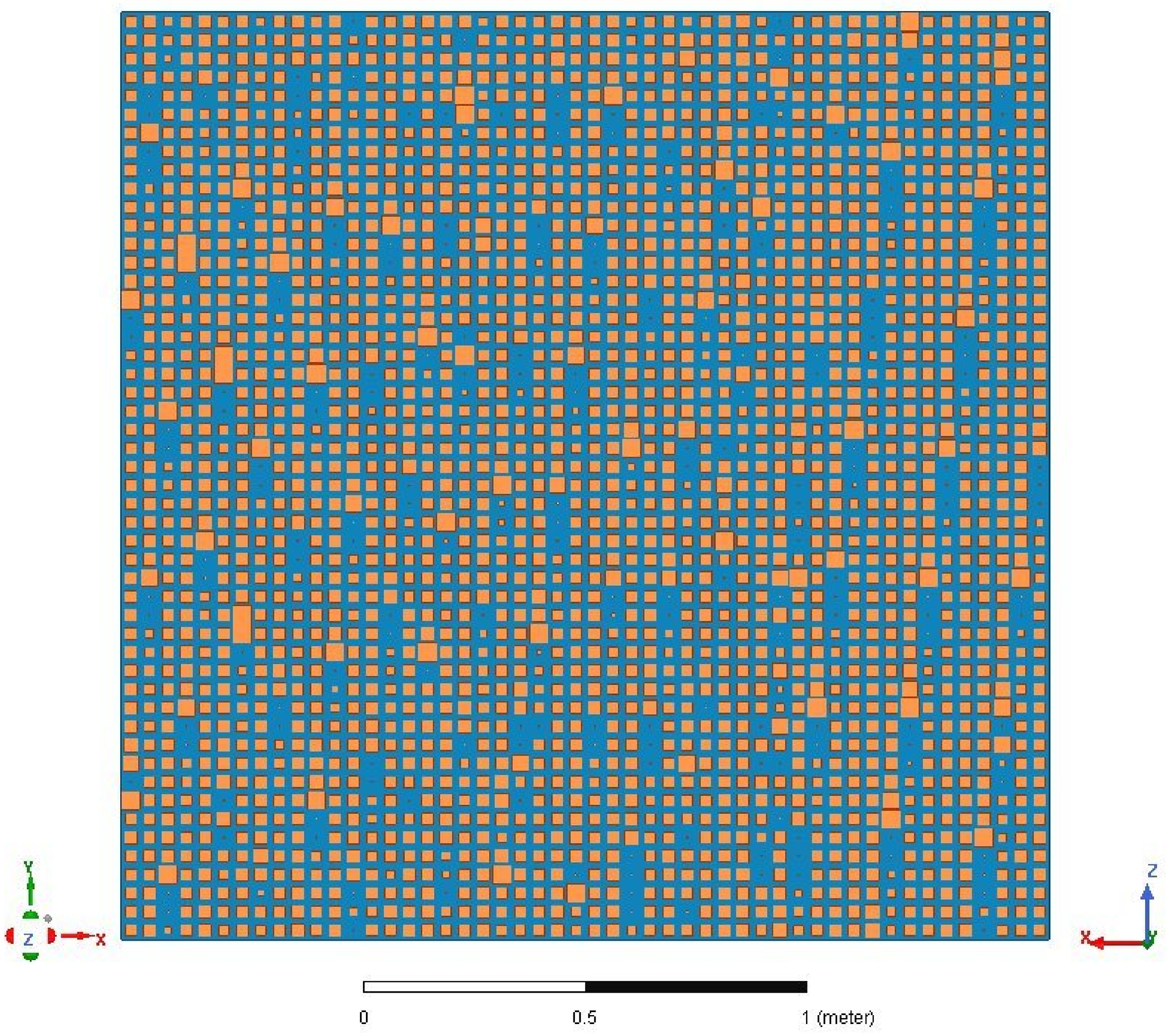}\tabularnewline
(\emph{a})&
(\emph{b})\tabularnewline
\includegraphics[%
  clip,
  width=0.48\columnwidth,
  keepaspectratio]{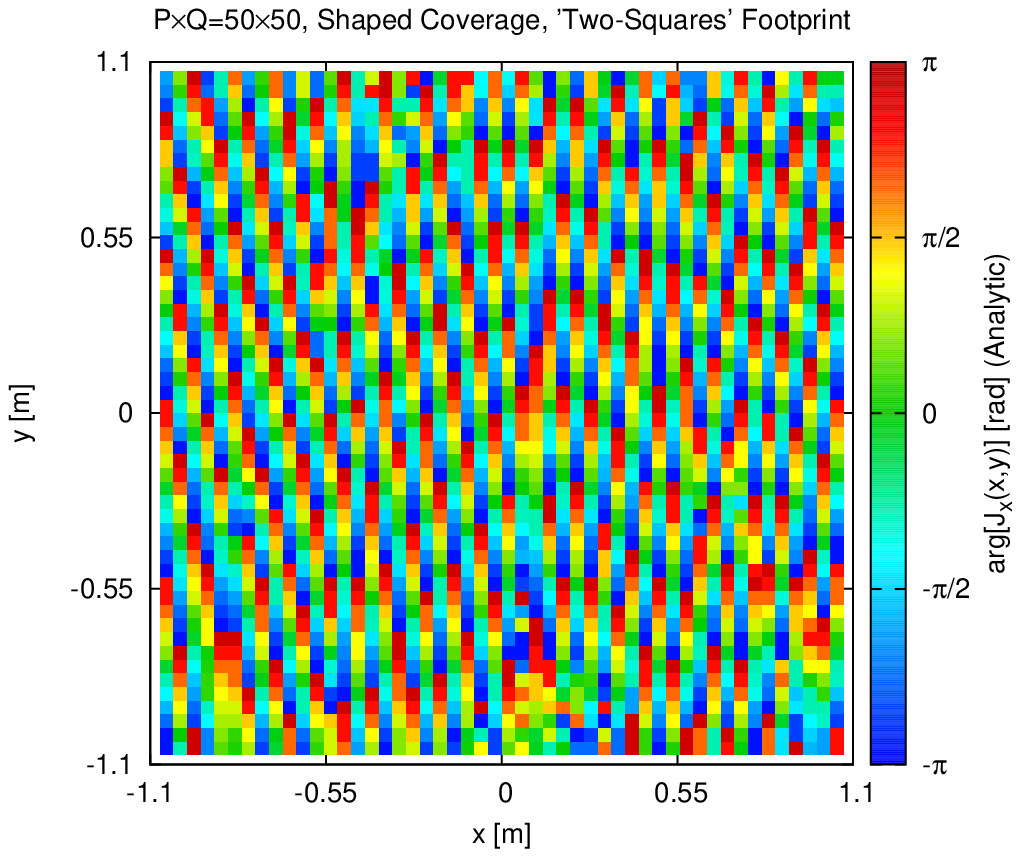}&
\includegraphics[%
  clip,
  width=0.48\columnwidth,
  keepaspectratio]{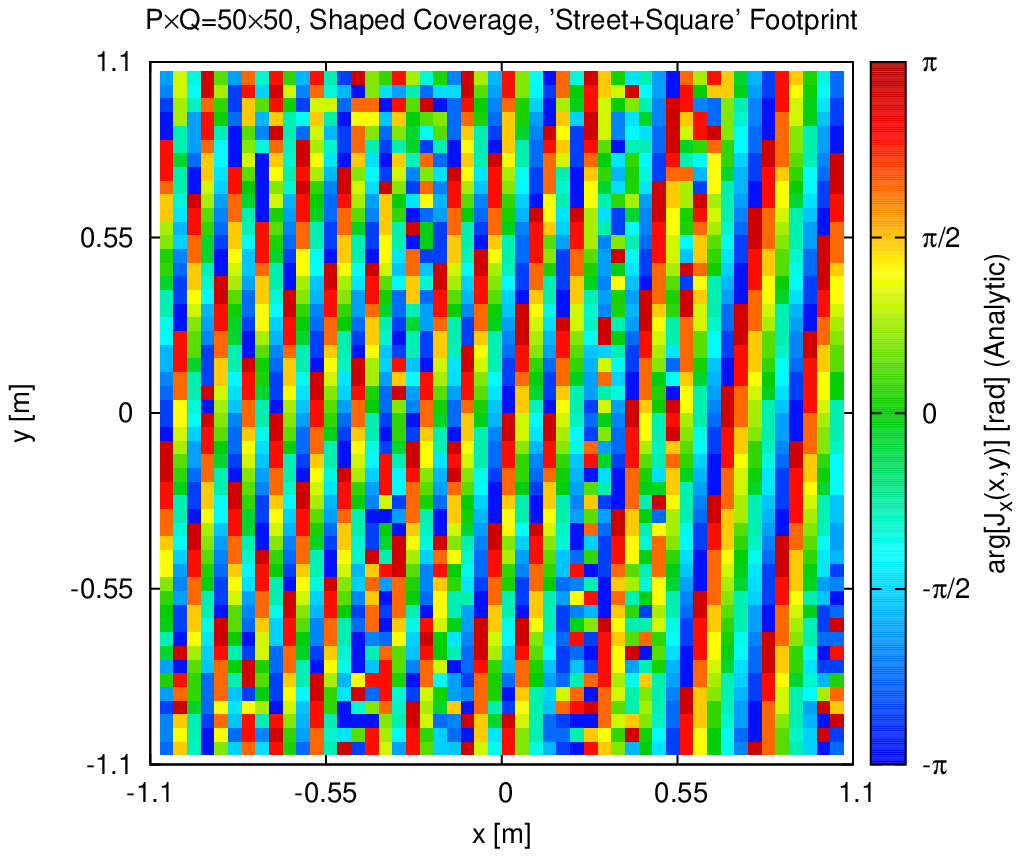}\tabularnewline
(\emph{c})&
(\emph{d})\tabularnewline
\includegraphics[%
  clip,
  width=0.50\columnwidth,
  keepaspectratio]{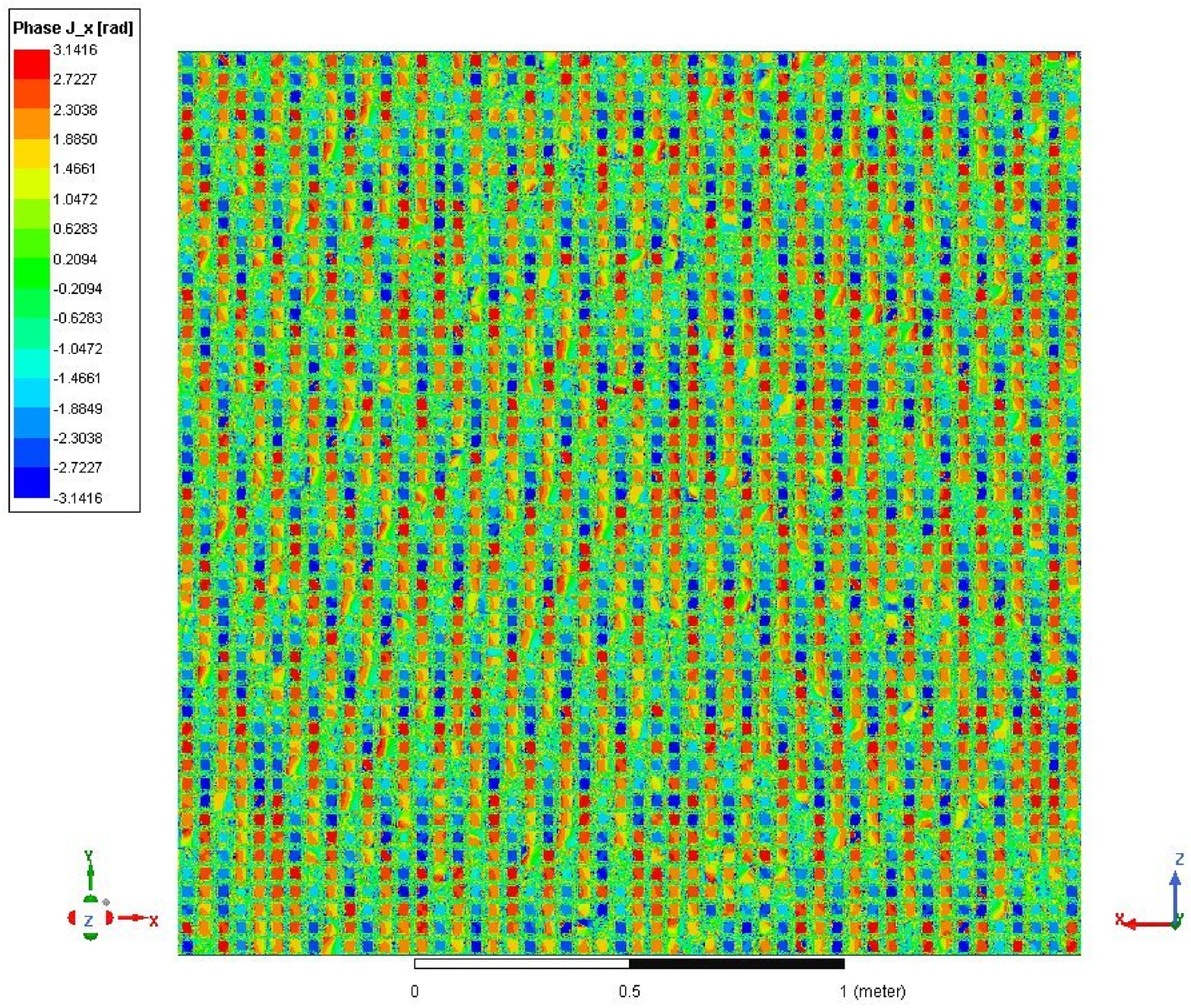}&
\includegraphics[%
  clip,
  width=0.48\columnwidth,
  keepaspectratio]{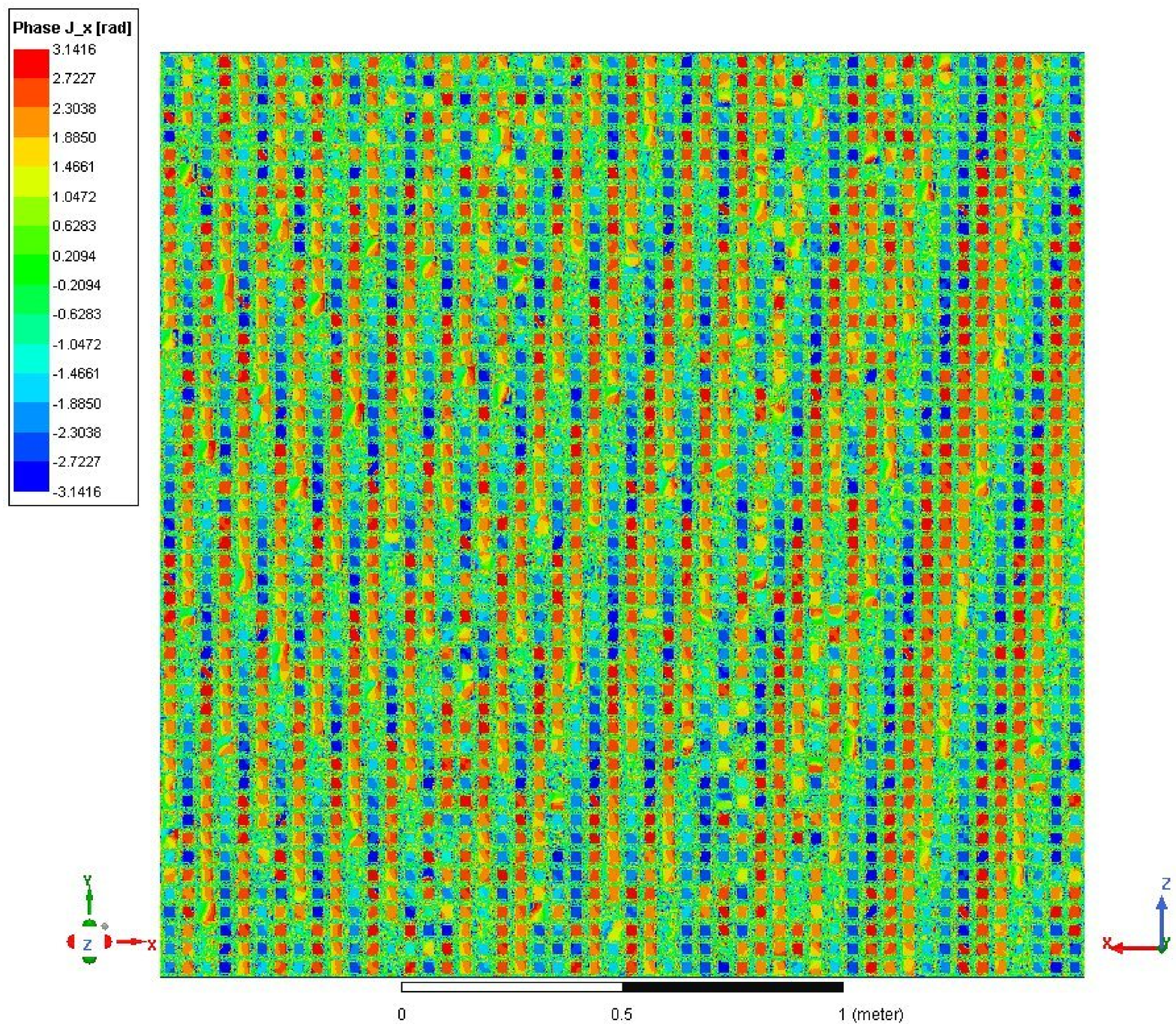}\tabularnewline
(\emph{e})&
(\emph{f})\tabularnewline
\end{tabular}\end{center}

\begin{center}\textbf{Fig. 10 - G. Oliveri et} \textbf{\emph{al.}}\textbf{,}
{}``Building a Smart \emph{EM} Environment - \emph{AI}-Enhanced Aperiodic
...''\end{center}
\newpage

\begin{center}~\vfill\end{center}

\begin{center}\begin{tabular}{cc}
\includegraphics[%
  clip,
  width=0.48\columnwidth,
  keepaspectratio]{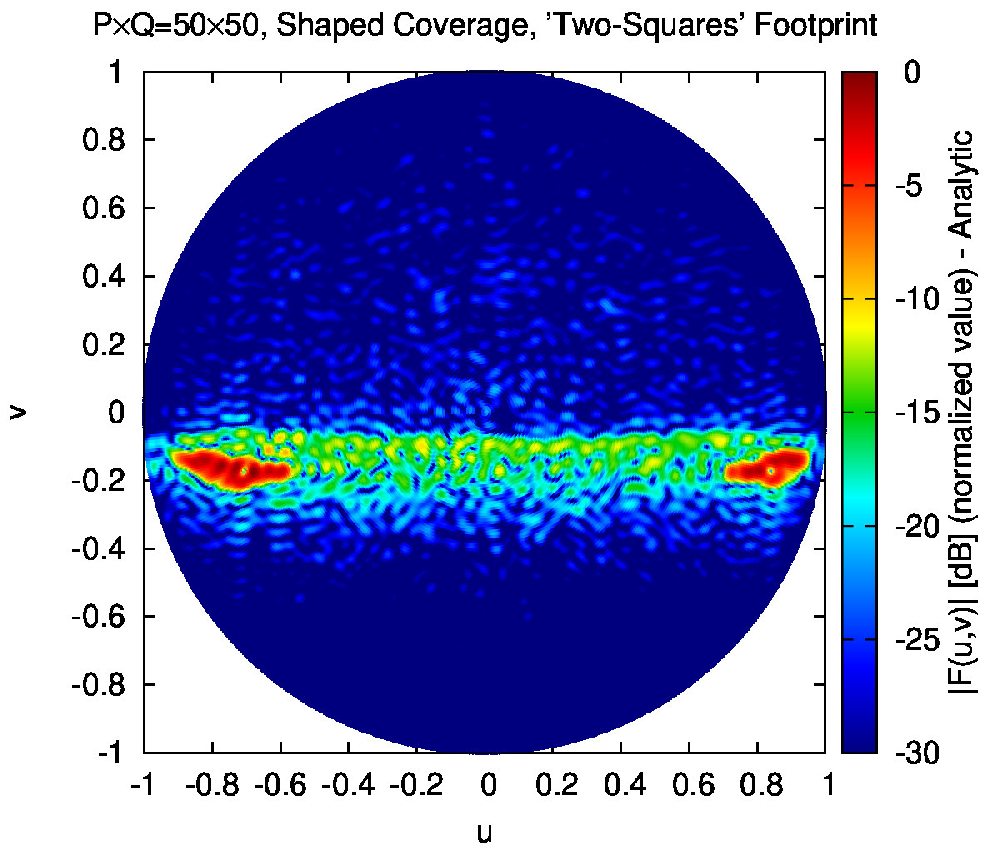}&
\includegraphics[%
  clip,
  width=0.48\columnwidth,
  keepaspectratio]{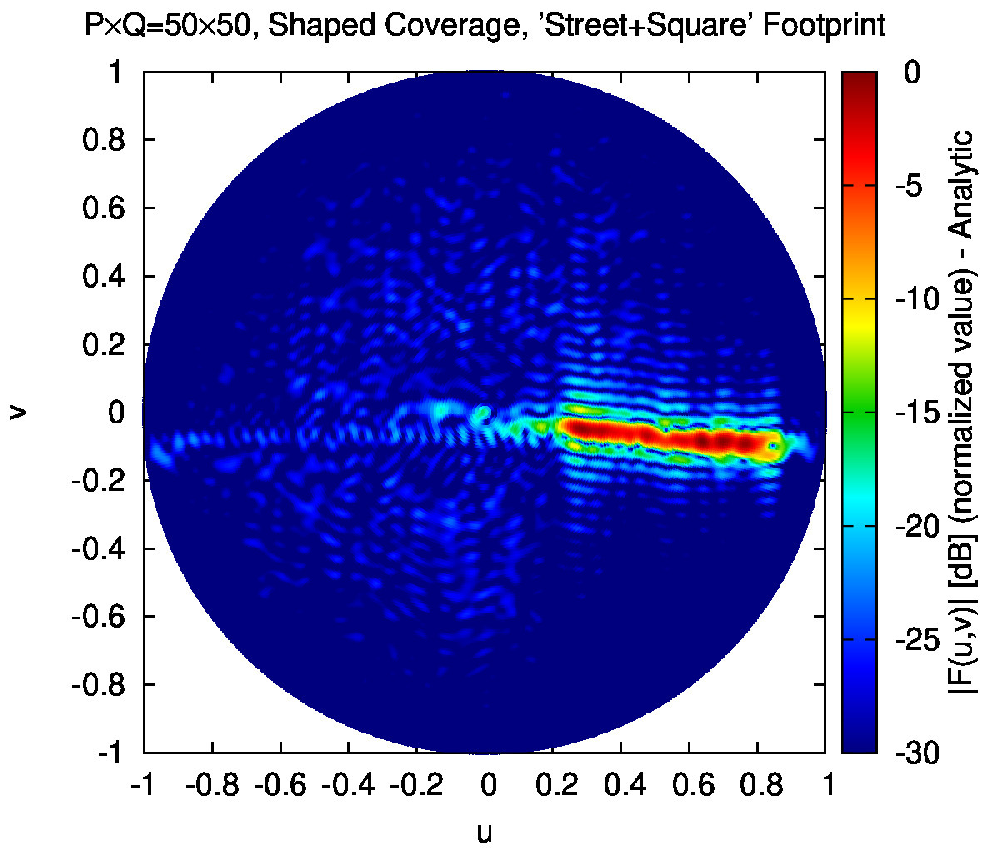}\tabularnewline
(\emph{a})&
(\emph{b})\tabularnewline
&
\tabularnewline
\includegraphics[%
  clip,
  width=0.48\columnwidth,
  keepaspectratio]{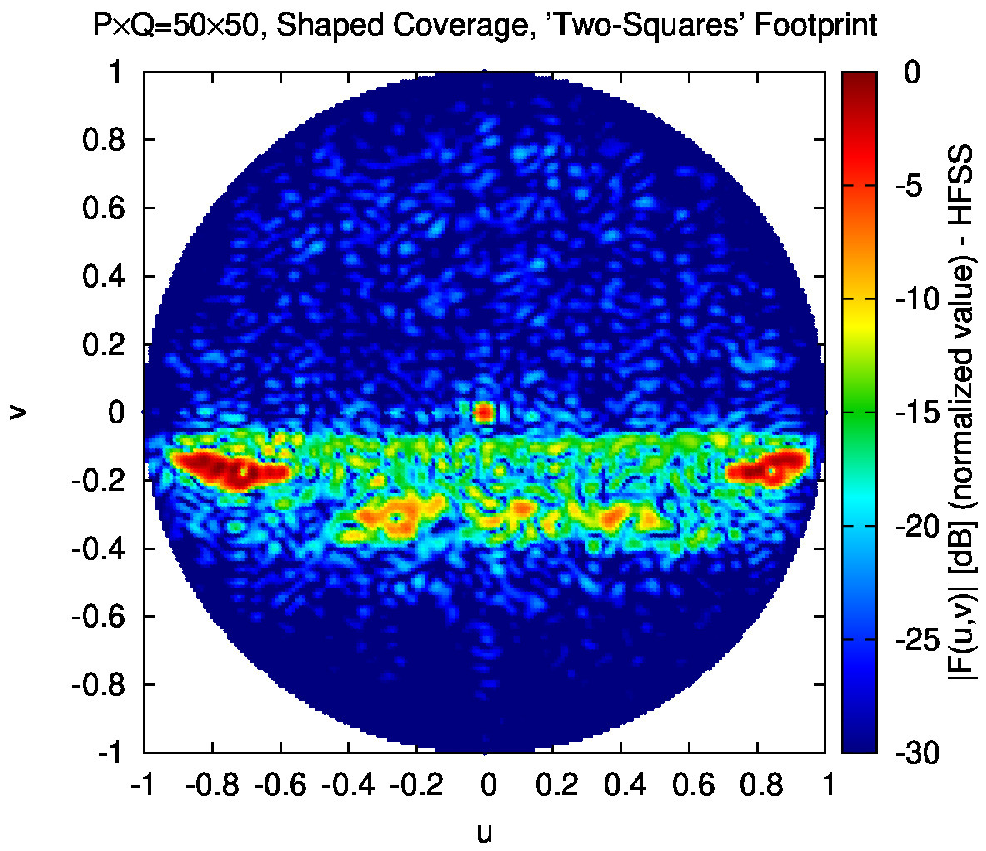}&
\includegraphics[%
  clip,
  width=0.48\columnwidth,
  keepaspectratio]{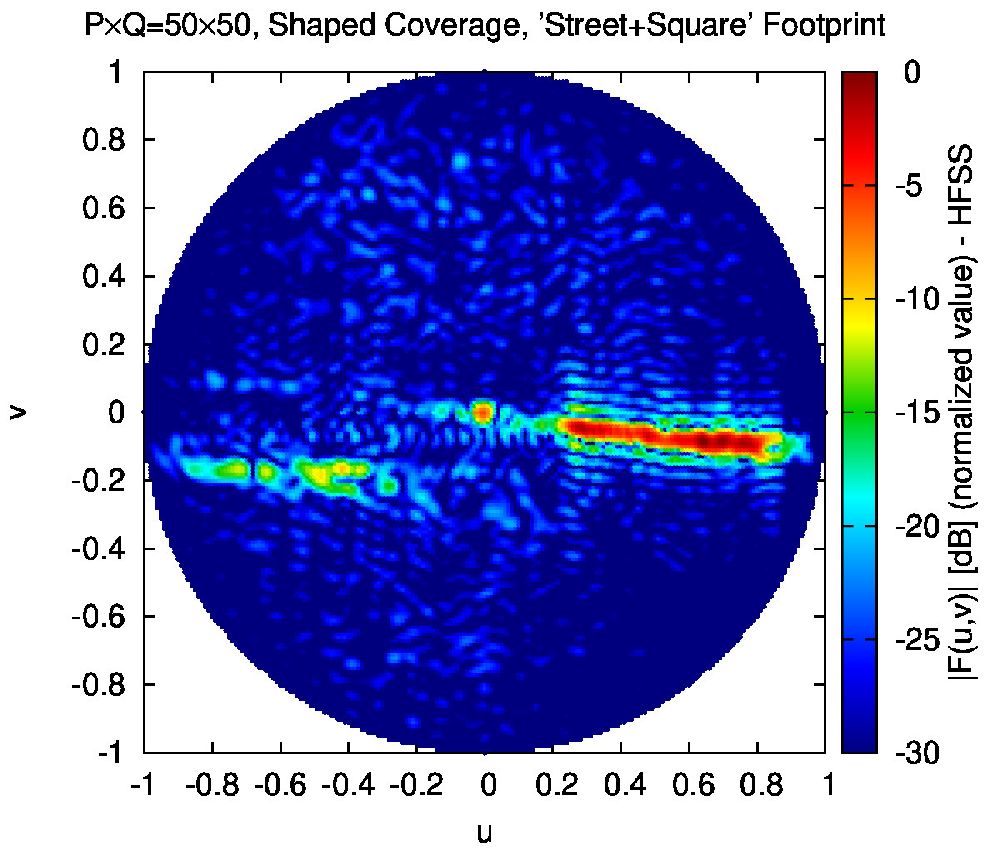}\tabularnewline
(\emph{c})&
(\emph{d})\tabularnewline
\end{tabular}\end{center}

\begin{center}~\vfill\end{center}

\begin{center}\textbf{Fig. 11 - G. Oliveri et} \textbf{\emph{al.}}\textbf{,}
{}``Building a Smart \emph{EM} Environment - \emph{AI}-Enhanced Aperiodic
...''\end{center}
\newpage

\begin{center}\begin{sideways}
\begin{tabular}{cc}
\includegraphics[%
  clip,
  width=0.70\columnwidth,
  keepaspectratio]{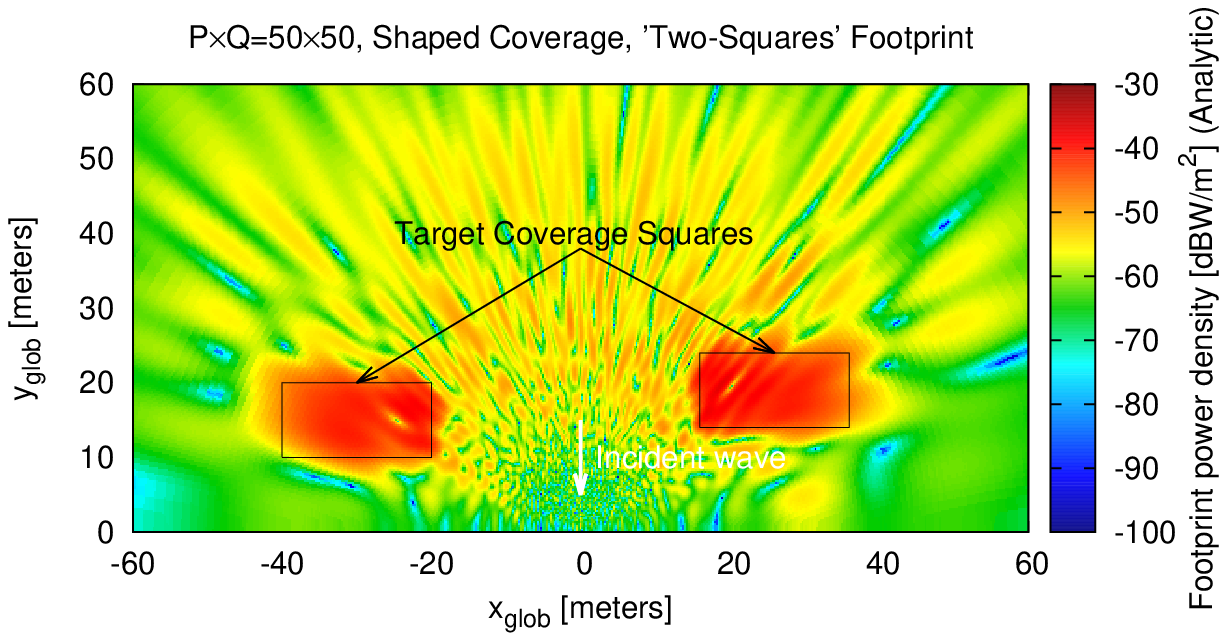}&
\includegraphics[%
  clip,
  width=0.70\columnwidth,
  keepaspectratio]{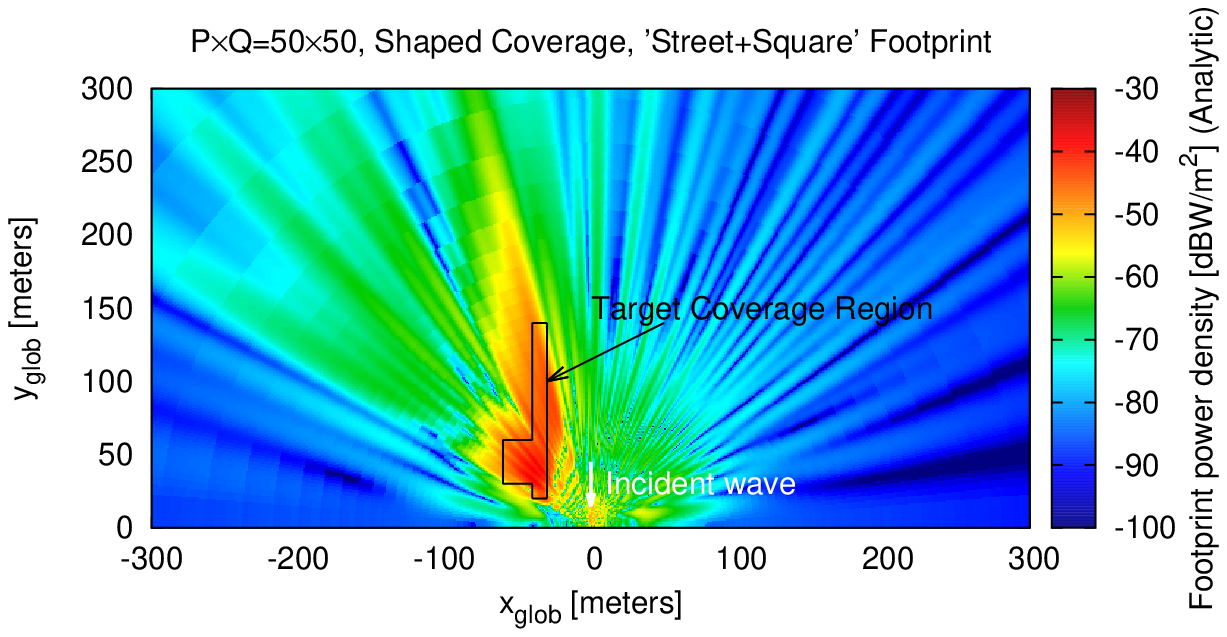}\tabularnewline
(\emph{a})&
(\emph{b})\tabularnewline
\includegraphics[%
  clip,
  width=0.70\columnwidth,
  keepaspectratio]{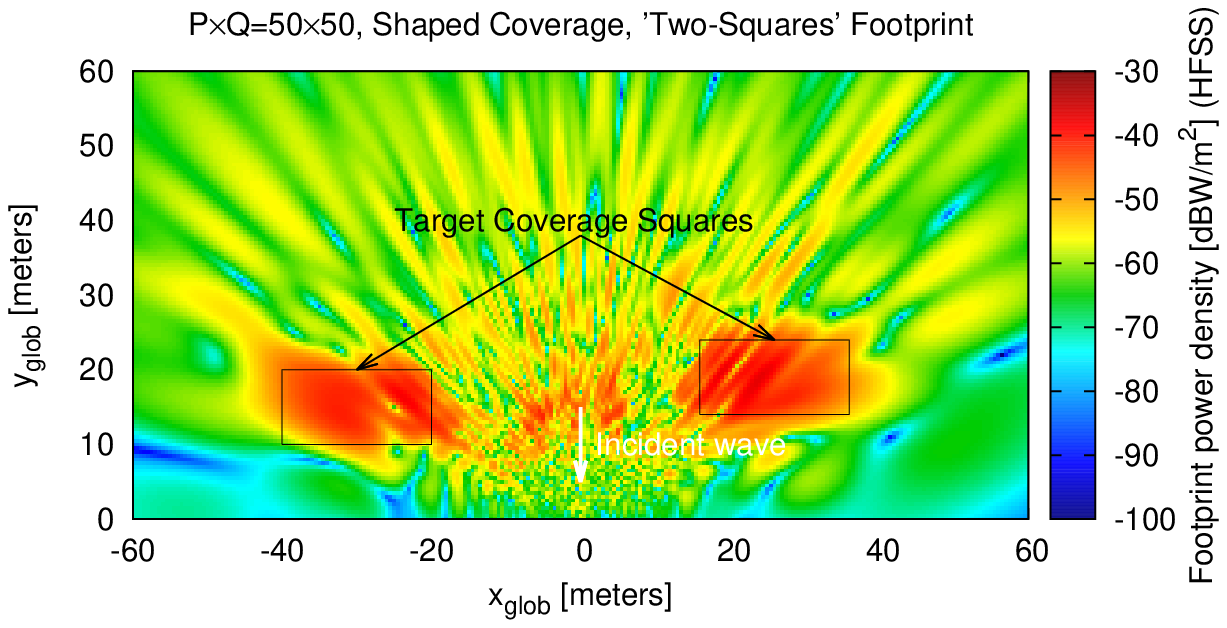}&
\includegraphics[%
  clip,
  width=0.70\columnwidth,
  keepaspectratio]{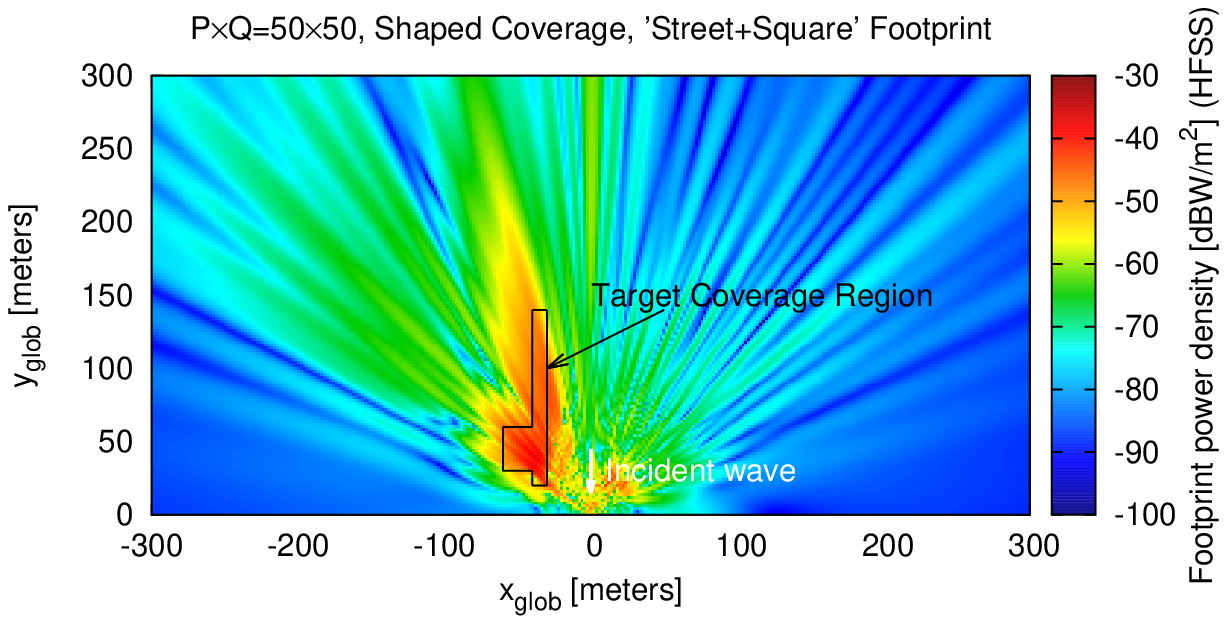}\tabularnewline
(\emph{c})&
(\emph{d})\tabularnewline
\end{tabular}
\end{sideways}\end{center}

\begin{center}\textbf{Fig. 12 - G. Oliveri et} \textbf{\emph{al.}}\textbf{,}
{}``Building a Smart \emph{EM} Environment - \emph{AI}-Enhanced Aperiodic
...''\end{center}
\newpage

\begin{center}~\vfill\end{center}

\begin{center}\begin{tabular}{c}
\includegraphics[%
  clip,
  width=0.70\columnwidth,
  keepaspectratio]{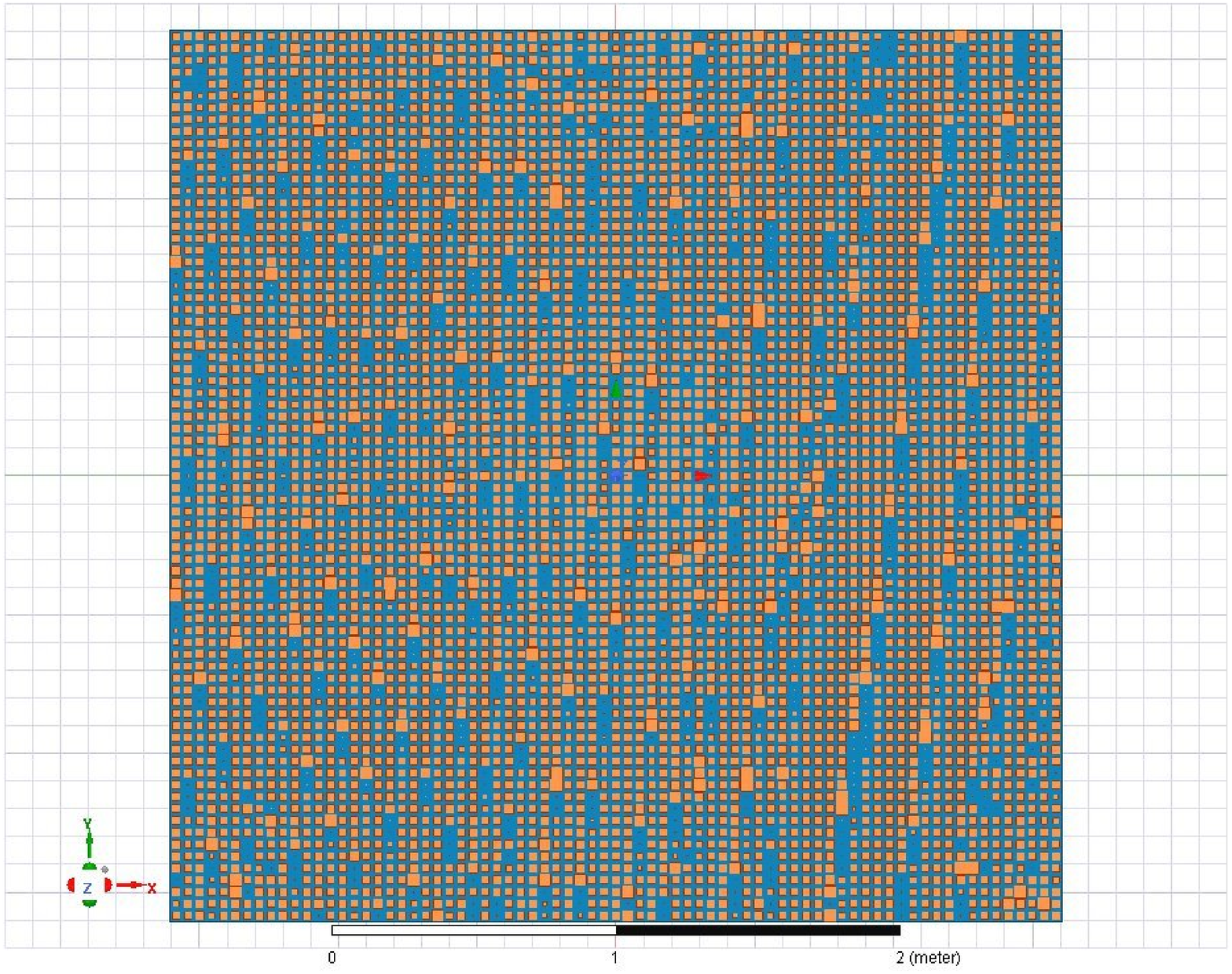}\tabularnewline
(\emph{a})\tabularnewline
\tabularnewline
\includegraphics[%
  clip,
  width=0.80\columnwidth,
  keepaspectratio]{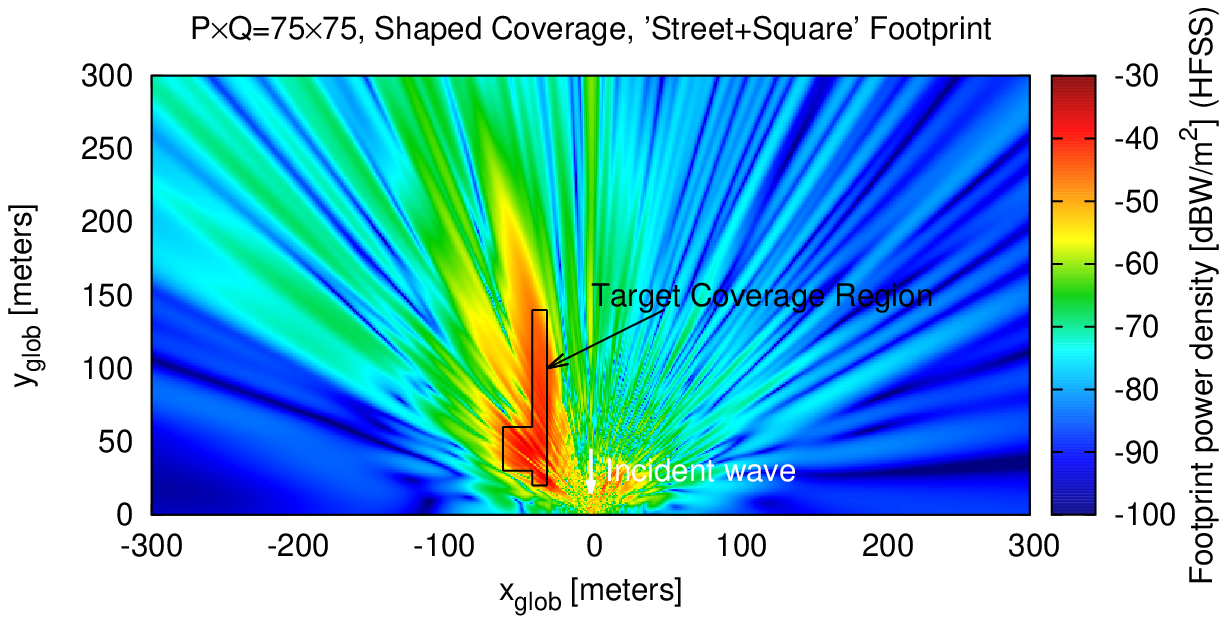}\tabularnewline
(\emph{b})\tabularnewline
\end{tabular}\end{center}

\begin{center}~\vfill\end{center}

\begin{center}\textbf{Fig. 13 - G. Oliveri et} \textbf{\emph{al.}}\textbf{,}
{}``Building a Smart \emph{EM} Environment - \emph{AI}-Enhanced Aperiodic
...''\end{center}
\newpage

\begin{center}~\vfill\end{center}

\begin{center}\begin{tabular}{c}
\includegraphics[%
  clip,
  width=0.70\columnwidth,
  keepaspectratio]{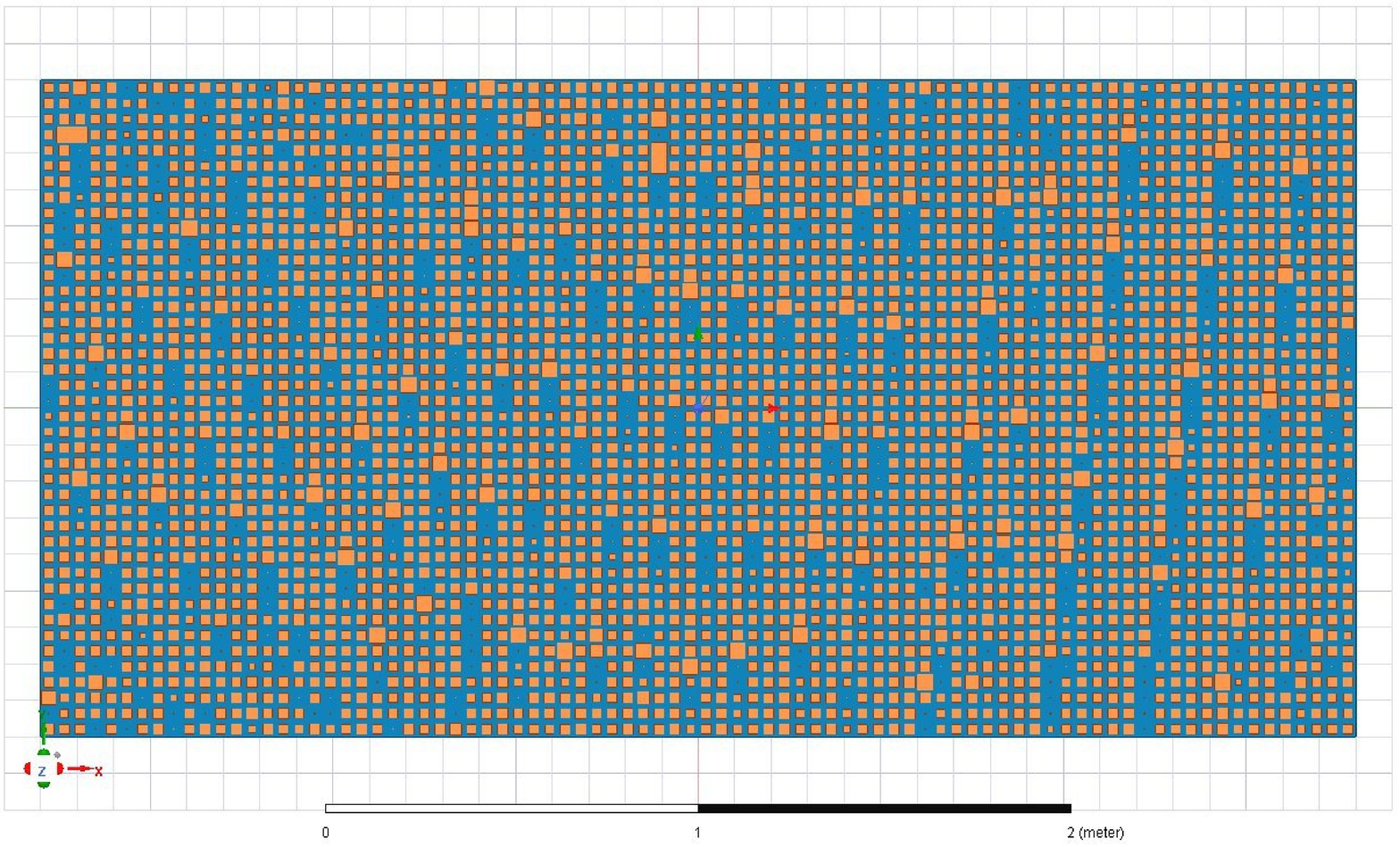}\tabularnewline
(\emph{a})\tabularnewline
\tabularnewline
\includegraphics[%
  clip,
  width=0.80\columnwidth,
  keepaspectratio]{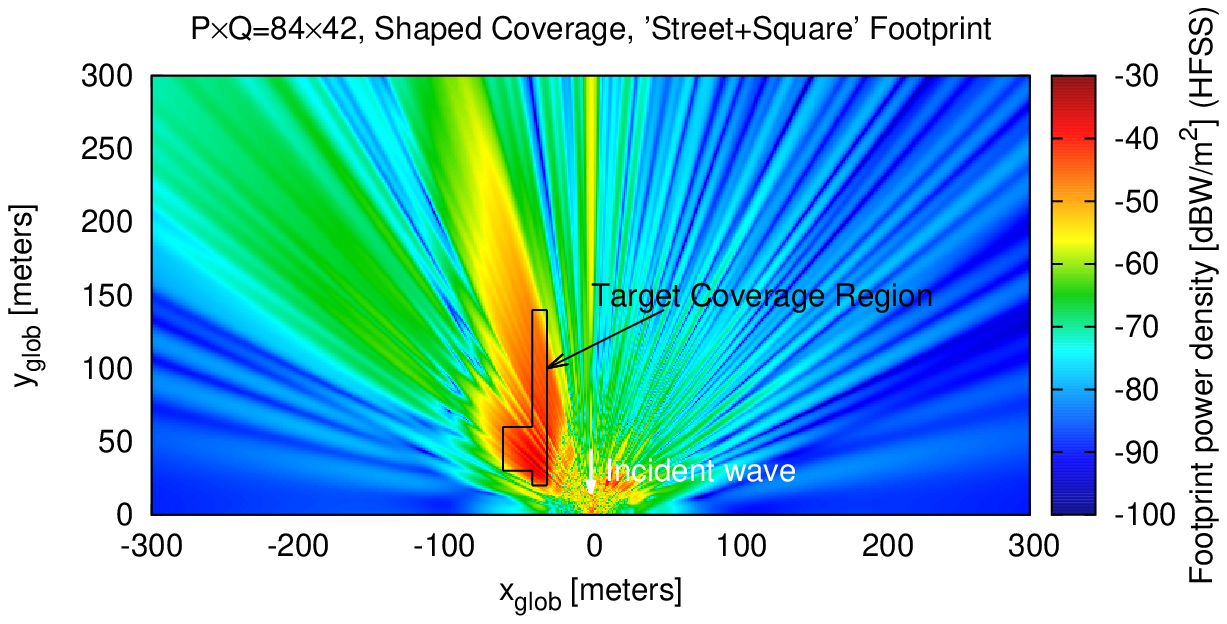}\tabularnewline
(\emph{b})\tabularnewline
\end{tabular}\end{center}

\begin{center}~\vfill\end{center}

\begin{center}\textbf{Fig. 14 - G. Oliveri et} \textbf{\emph{al.}}\textbf{,}
{}``Building a Smart \emph{EM} Environment - \emph{AI}-Enhanced Aperiodic
...''\end{center}
\newpage

\begin{center}~\vfill\end{center}

\begin{center}\begin{tabular}{c}
\includegraphics[%
  clip,
  width=0.70\columnwidth,
  keepaspectratio]{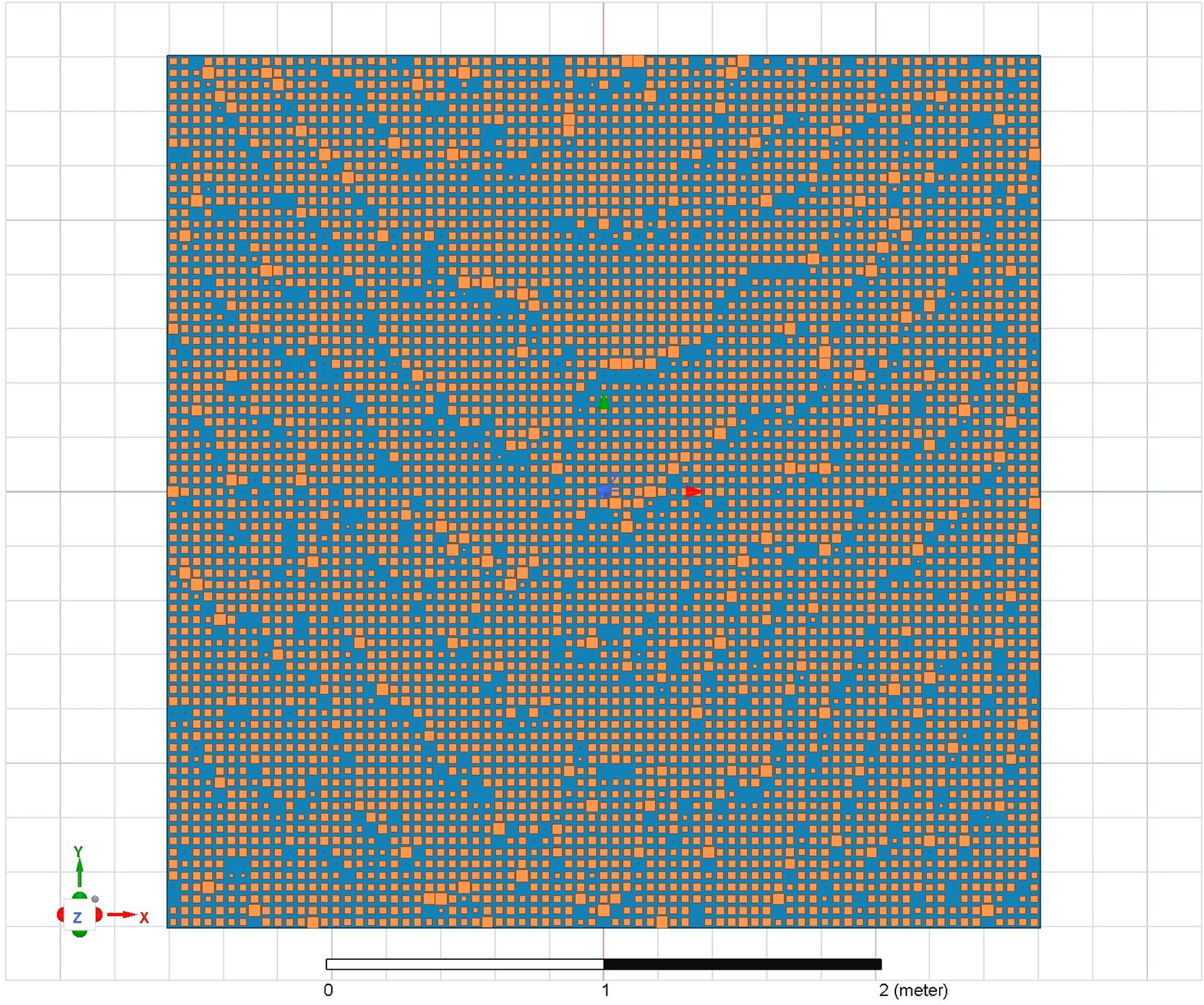}\tabularnewline
(\emph{a})\tabularnewline
\tabularnewline
\includegraphics[%
  clip,
  width=0.80\columnwidth,
  keepaspectratio]{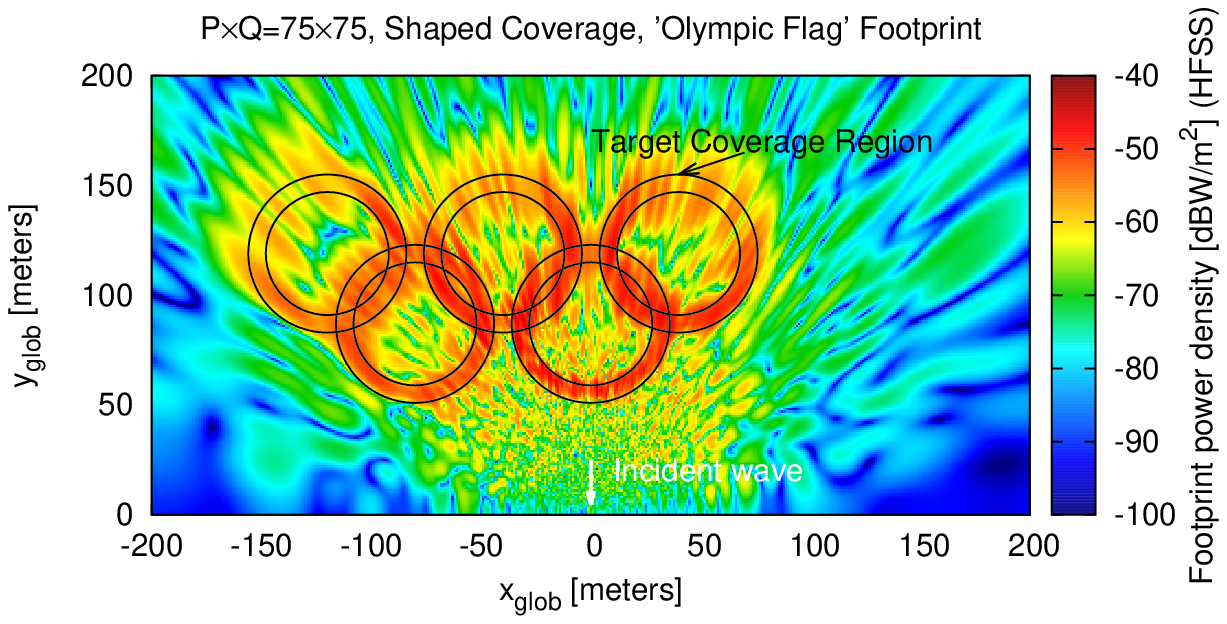}\tabularnewline
(\emph{b})\tabularnewline
\end{tabular}\end{center}

\begin{center}~\vfill\end{center}

\begin{center}\textbf{Fig. 15 - G. Oliveri et} \textbf{\emph{al.}}\textbf{,}
{}``Building a Smart \emph{EM} Environment - \emph{AI}-Enhanced Aperiodic
...''\end{center}
\end{document}